\documentclass[bimj,fleqn]{w-art}
\usepackage{times}
\usepackage{w-thm}
\usepackage[authoryear]{natbib}
 \usepackage{amsmath, amssymb}
 \usepackage{graphicx, subfigure} 
 \usepackage{enumerate}
 \usepackage{url, hyperref} 
 \usepackage{setspace}
\usepackage{xcolor}

\newcommand{\revisioncolor}{black}
\newcommand{\mytextcolor}[1]{\textcolor{\revisioncolor}{#1}}

\chardef\bslash=`\\ 

\newcommand{\bOmega} {\mbox{\boldmath $\Omega$}}

\newcommand{\bbeta} {\mbox{\boldmath $\beta$}}

\newcommand{\bSigma} {\mbox{\boldmath $\Sigma$}}

\newcommand{\bb} {\mbox{\boldmath $b$}}

\newcommand{\bx} {\mbox{\boldmath $x$}}

\newcommand{\bX} {\mathbf {X}}

\newcommand{\bS} {\mathbf {S}}

\newcommand{\calD}{\mathcal{D}}
\newcommand{\bzero} {\mbox{\boldmath $0$}}

\def\T{{ \mathrm{\scriptscriptstyle T} }}

\numberwithin{equation}{section}

\begin{document}

\DOIsuffix{bimj.200100000}
\Volume{xx}
\Issue{xx}
\Year{2020}
\pagespan{1}{}
\keywords{Beta distribution; Generalized biparabolic distribution; Linear predictor; Link function; Maximum likelihood\\[1pc]
\noindent\hspace*{-4.2pc} 
\textit{This reprint is in press at Biometrical Journal and may differ from the published version in typographic detail.}
}  

\title[Mode regression for bounded responses]{Parametric mode regression for bounded responses}

\author[Zhou {\it{et al.}}]{Haiming Zhou\footnote{Corresponding author: {\sf{e-mail: zhouh@niu.edu}}, Phone: +1-815-753-6714}\inst{,1}} 
\address[\inst{1}]{Department of Statistics and Actuarial Science, Northern Illinois University, DeKalb, Illinois 60115}
\author[]{Xianzheng Huang\inst{2}}
\address[\inst{2}]{Department of Statistics, University of South Carolina, Columbia, South Carolina 29208}
\author[]{For the Alzheimer's Disease Neuroimaging Initiative\footnote{	{\scriptsize{Data used in preparation of this article were obtained from the Alzheimer's Disease Neuroimaging Initiative (ADNI) database (adni.loni.usc.edu). As such, the investigators within the ADNI contributed to the design and implementation of ADNI and/or provided data but did not participate in analysis or writing of this report. A complete listing of ADNI investigators can be found at: \url{http://adni.loni.usc.edu/wp-content/uploads/how_to_apply/ADNI_Acknowledgement_List.pdf}.}}}\inst{}}

\Receiveddate{zzz} \Reviseddate{zzz} \Accepteddate{zzz} 

\begin{abstract}
We propose new parametric frameworks of regression analysis with the conditional mode of a bounded response as the focal point of interest. \mytextcolor{Covariate} effects estimation and prediction based on the maximum likelihood method under two new classes of regression models are demonstrated. We also develop graphical and numerical diagnostic tools to detect various sources of model misspecification. Predictions based on different central tendency measures inferred using various regression models are compared using synthetic data in simulations. Finally, we conduct regression analysis for data from the Alzheimer's Disease Neuroimaging Initiative to demonstrate practical implementation of the proposed methods. Supplementary materials that contain technical details, and additional simulation and data analysis results are available online.
\end{abstract}
 
\maketitle

\section{Introduction}
\begin{figure} 
	\centering
	\setlength{\linewidth}{0.5\textwidth}
	\includegraphics[width=\linewidth]{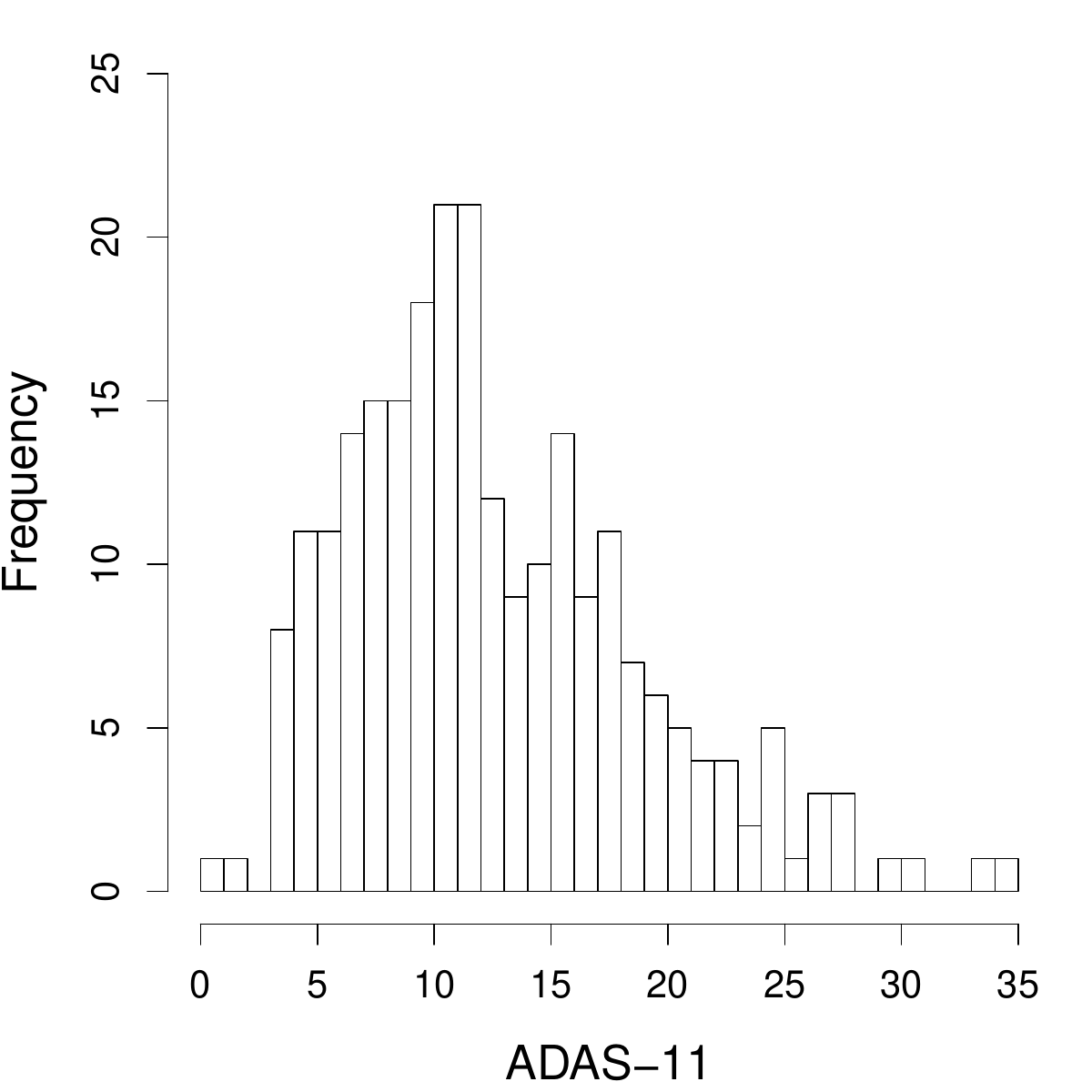} 	
	\caption{Histogram of ADAS-11 scores at month-12 of subjects in ADNI Phase 1 study.}
	\label{f:scores}
\end{figure}

The statistical models and methodology presented in this article are motivated by the Alzheimer's Disease Neuroimaging Initiative (ADNI) launched in 2003 and led by Principal Investigator Michael W. Weiner, MD. It is an ongoing study with a public-private partnership in the United States and Canada that gathers and analyzes thousands of subjects' brain scans, genetic profiles, and biomarkers in blood and cerebrospinal fluid. The main goal of ADNI is to understand relationships among the clinical, cognitive, imaging, genetic and biochemical biomarker characteristics of the entire spectrum of Alzheimer's diseases (AD). Ultimately, the hope is to achieve early detection of AD in preparation for early intervention of the disease progression, and also to help recruiting appropriate individuals in clinical trials. Clinical outcomes for assessing one's cognitive function in ADNI are bounded scores from well-established neuropsychological tests, such as the Alzheimer's disease assessment scale \citep[ADAS,][]{rosen1984new,  kueper2018alzheimer}, mini-mental state examination \citep{tombaugh1992mini}, and Rey auditory verbal learning test \citep{schmidt1996rey}. Distributions of these test scores from the ADNI cohort are typically heavy-tailed and skewed. As an example, Figure~\ref{f:scores} presents histogram of the ADAS-cognition sub-scale scores, also referred to as ADAS-11, of subjects at month 12 who were diagnosed with late mild cognitive impairment (LMCI) when they entered the ADNI Phase 1 study.

In order to effectively reveal the association between one's cognitive skill and potential influential biomarkers, we formulate parametric mode regression models tailored for heavy-tailed, skewed, and bounded response data, with efficient prediction as our goal of statistical inference besides identifying influential biomarkers for AD. Under the nonparametric framework, \citet{NIPS2017_6743} carried out mode regression analysis of ADNI data to predict cognitive impairment using neuroimaging data. They noted that mean regression analysis for studying the association between the cognitive assessment of an individual and the individual's neuroimaging features failed to yield scientifically meaningful results due to the heavy-tailed and skewed noise presented in data typically arising in this application. In addition to biomedical applications, mode regression for association study has been routinely used in econometrics \citep{lee1989mode, lee1993quadratic, kemp2012regression, damien2017bayesian}, astronomy \citep{bamford2008revealing}, and traffic engineering \citep{einbeck2006modelling}. \citet{kemp2012regression} argued that the mode is the most intuitive measure of central tendency for positively skewed data found in many econometric applications such as wages, prices, and expenditures. More generally, the conditional mode serves as a more informative summary for associations between a response $Y$ and covariates $\bX$ than the conditional mean or median when the distribution of $Y$ given $\bX$ is heavy-tailed or skewed. When comparing with predictions based on conditional means or medians, predictions based on conditional modes can provide more meaningful estimated outcomes. \citet{yao2014new} showed that, when the interval width is fixed, a mode-based prediction interval tends to have a higher coverage probability than a mean-based prediction interval. 

Most existing works on mode regression involve nonparametric components \citep{yao2014new,chen2016nonparametric,zhang2013robust,zhao2014robust,liu2013robust,yang2014robust}. These nonparametric and semiparametric approaches are developed under the frequentist framework. A few developments under the Bayesian framework include \citet{yu2012bayesian} and \citet{damien2017bayesian}.  Even though nonparametric methods and semiparametric methods can protect against misleading inference caused by inadequate parametric assumptions, often at the price of low statistical efficiency, it is not unreasonable to believe that an inference procedure may still provide reliable inference for the mode of a distribution even when certain aspects, such as tails, of the distribution are not well estimated by this procedure \citep{hall1992global, zhou2018bandwidth}. Hence, with the potential gain in efficiency, parametric regression models can be useful in studying the association between a response and covariates via inferring the conditional mode. 

Assuming a unimodal conditional distribution for the response, we formulate in Section~\ref{s:modelslkh} two new classes of mode regression models for a bounded response. Bounded response data are ubiquitous in practice, with the ADAS-11 score as one example. Other examples include rates or proportions, such as a disease prevalence, the fraction of household income spent on food, and the proportion of food and hygienic waste in residential solid waste. Although, technically, one can often map a bounded response to a new response whose support is the entire real line, say, via a logit transformation for a rate response, and then carry out regression analysis on the new response, it is practically more appealing to directly study the association between the original response and covariates. This practical consideration, along with the observation that many proportion responses encountered in practice are asymmetrically distributed, motivated the beta regression model proposed by \citet{ferrari2004beta}, with the mean of the beta distribution depending on covariates. \citet{smithson2006better} followed a similar strategy to formulate a regression model for a response bounded on [0, 1], where they specified a mean model and a variance model as functions of two sets of covariates separately. They later generalized this regression model by using a mixture of beta distributions \citep{verkuilen2012mixed}. Starting from a beta regression model, \citet{guolo2014beta} incorporated the serial dependence between responses via a Gaussian copula to model time series data bounded on the unit interval. Following the construction of their regression models, all the aforementioned works carry out frequentist inference, mostly based on maximum likelihood. Under the Bayesian inferential framework, \citet{bayes2012new} replaced the beta distribution with the beta rectangular distribution, defined as the mixture of a beta distribution and a uniform distribution, to achieve more robustness to outliers of a proportion response. \citet{figueroa2013mixed} introduced mixed Bayesian regression models by incorporating random effects in the linear predictor when specifying the mean function. Also considering Bayesian regression analysis for bounded data, \citet{migliorati2018new} proposed a flexible \mytextcolor{beta} distribution for the response given covariates based on a special mixture of two beta distributions to balance between flexibility and tractability.  Unlike all the above regression models which focus on inferring the conditional mean of a bounded response, \citet{bayes2017quantile} developed quantile regression models for bounded responses built upon on beta distributions. \citet{barrientos2012fully} took on a fully nonparametric Bayesian approach to model the covariates-dependent distribution of a bounded response. One major feature of our work that distinguishes it from these existing works on regression analysis for bounded data is that the conditional mode is the focal point of inference. This very key feature motivates our construction of the regression models presented in Section~\ref{s:modelslkh}

Following the model formulation, we outline maximum likelihood estimation of parameters in these models in Section ~\ref{s:modelslkh}. In Section~\ref{s:diag} we propose graphical and numerical diagnostics methods for detecting various sources of model misspecification when one draws inference based on an assumed model in the two proposed families. Section~\ref{s:empirical} presents simulation studies where we carry out mode regression analysis using these assumed models based on data generated from models that may or may not belong to the two families. In these simulation experiments, we report maximum likelihood estimates (MLEs) for covariate effects, operating characteristics of the proposed diagnostics methods, and prediction intervals constructed based on the proposed mode regression models. In Section~\ref{s:realdata} we carry out mean and mode regression analysis for a dataset from ADNI. Finally, we summarize contributions of this work and discuss follow-up research in Section~\ref{s:discussion}.

\section{Two families of regression models and maximum likelihood estimation}
\label{s:modelslkh}
\subsection{Regression models}
\label{s:models}
Without loss of generality, we assume from now on that the response variable has support on $[0,1]$, since any other bounded support can be rescaled to the unit interval. Inspired by the existing mean and quantile regression models for bounded data originating from beta regression, we first formulate a beta mode regression model for a bounded response. Recall that, for a random variable $V$ that follows a beta distribution, its probability density function (pdf) is given by 
\begin{align}
f_{\hbox {\tiny beta}}(v; \alpha_1, \alpha_2)=\frac{\Gamma(\alpha_1+\alpha_2)}{\Gamma(\alpha_1)\Gamma(\alpha_2)}v^{\alpha_1-1}
(1-v)^{\alpha_2-1}, \, \textrm{ for $v\in [0, 1]$}, 
\label{eq:betapdf}
\end{align}
where $\Gamma(t)$ is the gamma function, $\alpha_1$ and $\alpha_2$ are two positive shape parameters. When $\alpha_1, \alpha_2>1$, there is a unique mode for the beta distribution given by $\theta=(\alpha_1-1)/(\alpha_1+\alpha_2-2)$. Directly including the mode in the parameterization makes it more convenient to draw inference for the mode. For this reason, we set $\alpha_1=1+m \theta$ and $\alpha_2=1+m (1-\theta)$, where $m>0$. This parameterization not only signifies the parameter of central interest, $\theta$, but also makes $\alpha_1$ and $\alpha_2$ larger than one, ensuring the existence of a unique mode. The variance of the beta distribution under this parameterization is $(1+m\theta)\{1+m(1-\theta)\}/\{(2+m)^2(3+m)\}$, suggesting a smaller variance as $m$ increases. 

\begin{figure} 
	\centering
		\setlength{\linewidth}{0.5\textwidth}
	\includegraphics[width=\linewidth]{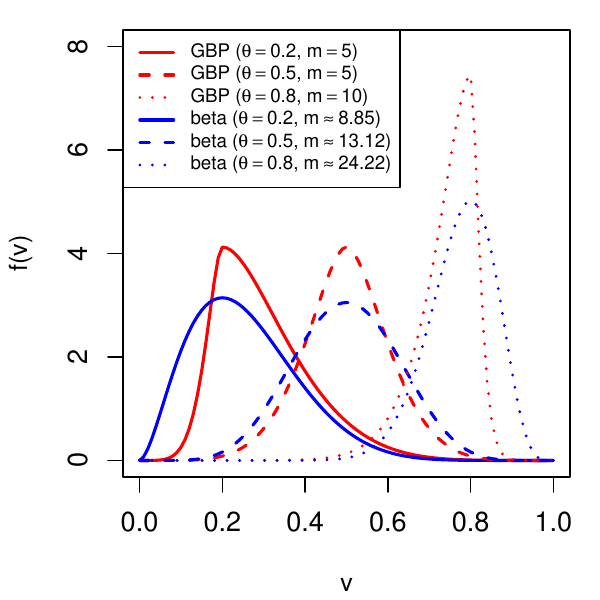} 	
	\caption{\label{f:GBPBeta}Probability density functions of GBP distributions (red lines) with $(\theta, m)=(0.2, 5)$ (solid line), (0.5, 5) (dashed line), and (0.8, 10) (dotted line), respectively, and beta density curves (blue lines) with the same mode and variance as those of the GBP densities depicted in the same line type.}
\end{figure}

With the mode as our choice of central tendency measure for the bounded response, it is desirable to include the mode as one of the canonical parameters in the response  distribution without additional reparameterization as we do above for a beta distribution. To the best of our knowledge, the family of generalized biparabolic distributions \citep[GBP,][]{garcia2009generalized} is the only named distribution family that, first, are defined on a bounded support, second, includes  symmetric and asymmetric distributions, and third, has the mode as the sole location parameter appearing in the pdf. More specifically, if $V$ follows a GBP distribution on the support $[0, 1]$, its pdf is given by 
\begin{align}
f_{\hbox {\tiny GBP}}(v; \theta, m)=\frac{(2m+1)(m+1)}{(3m+1)}d^m (2-d^m), 
\label{eq:GBPpdf}
\end{align}
where $m$ is a positive shape parameter, and $d=I(0< v \le \theta)v/\theta+I(\theta< v \le 1)(1-v)/(1-\theta)$, in which $\theta\in (0, 1)$ is the mode of the distribution, and $I(\cdot)$ is the indicator function. A larger $m$ leads to a GBP distribution more concentrated around the mode with a smaller variance. Figure~\ref{f:GBPBeta} depicts three GBP density functions, in comparison with three beta density functions that share the same mode and variance as the corresponding depicted GBP distributions. This figure shows the general pattern that, with the mode and variance fixed, a GBP density displays a sharper drop toward zero on both sides of the mode than that for a beta density.  

To complete the formulation of a regression model, we assume that, given $\bX$, the mode of $Y$ relates to a linear predictor $\eta(\bX)=\beta_0+\bbeta_1^\T \bX$ via a link function $g(t)$, that is, 
\begin{align}
\textrm{Mode}(Y|\bX)=\theta(\bX)=g\{\eta(\bX)\}.\label{eq:Mx}
\end{align}
Commonly employed link functions include logit, probit, log-log, and complementary log-log. 
To this end, we have two regression models for $Y$ once a link function is chosen, written succinctly as 
\begin{align*}
Y|\bX & \sim \textrm{\mytextcolor{beta}}(1+m \theta(\bX), \, 1+m \{1-\theta(\bX)\}), \\
Y|\bX & \sim \textrm{GBP}(\theta(\bX), \, m), 
\end{align*}
which are henceforth referred to as the beta mode model and the GBP mode model, respectively. Note that neither regression model is a generalized linear model \citep{mccullagh2019generalized} because neither (\ref{eq:betapdf}) nor (\ref{eq:GBPpdf}) is in the form of a density corresponding to an exponential dispersion distribution \citep{jorgensen1987exponential} when they are parameterized via $\theta$ and $m$ as described above. Both regression models allow heteroscedasticity in the sense that the conditional variance of $Y$ depends on covariates. For example, under the beta mode model, $\textrm{Var}(Y|\bX)=\{1+m\theta(\bX)\}[1+m\{1-\theta(\bX)\}]/\{(2+m)^2(3+m)\}$. 

\subsection{Maximum likelihood estimation}
Given a random sample of size $n$, $\calD=\{(Y_i, \bX_i), i=1,\ldots, n\}$, the log-likelihood function associated with a beta mode model is 
\begin{align*}
\ell_{\hbox {\tiny beta}}(\bOmega; \calD) &= n \log \Gamma(2+m)-\sum_{i=1}^n \log\left(\Gamma\left\{1+m \theta(\bX_i) \right\} \Gamma\left[1+m\left\{1-\theta(\bX_i)\right\}\right]\right)\\
& \quad +m\sum_{i=1}^n \left[ \theta(\bX_i) \log Y_i +\left\{1-\theta(\bX_i)\right\}\log (1-Y_i)\right].
\end{align*}
Maximizing $\ell_{\hbox {\tiny beta}}(\bOmega; \calD)$ with respect to $\bOmega=(\bbeta^\T, m)^\T$ yields the MLE for $\bOmega$ under this model, where $\bbeta=(\beta_0, \bbeta_1^\T)^\T$. Because the beta family is an exponential family, the corresponding likelihood function is concave, suggesting the existence of a unique MLE for $\bOmega$. Furthermore, regularity conditions required for the MLE to be consistent and asymptotically normal can also be easily verified for this regression model. 

When a GBP mode model is assumed, by (\ref{eq:GBPpdf}), the log-likelihood function is given by 
\begin{equation} 
\ell_{\hbox {\tiny GBP}}(\bOmega; \calD)=n\log\left\{\frac{(2m+1)(m+1)}{3m+1}\right\}+m\sum_{i=1}^n \log d_i +\sum_{i=1}^n \log\left(2-d_i^m\right),
\label{eq:loglkhGBP}
\end{equation}
where $d_i = I\{0<Y_i \le \theta(\bX_i)\}\{Y_i/\theta(\bX_i)\}+I\{\theta(\bX_i)< Y_i < 1\}\{(1-Y_i)/\{1-\theta(\bX_i)\}$. Maximizing $\ell_{\hbox {\tiny GBP}}(\bOmega; \calD)$ with respect to $\bOmega$ yields the MLE for $\bOmega$ under the GBP mode regression model. Unlike the beta family, the GBP  family is not an exponential family. For simplicity, let us assume $m$ known in (\ref{eq:GBPpdf}) and focus on the density as a function of $\theta$ for now. It can be shown that $\lim_{\theta\to v^+} (\partial^2/\partial\theta^2)\log f_{\hbox {\tiny GBP}}(v; \theta,m)=-2m^2/v^2$, whereas $\lim_{\theta\to v^-} (\partial^2/\partial\theta^2)\log f_{\hbox {\tiny GBP}}(v; \theta,m)=-2m^2/(1-v)^2$, indicating that the Hessian function is discontinuous at any realization of the distribution except for $v=0.5$. It can also be shown that, the GBP log-likelihood is concave in a neighborhood of the truth almost surely. Moreover, regularity conditions \citep[][page 281]{cox1979theoretical} for the consistency of MLE as the maximizer of (\ref{eq:loglkhGBP}) are satisfied for the GBP regression model, but additional conditions needed to establish asymptotic normality for MLE are not.

\section{Model diagnostics}
\label{s:diag}
Basing statistical inference on a specific parametric model raises the concern of model misspecification that can lead to misleading inference results. This concern motivates the diagnosis tools we develop in this section. 

\subsection{Graphical diagnosis}
\label{s:envy}
Half-normal residual plots with simulated envelopes \citep{Atkinson1987} are useful graphical tools for checking the goodness-of-fit of a model with complex response distributions. Let $\hat{\mu}(\bx)$ and $\hat{\sigma}^2(\bx)$ denote the MLEs for the mean and variance of $Y$ given $\bX=\bx$, respectively, resulting from an assumed regression model. Define the absolute standardized residual as $r_i = |Y_i  - \hat{\mu}(\bX_i)|/\hat{\sigma}(\bX_i)$. Given data $\calD=\{(Y_i, \bX_i), i=1,\ldots, n\}$, the algorithm below describes how to obtain a half-normal residual plot with a simulated envelope. 
\begin{description}
	\item[{\it Step 1}] Fit the assumed regression model to data $\calD$, calculate the absolute standardized residuals, then order the residuals from smallest to largest, denoted as $\{r_{(i)}, i=1, \ldots, n\}$.  Plot $r_{(i)}$ against the half-normal quantile $q_i = \Phi^{-1}\{ (i+n-0.125)/(2n+0.5)\}$, for $i=1,\ldots, n$, where $\Phi(\cdot)$ is the cumulative distribution function of $N(0, 1)$.
	\item[{\it Step 2}] For $k=1, \ldots, \mytextcolor{K}$, conditioning on $\bX_i$, generate a new response $Y_i^{*(k)}$ from the estimated regression model resulting from {\it Step 1}, for $i=1, \ldots, n$. This produces new data $\calD^{*(k)}=\{(Y_i^{*(k)}, \bX_i), i=1,\ldots, n\}$, for $k=1, \ldots, \mytextcolor{K}$. 
	\item[{\it Step 3}] For $k=1, \ldots, \mytextcolor{K}$, fit the assumed regression model to data $\calD^{*(k)}$ and calculate the ordered absolute standardized residuals $\{r_{(i)}^{*(k)}, i=1, \ldots, n \}$. 
	\item[{\it Step 4}] Compute $r_i^{\hbox {\tiny $L$}} = \min_{1\leq k\leq \mytextcolor{K}}\{r_{(i)}^{*(k)}\}$ and $r_i^{\hbox {\tiny $U$}} = \max_{1\leq k\leq \mytextcolor{K}}\{r_{(i)}^{*(k)}\}$, for $i=1, \ldots, n$. Plot points $\{(x, y): (q_i, r_i^{\hbox {\tiny $L$}}), i=1, \ldots, n\}$ and $\{(x, y): (q_i, r_i^{\hbox {\tiny $U$}}), i=1, \ldots, n\}$ on the same plot obtained in {\it Step 1} to form the envelope. 
\end{description}
\mytextcolor{Here we set $K=19$ as suggested by \cite{Atkinson1987}. This way, the resulting envelope has a probability approximately equal to $0.95$ to cover the ordered residuals $r_{(i)}$ obtained from the original data if the assumed model agrees with the true model.} A substantially larger proportion of residuals falling outside the envelope indicates a lack-of-fit of the assumed model. 

\subsection{Score tests for model diagnosis}
\label{s:test}
To assess the adequacy of an assumed regression model quantitatively, we develop tests using score functions constructed based on matching moments. The proposed score tests exploit certain moments of the response variable or functions of it that are special in some way so that they are difficult to be estimated well via maximizing a misspecified likelihood function. 

When the assumed model is a beta mode model, we construct a bivariate score function based on the following results relating to a beta random variable $V$, 
\begin{align*}
E(\log V)& =\psi(\alpha_1)-\psi(\alpha_1+\alpha_2), \\
E(V\log V)& =\frac{\alpha_1\{\psi(\alpha_1+1)-\psi(\alpha_1+\alpha_2+1)\}}{\alpha_1+\alpha_2},
\end{align*}
where $\psi(t)=\{(d/dt)\Gamma(t)\}/\Gamma(t)$ is the digamma function. Matching these two expectations with their sample counterparts, we formulate the following score function evaluated at $(Y_i, \bX_i)$ for model diagnosis when the assumed regression model is a beta mode model, 
\begin{equation}
\bS_{i,\hbox{\tiny beta}}(\bOmega)=
\begin{bmatrix}
\log Y_i - \psi\{1+m\theta(\bX_i)\}+\psi(2+m)\\
\displaystyle{Y_i \log(Y_i)-\frac{\{1+m \theta(\bX_i)\}\left[\psi\{2+m\theta(\bX_i)\}-\psi(3+m)\right]}{2+m}}
\end{bmatrix}.
\label{eq:betascore}
\end{equation}

If the assumed model is a GBP mode model, we formulate a bivariate score function based on matching the first two moments of $Y|\bX\sim \textrm{GBP}(\theta(\bX), \,m)$ \citep{{garcia2009generalized}}, 
\begin{align*}
E(Y|\bX)&=\frac{6m^2\theta(\bX)+7m+2}{6m^2+14m+4}, \\
\textrm{Var}(Y|\bX)& =\left\{4(3m+1)^2(m+2)^2(2m+3)(m+3)\right\}^{-1} \left[4m^2(37m^2+61m\right.\\
& \quad \left. +10)\theta(\bX)\{\theta(\bX)-1\}+82m^4+247m^3+247m^2+96m+12\right].
\end{align*}
That is, the score function evaluated at $(Y_i, \bX_i)$ for assessing the adequacy of an assumed GBP mode model is 
\begin{equation}
\bS_{i, \hbox{\tiny GBP}}(\bOmega)= 
\begin{bmatrix}
Y_i-E(Y_i|\bX_i)\\
Y_i^2-\textrm{Var}(Y_i|\bX_i)-\{E(Y_i|\bX_i)\}^2
\end{bmatrix}.
\label{eq:gbpscore}
\end{equation}

Generically denote by $\bS_i(\bOmega)$ the score function in (\ref{eq:betascore}) or (\ref{eq:gbpscore}), depending on whether one assumes a beta mode model or a GBP mode model. We mimic the Hotelling's $T^2$ statistic \citep{hotelling1931} to define the following test statistic, 
\begin{equation}
Q(\hat \bOmega; \calD)=\frac{n-2}{2(n-1)}\overline \bS^\T \hat \bSigma^{-1} \overline \bS,
\label{eq:teststat}
\end{equation}
where $\hat \bOmega$ is the MLE of $\bOmega$ under the assumed model, $\overline \bS=n^{-1}\sum_{i=1}^n \bS_i(\hat\bOmega)$, and $\hat \bSigma=\{n(n-1)\}^{-1}\sum_{i=1}^n\{\bS_i(\hat\bOmega)-\overline \bS\}\{\bS_i(\hat\bOmega)-\overline \bS\}^\T$ is an estimator for the variance-covariace of $\overline \bS$. Under the null hypothesis that the assumed model is the true model, one has $E(\overline \bS)=\bzero$ when evaluating $\hat\bOmega$ at the truth, and thus a small value for $Q(\hat\bOmega; \calD)$ is expected under the null. In contrast, when the assumed model differs from the true model to the extent that $E(\overline \bS)$ substantially deviates from zero, a large realization of $Q(\hat\bOmega; \calD)$ is expected. According to \citet{hotelling1931}, if $\bS_i(\bOmega)$ is a bivariate normal random variable, then $Q(\bOmega; \calD)\sim F_{2, n-2}$ under the null. With a response supported on $[0, 1]$, a bivariate normal is not likely to approximate well the distributions of the scores in (\ref{eq:betascore}) and (\ref{eq:gbpscore}), although a large $Q(\hat\bOmega; \calD)$ still implies poor fit for relevant moments and thus casts  doubt on the assumed model. To accurately approximate certain percentiles of the null distribution of $Q(\hat\bOmega; \calD)$, we use a parametric bootstrap procedure that leads to an estimated $p$-value associated with the test statistic. The algorithm in supplementary Section S1 outlines the bootstrap procedure under the null stating that the true model is a GBP mode model. A similar bootstrap procedure is used to estimate the $p$-value of the test statistic when one assumes a beta mode model. 

Empirical evidence from simulation studies \mytextcolor{(supplementary Figure~S3)} suggest that this bootstrap procedure can estimate the tail of the null distribution of $Q(\hat \bOmega; \calD)$ well enough to preserve the right size of the proposed score tests. Besides how well one can estimate certain percentiles of a null distribution, operating characteristics of the score tests also depend on the extent of distortion on moment estimation when an inadequate model is assumed. More empirical evidence on this aspect are presented next, along with the performance of maximum likelihood estimation and predictions based on synthetic data generated from various regression models.

\section{Simulation study}
\label{s:empirical}
\mytextcolor{Source code to reproduce the results in this section is available as Supporting Information
on the journal's web page (\url{http://onlinelibrary.wiley.com/doi/xxx/suppinfo}).}
\subsection{Design of simulation experiments}
In all experiments, we simulate a bivariate covariate, $\bX=(X_1, X_2)^\T$, as the predictor in a regression model. When carrying out regression analysis, we assume a linear predictor, $\eta(\bX)=\beta_0+\beta_1 X_1+\beta_2 X_2$, and the logit link $g(t)=1/(1+e^{-t})$ in (\ref{eq:Mx}), despite the true data generating process. 

When conducting regression analysis assuming a beta mode model, we first simulate $X_2$ from $\textrm{Bernoulli}(0.5)$, and then generate data for $X_1$ according to $N(I(X_2=1)-I(X_2=0), \,1)$. Given  covariates data, responses are generated from each of the following four conditional distributions:
\begin{itemize}
\item[(B1)] $Y|\bX \sim \textrm{\mytextcolor{beta}}(1+m\theta(\bX), \, 1+m\{1-\theta(\bX)\})$, where $\theta(\bX)=1/[1+\exp\{-\eta(\bX)\}]$, with $\eta(\bX)=\beta_0+\beta_1 X_1+\beta_2 X_2$;
\item[(B2)] $Y|\bX \sim \textrm{\mytextcolor{beta}}(1+m\theta(\bX), \, 1+m\{1-\theta(\bX)\})$, where $\theta(\bX)=1/[1+\exp\{-\eta(\bX)\}]$, with $\eta(\bX)=\beta_0+\beta_1 X_1+\beta_2 X_1^2+ \beta_3 X_2$;
\item[(B3)] $Y|\bX \sim \textrm{\mytextcolor{beta}}(1+m\theta(\bX),\,1+m\{1-\theta(\bX)\})$, where $\theta(\bX)=0.5 \Phi[2\{\eta(\bX)+2\}]+0.5 \Phi[2\{\eta(\bX)-2\}]$, with $\eta(\bX)=\beta_0+\beta_1 X_1+ \beta_2 X_2$; 
\item[(B4)] $Y|\bX \sim \textrm{GBP}(\theta(\bX), \, m)$, where $\theta(\bX)=1/[1+\exp\{-\eta(\bX)\}]$, with $\eta(\bX)=\beta_0+\beta_1 X_1+\beta_2 X_2$.
\end{itemize}

When a GBP mode model is assumed for regression analysis, we consider the following four regression models according to which responses are generated after data for $X_1$ are simulated from $N(0, 1)$ and data for $X_2$ are simulated from $\textrm{Bernoulli}(0.5)$: 
\begin{itemize}
\item[(G1)] $Y|\bX\sim \textrm{GBP}(\theta(\bX), \, m)$, where $\theta(\bX)=1/[1+\exp\{-\eta(\bX)\}]$, with $\eta(\bX)=\beta_0+\beta_1 X_1+\beta_2 X_2$; 
\item[(G2)] $Y|\bX\sim \textrm{GBP}(\theta(\bX), \, m)$, where $\theta(\bX)=1/[1+\exp\{-\eta(\bX)\}]$, with $\eta(\bX)=\beta_0+\beta_1 X_1+\beta_2 X_1^2+\beta_3 X_2$; 
\item[(G3)] $Y|\bX\sim \textrm{GBP}(\theta(\bX), \, m)$, where $\theta(\bX)=0.5 \Phi[2\{\eta(\bX)+2\}]+0.5 \Phi[2\{\eta(\bX)-2\}]$, in which $\eta(\bX)=\beta_0+\beta_1 X_1+\beta_2 X_2$; 
\item[(G4)] $Y|\bX\sim \textrm{\mytextcolor{beta}}(1+m\theta(\bX), \, 1+m\{1-\theta(\bX)\})$, where $\theta(\bX)=1/[1+\exp\{-\eta(\bX)\}]$, with $\eta(\bX)=\beta_0+ \beta_1 X_1+\beta_2 X_2$.
\end{itemize}

\mytextcolor{These simulation settings are designed to cover a wide range of scenarios that are of theoretical and practical interest. For instance, cases (B1)--(B4) allow correlated covariates, whereas covariates in (G1)--(G4) are independent. More importantly, we include three sources of model misspecification in this experiment that are frequently discussed in the literature on parametric regression models. In particular,} (B1) and (G1) create scenarios where the assumed model coincides with the true model, and the other cases give rise to scenarios where we implement maximum likelihood estimation under a misspecified model. Under (B2) and (G2), the assumed models misspecify the linear predictor; under (B3) and (G3), the assumed models involve a misspecified link function for the mode; and under (B4) and (G4), the assumed conditional distribution of $Y$ given covariates is not from the same family that the true conditional distribution belongs to. 

\subsection{\mytextcolor{Covariate} effects estimation}
Given each of the above data generating processes, we generate data sets $\{(Y_i, X_{1i}, X_{2i})\}_{i=1}^n$ with $n=50, 100$. Under each simulation setting, we repeat maximum likelihood estimation using 300 Monte Carlo data sets. 

When the assumed model matches the true model, the MLE for $\bOmega$ is expected to be consistent estimator. Table~\ref{t:mles} provides summary statistics for these MLEs and the estimated standard deviations associated with these estimates based on data generated according to (B1) and (G1), respectively, with $m=10$ and $\bbeta=(1, 1, 1)^\T$, where the estimated standard deviations result from sandwich variance estimation for M-estimators \citep[][Section 7.2.1]{boos2013essential}. The close agreement between the MLE of $\bOmega$ and the truth, and the resemblance between the estimated standard deviations and the empirical standard deviations suggest that the first two moments of the asymptotic distribution of the MLEs are estimated reasonably well. 

\begin{table}[h]
\centering
\caption{\label{t:mles}\small{Averages of MLEs for parameters in a beta mode model and a GBP mode model across 300 Monte Carlo replicates generated according to (B1) and (G1), respectively, and averages of the corresponding estimated standard deviations ($\widehat{\textrm{s.d.}}$) in comparison with the empirical standard deviations (s.d.). Numbers in parentheses are $100\times$(Monte Carlo standard errors) associated with the averages. The true parameter values are $\bbeta=(\beta_0, \beta_1, \beta_2)^\T=(1, 1, 1)^\T$, and $\log m=\log 10\approx 2.303$.}}
\begin{tabular}{*{8}{c}}
\hline 
 & MLE & $\widehat{\textrm{s.d.}}$ & s.d. & & MLE & $\widehat{\textrm{s.d.}}$ & s.d. \\
\hline
(B1)	 & \multicolumn{3}{c}{$n=50$}      & &    \multicolumn{3}{c}{$n=100$}\\
				 \cline{2-4}  \cline{6-8} 
$\beta_0$ & 0.973 (1.40) & 0.225 (0.34) & 0.243 & & 1.011 (0.93) & 0.162 (0.15) & 0.161\\
$\beta_1$ & 0.983 (1.01) & 0.173 (0.27) & 0.174 & & 1.000 (0.80) & 0.123 (0.13) & 0.138\\
$\beta_2$ & 1.005 (2.49) & 0.411 (0.77) & 0.431 & & 0.982 (1.80) & 0.290 (0.34) & 0.311\\ 
$\log m$  & 2.413 (1.50) & 0.224 (0.04) & 0.260 & & 2.347 (0.83) & 0.159 (0.02) & 0.144\\
\hline
(G1)	 & \multicolumn{3}{c}{$n=50$}      & &  \multicolumn{3}{c}{$n=100$}\\
					 \cline{2-4}  \cline{6-8} 
$\beta_0$ & 0.994 (0.49) & 0.069 (0.12) & 0.085 & & 0.994 (0.31) & 0.050 (0.06) & 0.054\\
$\beta_1$ & 0.984 (0.44) & 0.063 (0.13) & 0.076 & & 0.993 (0.29) & 0.045 (0.07) & 0.049\\
$\beta_2$ & 0.986 (0.74) & 0.109 (0.19) & 0.128 & & 1.004 (0.47) & 0.079 (0.08) & 0.081\\ 
$\log m$  & 2.356 (0.81) & 0.142 (0.01) & 0.141 & & 2.322 (0.61) & 0.100 (0.01) & 0.105\\
\hline 
\end{tabular}
\end{table} 

\subsection{Performance of model diagnosis methods}\label{sec:sim:diagnosis}
\begin{figure} 
	\centering
	\setlength{\linewidth}{0.47\textwidth}
	\subfigure[]{ \includegraphics[width=\linewidth]{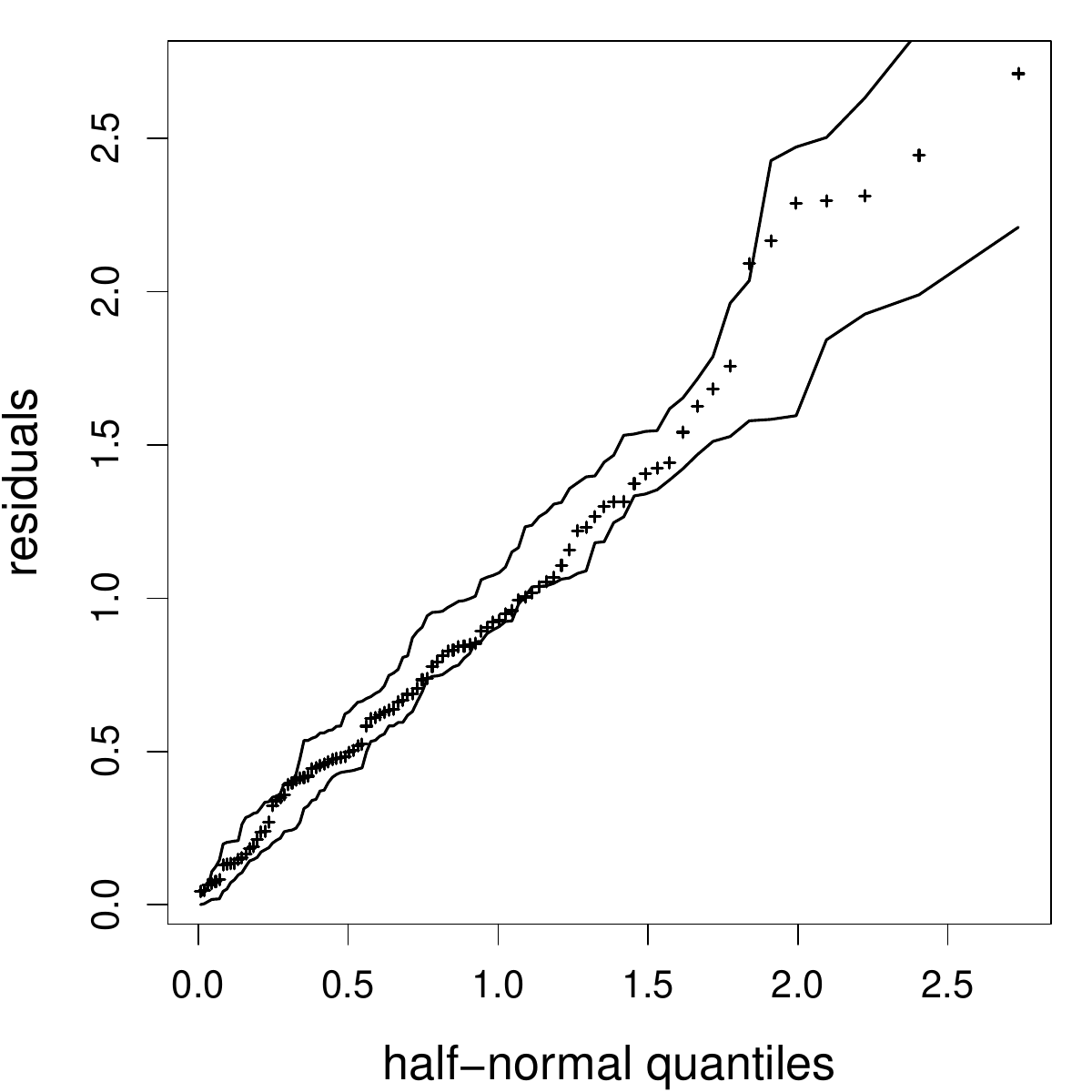} }
	\subfigure[]{ \includegraphics[width=\linewidth]{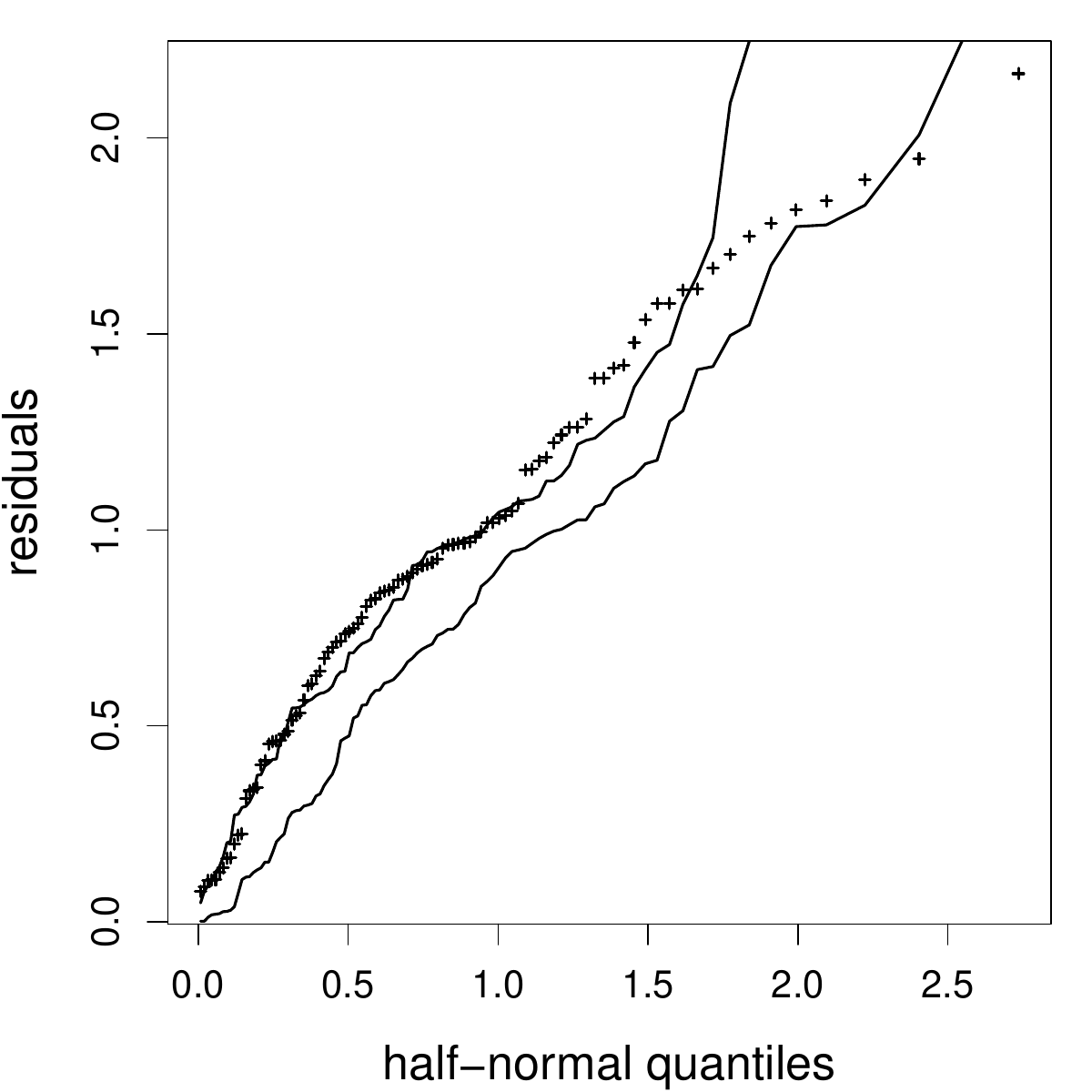} }\\
	\subfigure[]{ \includegraphics[width=\linewidth]{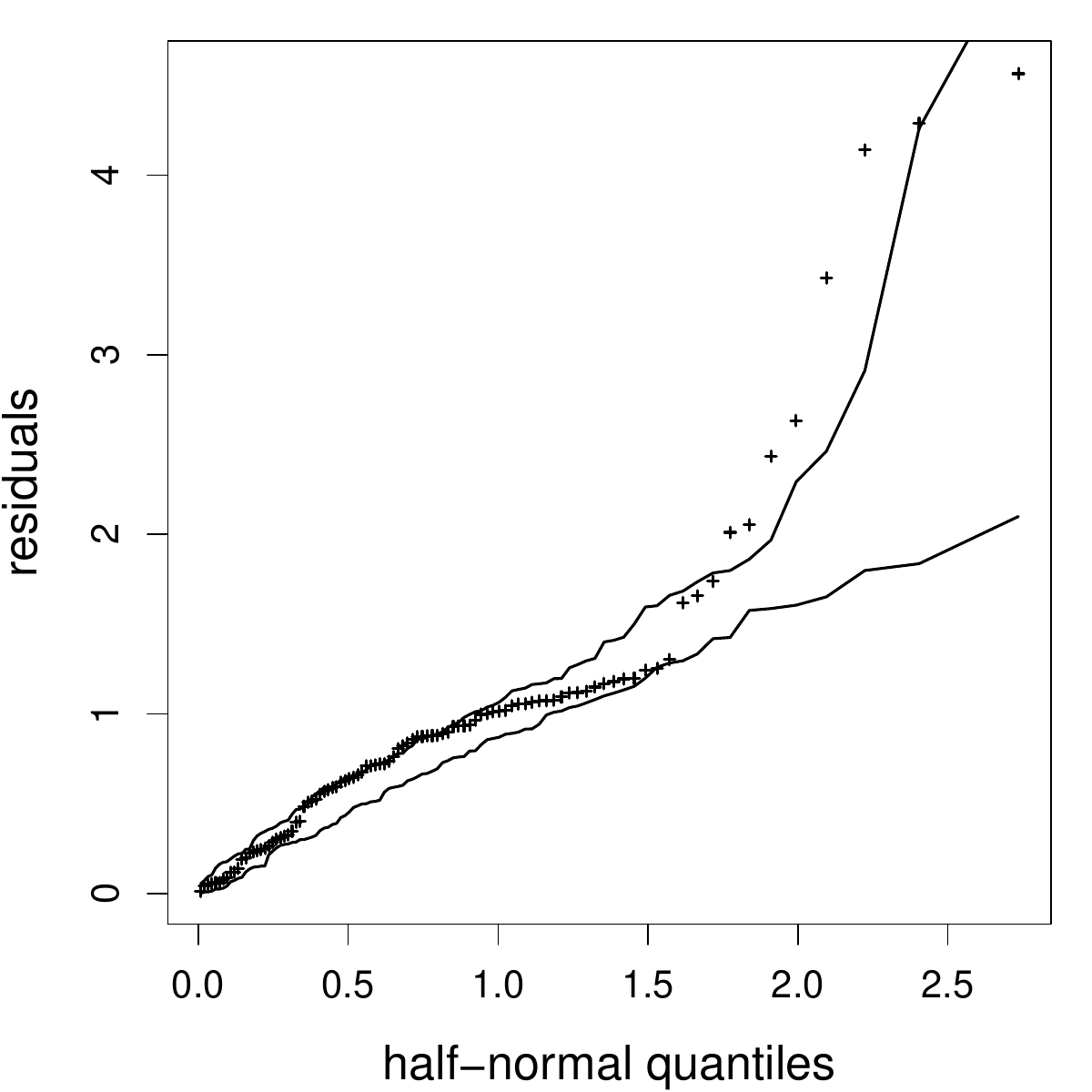} }
	\subfigure[]{ \includegraphics[width=\linewidth]{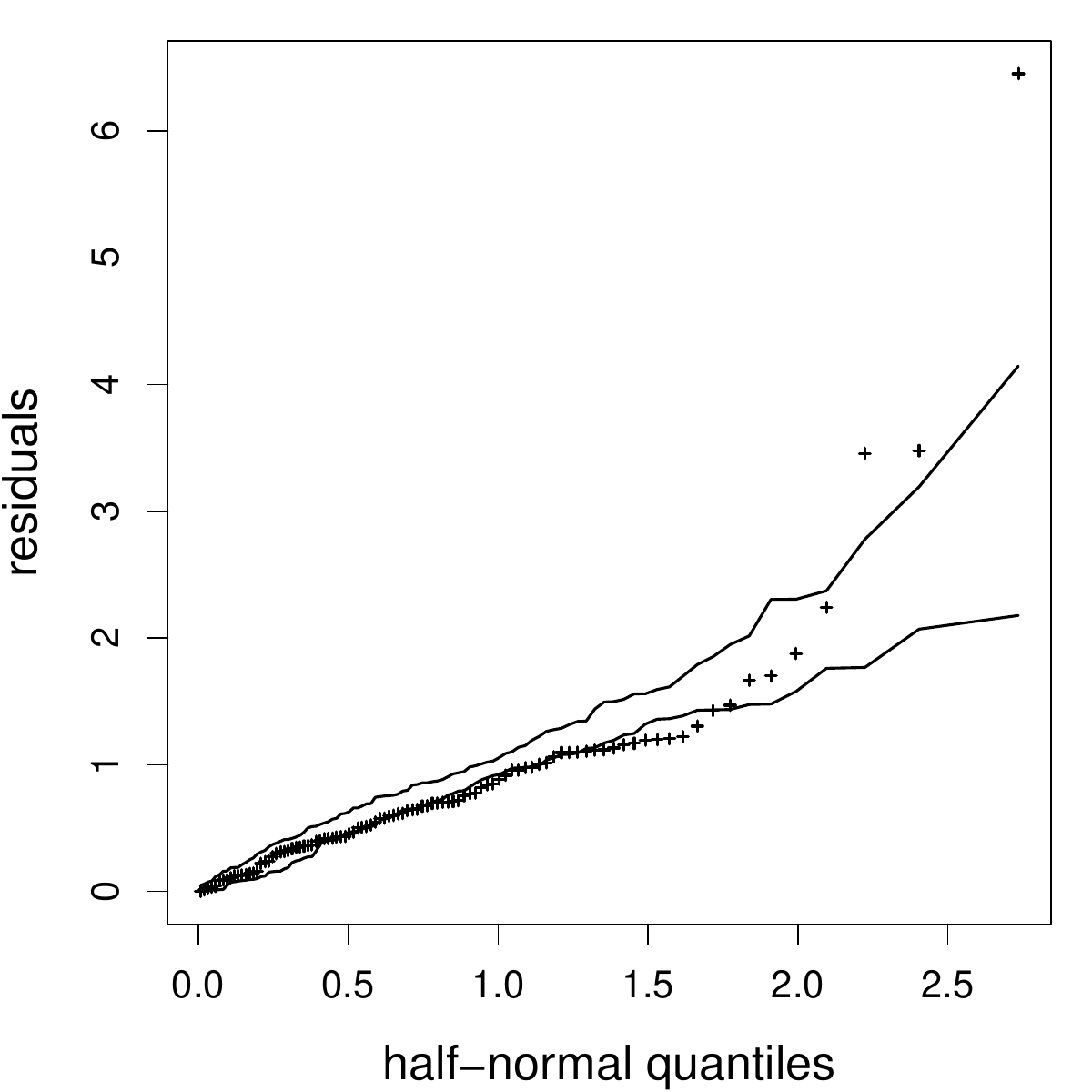} }
	\caption{Half-normal residual plots with simulated envelopes based on one random sample of size $n=100$ when one assumes a beta mode model for data generated from (B1)--(B4), shown in (a)--(d), respectively.}
	\label{Sim:envelope-beta}
\end{figure}

Besides covariate effects estimation, we also monitor operating characteristics of the model diagnostics tools proposed in Section~\ref{s:diag}. Assuming a beta mode model, Figure~\ref{Sim:envelope-beta} demonstrates the half-normal residual plots obtained based on one data set of size $n=100$ generated from each of (B1)--(B4), where $m=80$ in (B1)--(B3), and $m=10$ in (B4). \mytextcolor{Here setting $m=80$ for (B1)--(B3) is to make sure that the conditional variance under the beta model is similar to that under the GBP model with $m=10$.} Under (B1), where the assumed model matches the true data generating process, very few residuals fall outside of the envelope. In contrast, a large proportion of the residuals are outside of the envelope under (B2), where the linear predictor is misspecified. The plots under (B3) and (B4) also witness rather high proportions of residuals outside of the envelope. These empirical evidence indicate that the half-normal residual plot is an effective graphical indicator of various sources of model misspecification.

\begin{figure} 
	\centering
	\setlength{\linewidth}{0.47\textwidth}
	\subfigure[]{ \includegraphics[width=\linewidth]{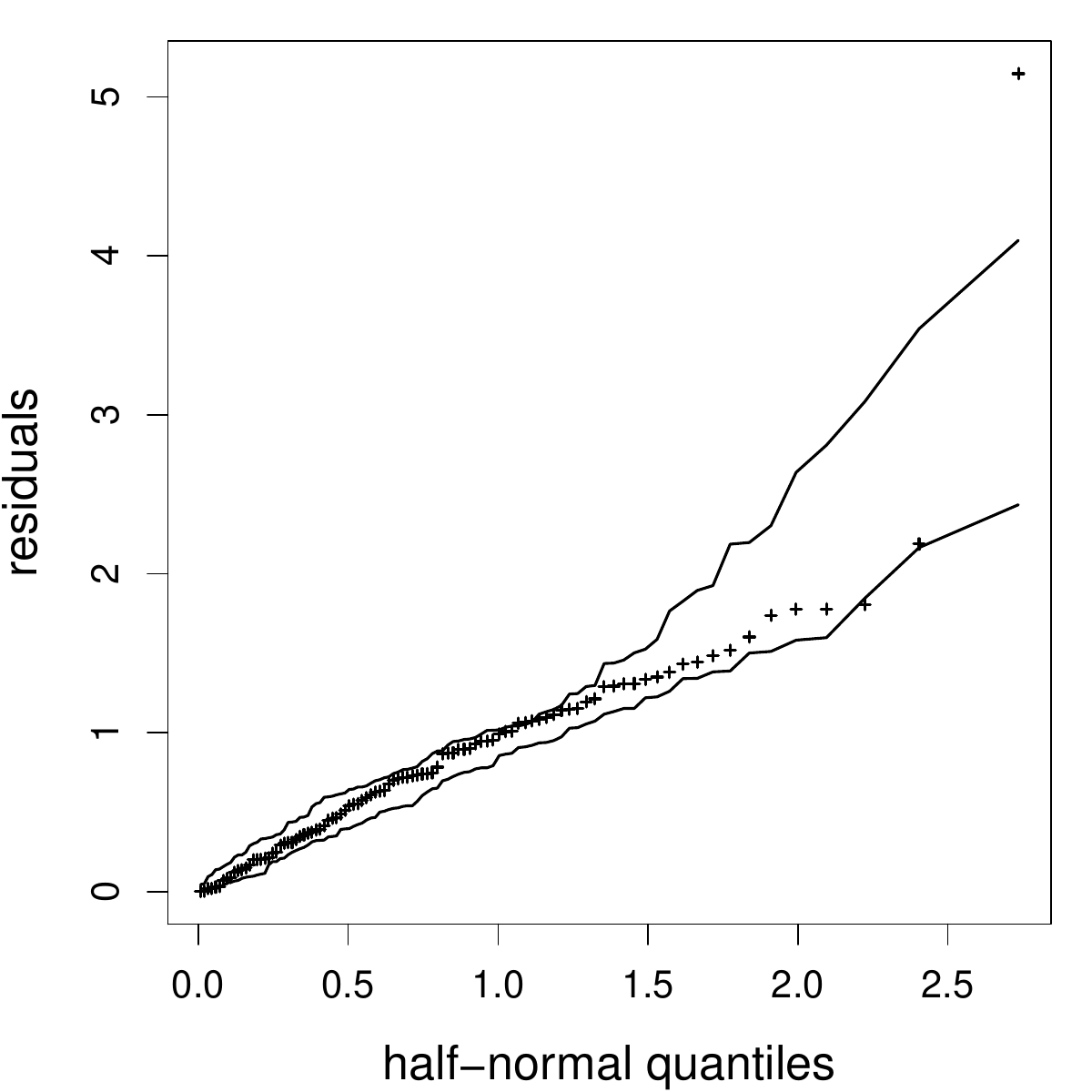} }
	\subfigure[]{ \includegraphics[width=\linewidth]{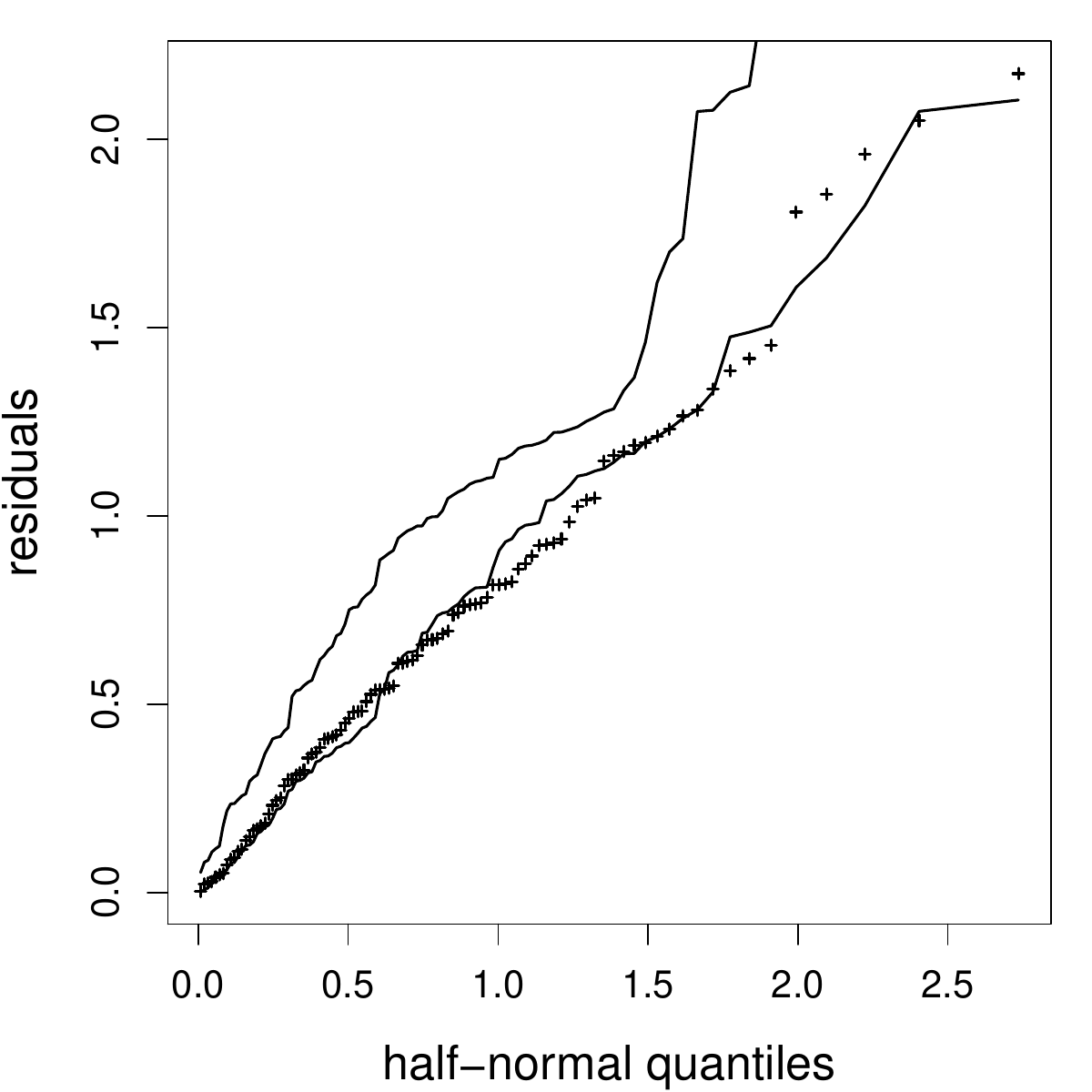} }\\
	\subfigure[]{ \includegraphics[width=\linewidth]{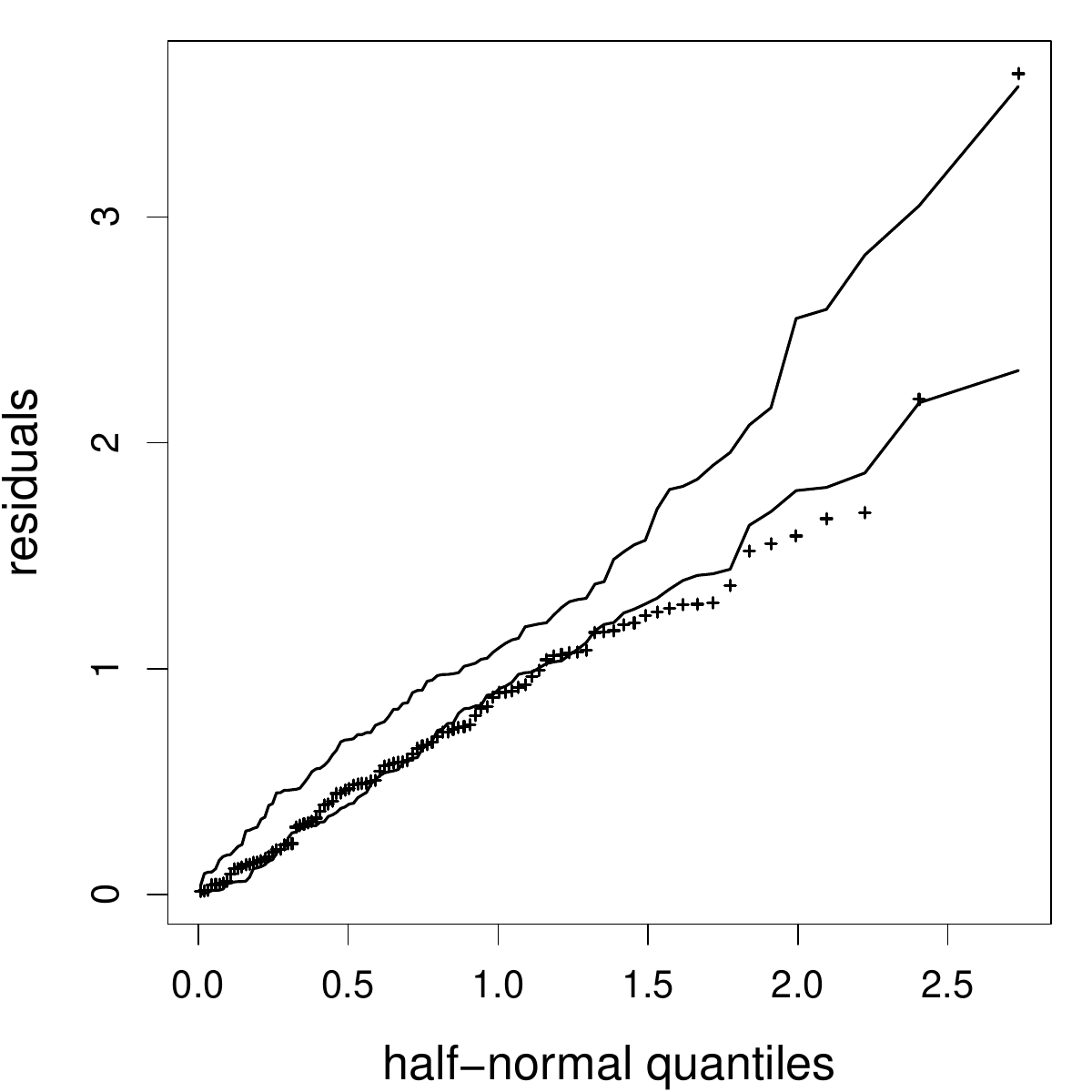} }
	\subfigure[]{ \includegraphics[width=\linewidth]{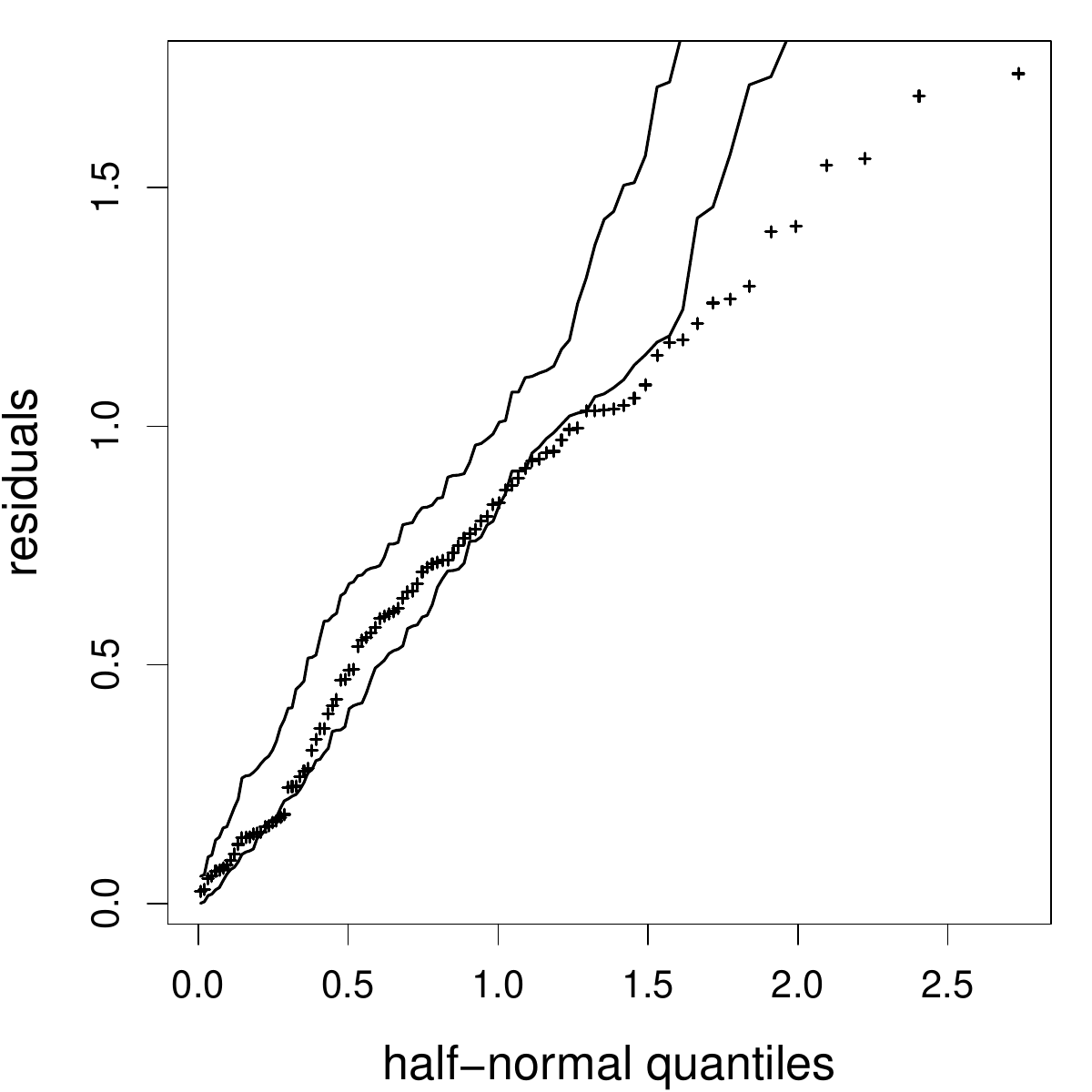} }
	\caption{Half-normal residual plots with simulated envelopes based on one random sample of size $n=100$ when one assumes a GBP mode model for data generated from (G1)--(G4), shown in (a)--(d), respectively.}
	\label{Sim:envelope-gbp}
\end{figure}

Assuming a GBP mode model, Figure~\ref{Sim:envelope-gbp} demonstrates the half-normal residual plots using one data set of size $n=100$ generated from each of (G1)--(G4), where $m=10$ in (G1)--(G3), and $m=80$ in (G4). \mytextcolor{Again setting $m=80$ for (G4) is to make sure that the conditional variance under the beta model is similar to that under the GBP model with $m=10$.} Similar to what one sees in the previous figure, in the absence of model misspecification as in (G1), most residuals are within the envelope. A much larger proportion of residuals fall outside of the envelope in the presence of linear predictor misspecification as in (G2). The plots also exhibit a moderate to high proportion of residuals outside of the envelopes under (G3) and (G4). Hence, the effectiveness of the half-normal residual plot for detecting various sources of model misspecification is also evident when one assumes a GBP mode model. 

\begin{figure} 
	\centering
	\setlength{\linewidth}{0.47\textwidth}
	\subfigure[]{ \includegraphics[width=\linewidth]{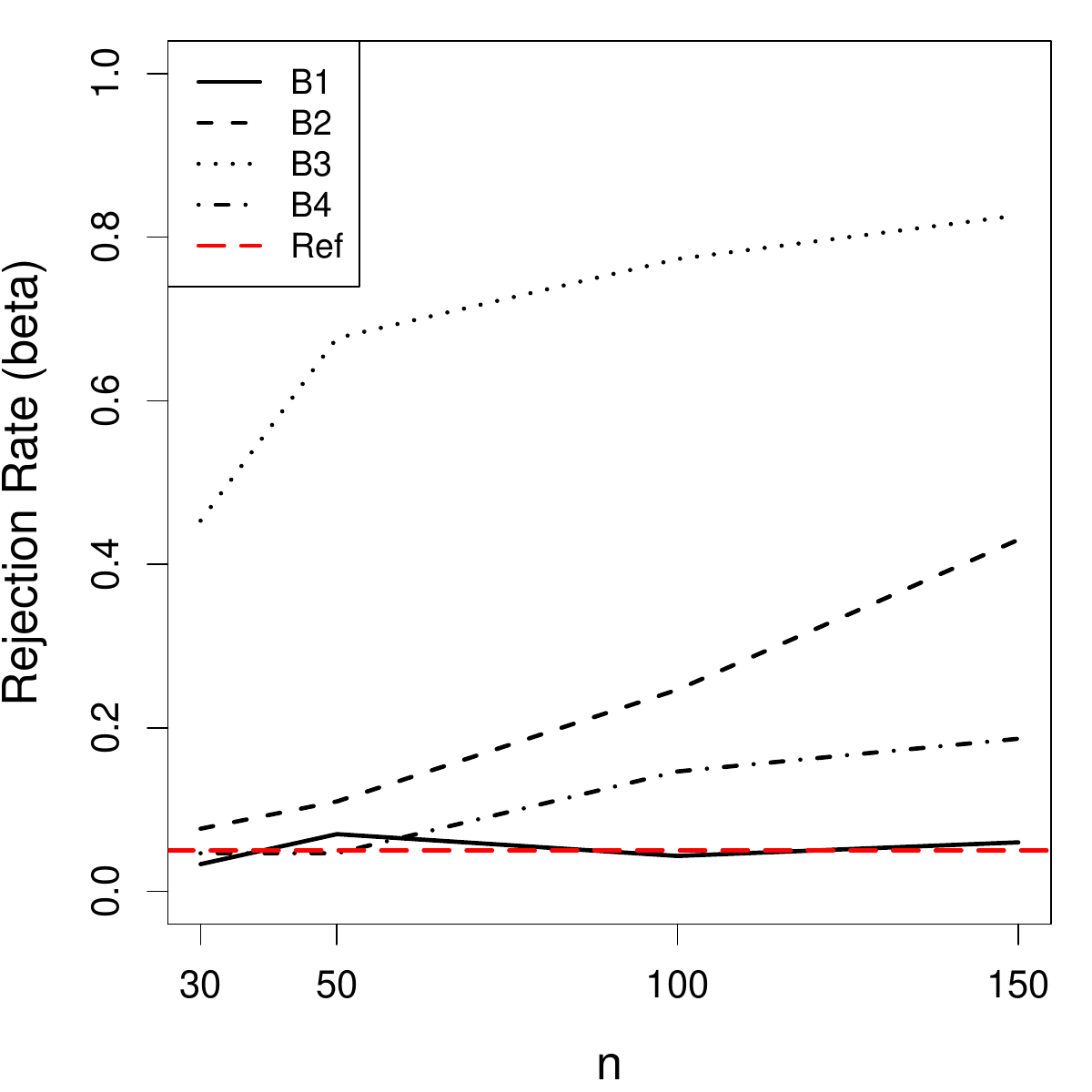} }
	\subfigure[]{ \includegraphics[width=\linewidth]{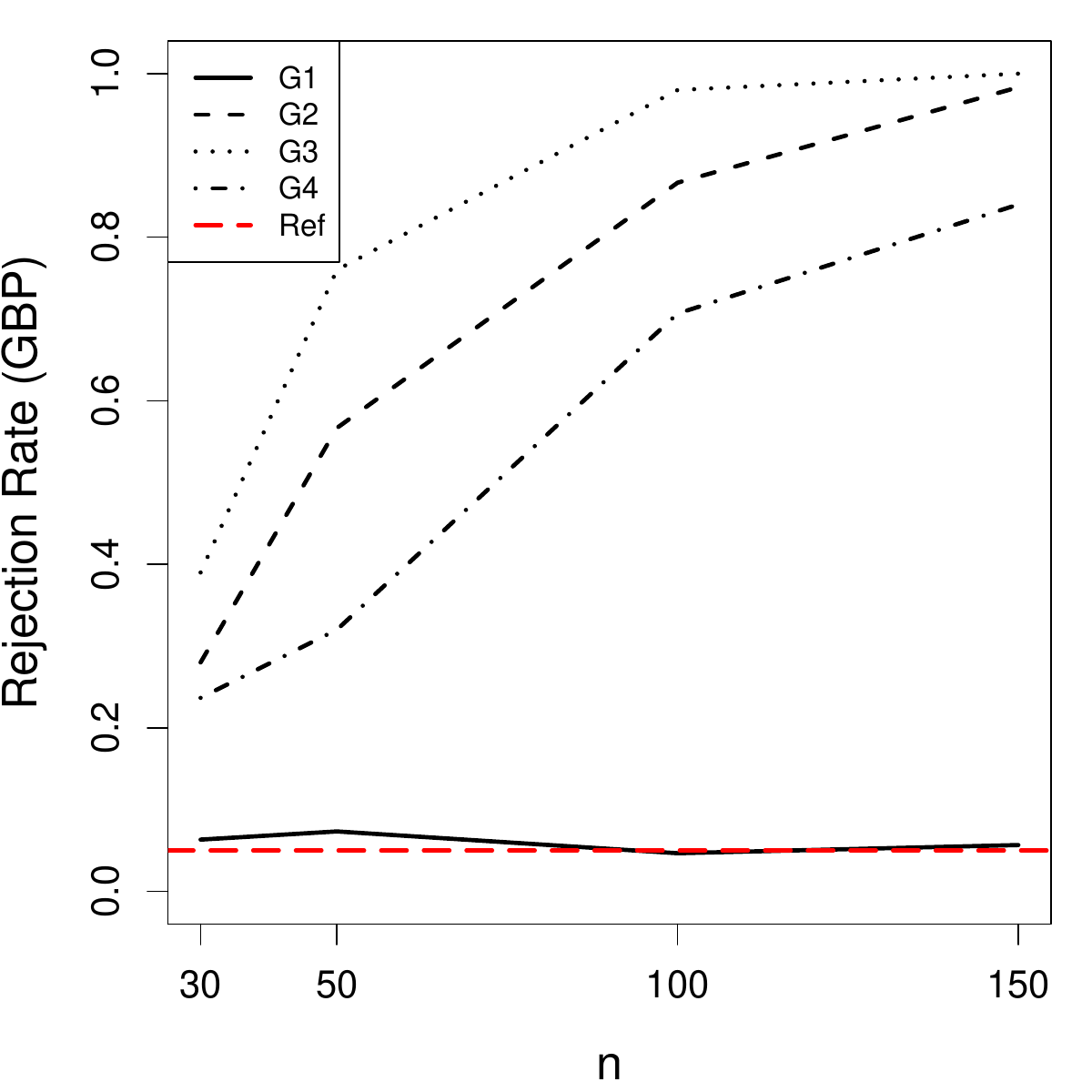} }
  \caption{\label{f:power}Rejection rates across 300 Monte Carlo replicates associated with the score tests under (B1)--(B4) when a beta mode model is assumed (in (a)), and under (G1)--(G4) when a GBP mode model is assumed (in the (b)). True model settings are: solid lines for (B1)\&(G1), dashed lines for (B2)\&(G2), dotted lines for (B3)\&(G3), and dashed-dotted lines for (B4)\&(G4). Red horizontal long-dashed lines are reference lines at nominal level 0.05.}
\end{figure}

Figure~\ref{f:power} presents the empirical power of the score tests proposed in Section~\ref{s:test} when a beta mode model or a GBP mode model is assumed, where the empirical power of a test is defined as the rejection rate of the test at significance level 0.05 across 300 Monte Carlo replicates under each true model specification. Given one simulated data set, we use 300 bootstrap samples to estimate the $p$-value associated with a score test. These empirical power indicate that the size of the score tests remain close to the nominal level in the absence of model misspecification, and, as the sample size grows, their power to detect any one of the three sources of model misspecification increases. Under each assumed mode model, the proposed score test has the highest power to detect a link misspecification, moderate power in response to a misspecified linear predictor, and the lowest power when the conditional distribution family is misspecified. The low power to detect the last type of model misspecification, especially when a beta mode model is assumed, may suggest that, given a GBP mode model, there exists a member in the family of beta mode models that can approximate the GBP mode model well enough to produce reasonable estimates for the first two moments. Lastly, these numerical evidence of model misspecification match nicely with the graphical evidence from half-normal residual plots in that a higher rejection rate observed for the score test under one scenario usually goes with a higher proportion of residuals outside of the envelope in the half-normal residual plot in the same scenario.

\subsection{Predictions}
\label{s:pred}
Predicting an outcome is often one of the ultimate goals in regression analysis, such as in ADNI where accurate prediction of AD progression is a major goal. Suppose $\bx$ is the covariate value at which one wishes to predict a bounded outcome, such as one's ADAS-11 score. In what follows, at a nominal coverage probability $q\in (0, 1)$, we first construct a prediction interval based on an estimated mode, denoted by $\mathcal{PI}_\theta(\bx, q)$, then we formulate a prediction interval based on an estimated mean, denoted by $\mathcal{PI}_\mu(\bx, q)$, and a prediction interval based on an estimated median, denoted by $\mathcal{PI}_\nu(\bx, q)$. Define $e=Y-\theta(\bx)$ as the mode residual, and denote by $f_e(e|\bx)$ the pdf of $e$ given $\bX=\bx$. 

Under an assumed mode regression model, following maximum likelihood estimation of $\bOmega$, one obtains the MLEs for $\theta(\bx)$ and $\mu(\bx)$, as well as an estimated pdf of $e$ given $\bX=\bx$. Denote these MLEs by $\hat \theta(\bx)$ and $\hat \mu(\bx)$, respectively, and denote by $\hat f_e(e|\bx)$ the estimated pdf. Then, based on these estimates, the narrowest $\mathcal{PI}_\theta(\bx, q)$ is $[\hat \theta(\bx)+e_1, \, \hat \theta(\bx)+e_2]$, where $e_1<0<e_2$ satisfy $\hat f_e(e_1|\bx)=\hat f_e(e_2|\bx)$ and $\int_{e_1}^{e_2} \hat f_e(e|\bx)\, de =q$.

To formulate a $(100\times q)\%$ mean-based prediction interval, we first make sure that $\hat \mu(\bx)\in \mathcal{PI}_\mu(\bx, q)$, then we construct an interval with the desired coverage probability that is close to $\hat \theta(\bx)$ as much as possible in order to achieve the narrowest $\mathcal{PI}_\mu(\bx, q)$. Clearly, if $\hat \mu(\bx)$ already falls in $\mathcal{PI}_\theta(\bx, q)$ constructed above, which is the narrowest by construction, then one may also use this interval as $\mathcal{PI}_\mu(\bx, q)$. Otherwise, we construct $\mathcal{PI}_\mu(\bx, q)$ with $\hat \mu(\bx)$ on one of the boundaries depending on how $\hat \mu(\bx)$ compares with $\hat \theta(\bx)$. In particular, if $\hat \mu(\bx)\ge \hat \theta(\bx)$, then we set $\mathcal{PI}_\mu(\bx, q)=[\hat \mu(\bx)-c, \, \hat \mu(\bx)]$, where $c>0$ is chosen such that $\int_{\hbox {\tiny $\hat \mu(\bx)-c$}}^{\hbox {\tiny $\hat \mu(\bx)$}} \hat f_{\hbox {\tiny $Y|\bX$}}(y|\bx) dy =q$, in which $\hat f_{\hbox {\tiny $Y|\bX$}}(y|\bx)$ is pdf of the assumed distribution of $Y$ given $\bX=\bx$ evaluated at $\hat \bOmega$. If $\hat\mu(\bx)<\hat\theta(\bx)$, then we let $\mathcal{PI}_\mu(\bx, q)=[\hat \mu(\bx), \, \hat \mu(\bx)+c]$, where $c>0$ satisfies $\int_{\hbox {\tiny $\hat \mu(\bx)$}}^{\hbox {\tiny $\hat \mu(\bx)+c$}} \hat f_{\hbox {\tiny $Y|\bX$}}(y|\bx) dy =q$. A $(100\times q)\%$ median-based prediction interval can be similarly constructed once an estimated median, denoted by $\hat \nu(\bx)$, is obtained. 

\begin{figure} 
\centering
\setlength{\linewidth}{0.45\textwidth}
\subfigure[]{ \includegraphics[width=\linewidth]{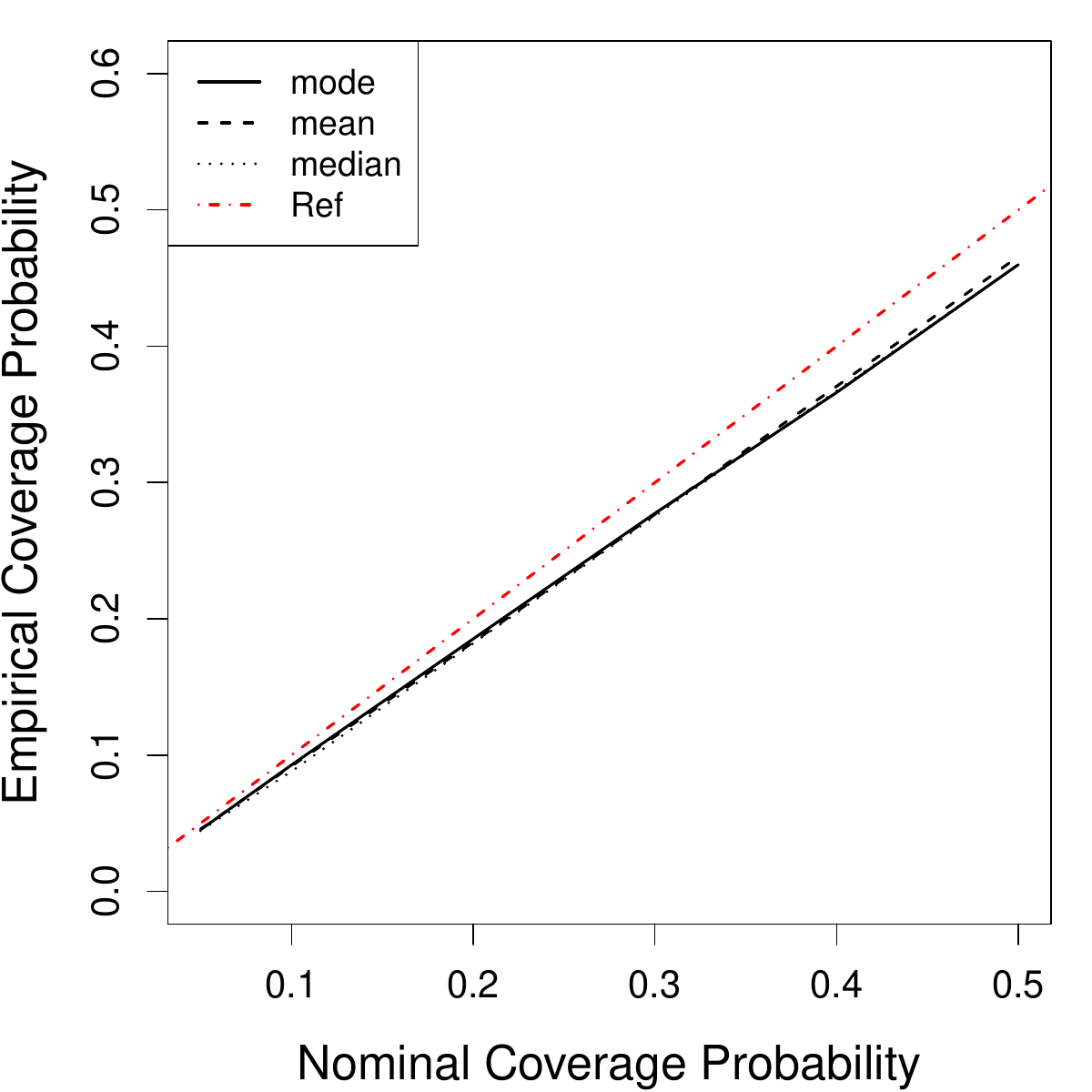} }
\subfigure[]{ \includegraphics[width=\linewidth]{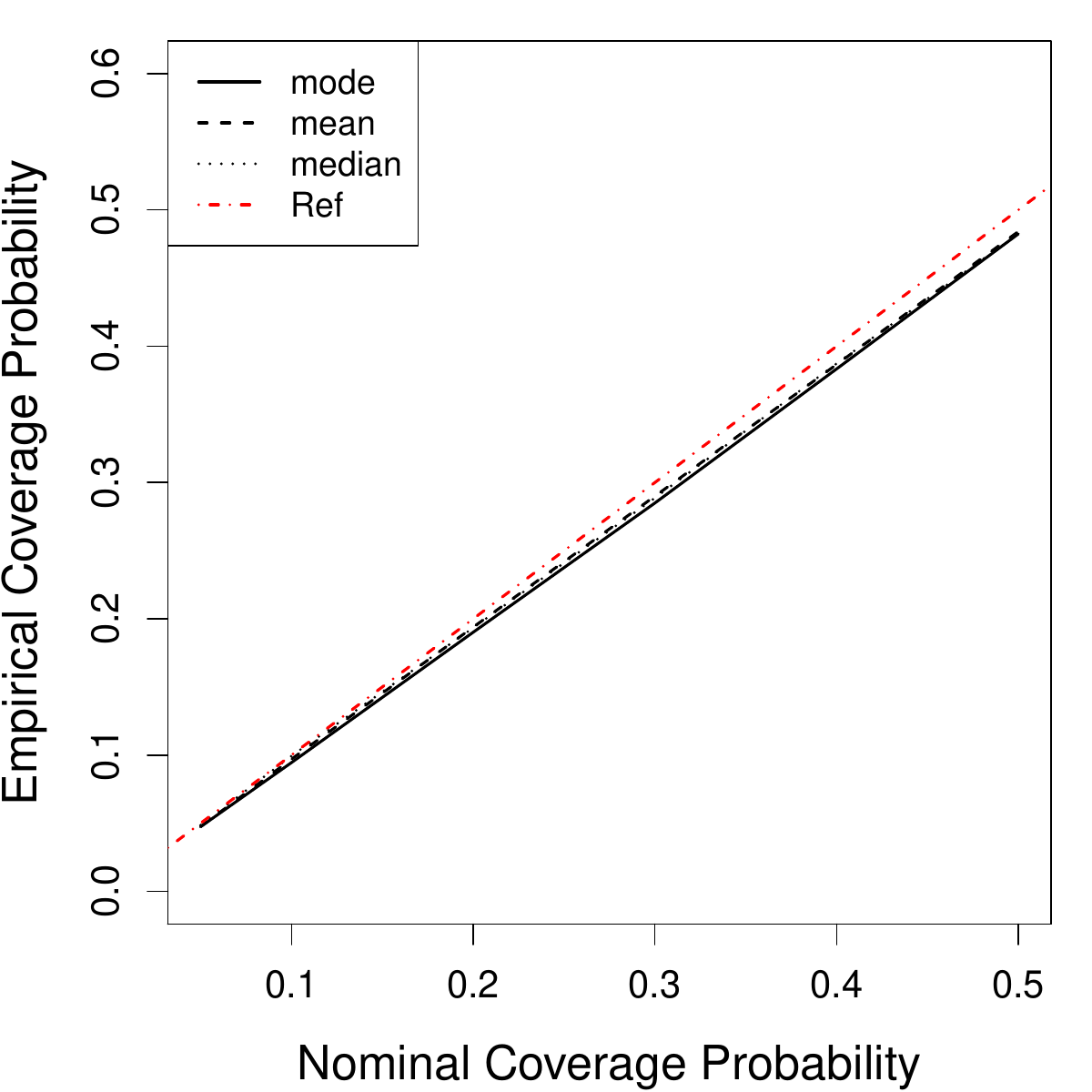} }\\
\subfigure[]{ \includegraphics[width=\linewidth]{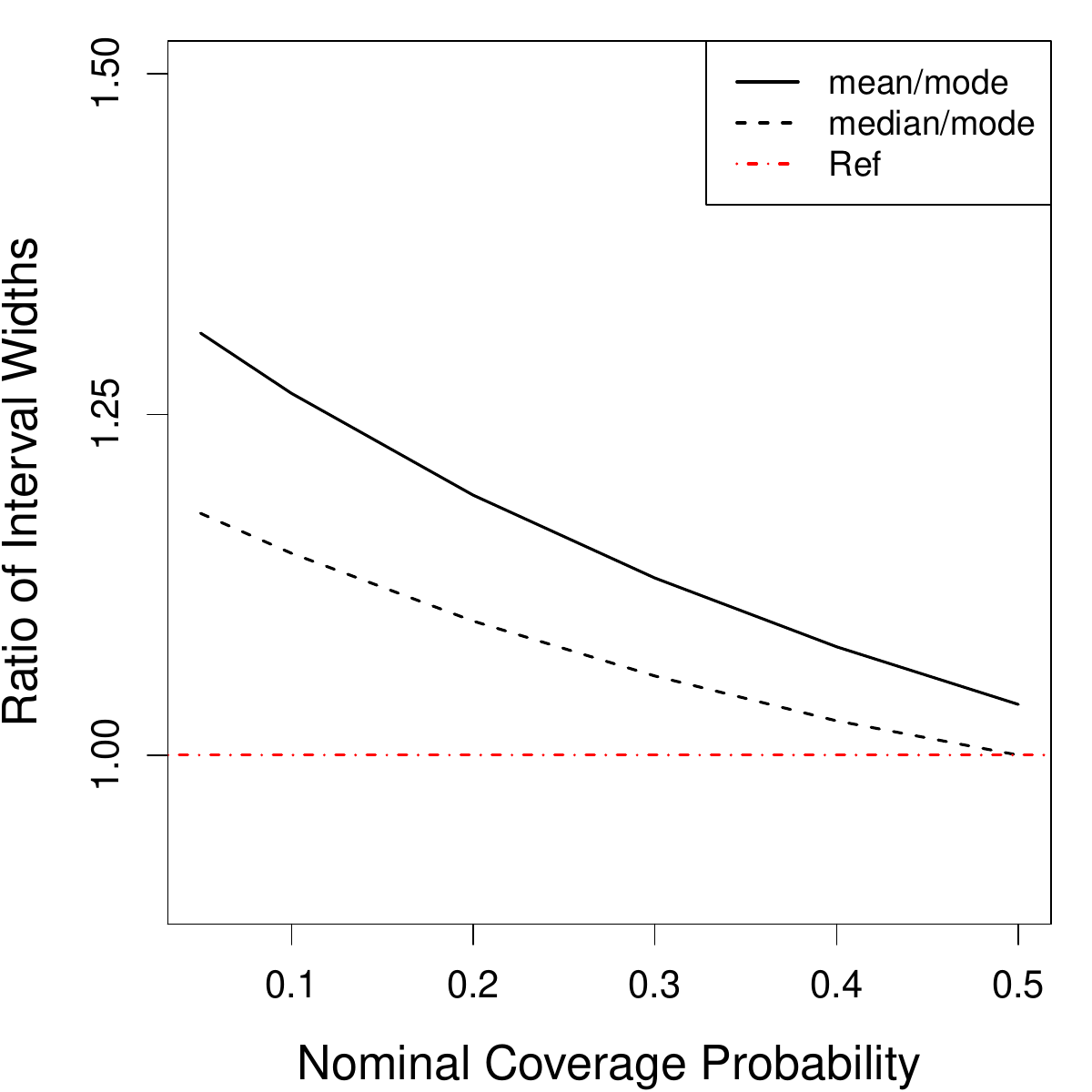} }
\subfigure[]{ \includegraphics[width=\linewidth]{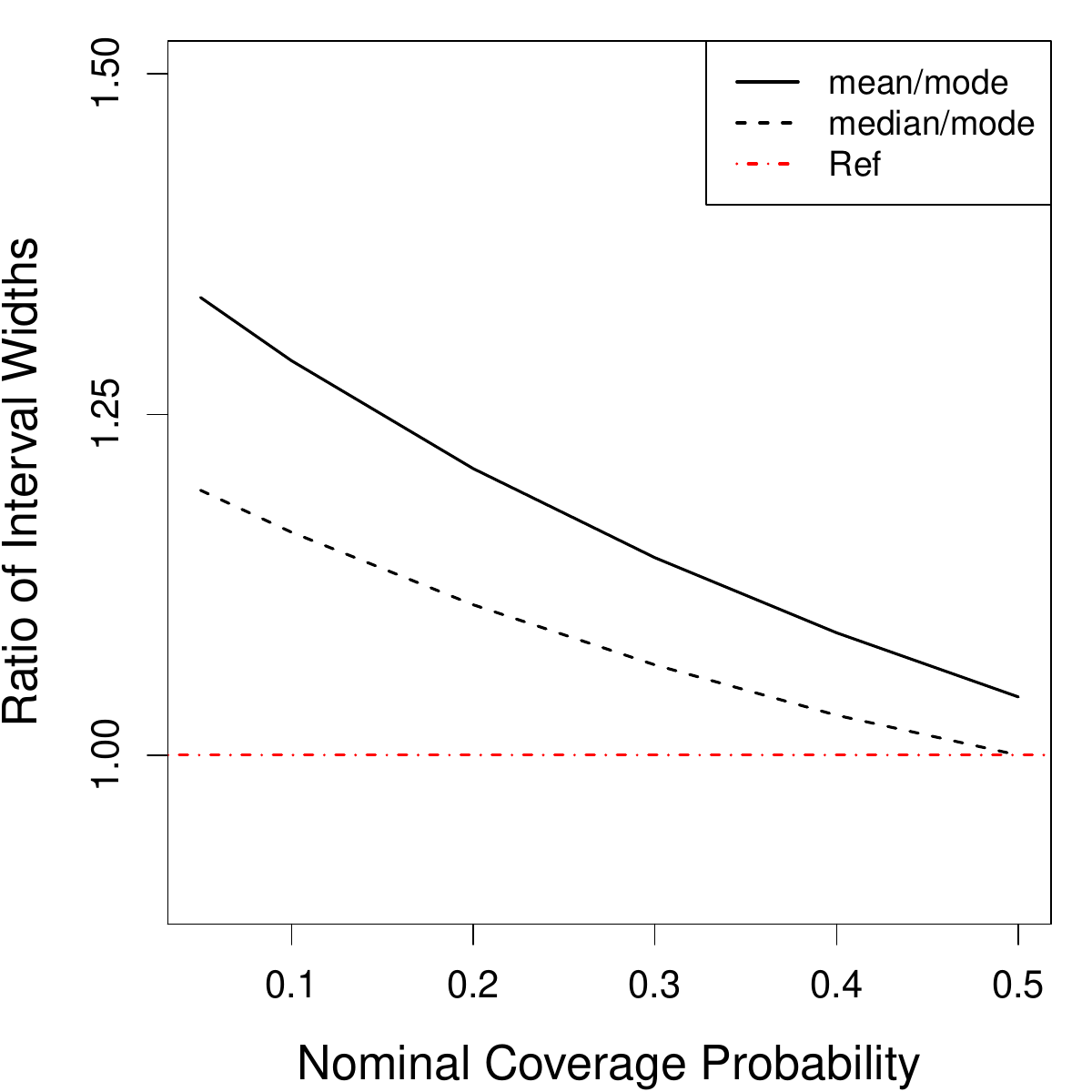} }
\caption{\label{f:ecpwidthbeta}{\small Prediction intervals based on data from the beta mode model in (B1). Top panels (a) and (b) depict average empirical coverage probabilities (across 300 Monte Carlo replicates) of mode-based prediction intervals (solid lines), those of mean-based prediction intervals (dashed lines), and those of median-based prediction intervals (dotted lines) versus nominal coverage probabilities. Red dash-dotted lines are $45^\circ$ reference lines. Lower panels (c) and (d) depict ratios of the average width of mean-based prediction intervals over that of mode-based prediction intervals (solid lines) and ratios of the average width of median-based prediction intervals over that of mode-based prediction intervals (dashed lines) versus nominal coverage probabilities. Red dash-dotted horizontal lines are reference lines at value one. Panels (a) and (c) are for $n=50$. Panels (b) and (d) are for $n=100$.}}
\end{figure}

With the assumed model being a beta mode model, Figure~\ref{f:ecpwidthbeta} depicts in upper panels averages empirical coverage probabilities of $\mathcal{PI}_\theta(\cdot, q)$, $\mathcal{PI}_\mu(\cdot, q)$, and $\mathcal{PI}_\nu(\cdot, q)$ versus nominal coverage probabilities $q$ when 300 Monte Carlo replicate data sets are generated from (B1) with $\bbeta=(3, 1, 1)$, $m=10$ and $n=50, \mytextcolor{100}$, where $q$ ranges from 0.05 to 0.5. The empirical coverage probability of each type of prediction intervals is obtained via five-fold cross validation. Take $\mathcal{PI}_\theta(\cdot, q)$ as an example, its empirical coverage probability based on one data set is defined as $n^{-1}\sum_{k=1}^5\sum_{i\in \mathcal{I}_k} I \{Y_i \in \mathcal{PI}^{(-k)}_\theta(\bX_i, q)\}$, where $\mathcal{PI}^{(-k)}_\theta(\bX_i, q)$ is the $(100\times q)\%$ mode-based prediction interval constructed using data excluding the $k$th testing data set corresponding to the index set $\mathcal{I}_k$, for $k=1, \ldots, 5$. The lower panels of Figure~\ref{f:ecpwidthbeta} compare the average width of the three types of prediction intervals. 

According to these figures, mode-based prediction intervals,  mean-based prediction intervals, and media-based prediction intervals achieve similar empirical coverage probabilities that become closer to the nominal coverage probability as $n$ increases. More importantly, the mode-based prediction interval tends to be narrowest among the three, and the mean-based prediction interval is the widest. By construction, it is expected that $\mathcal{PI}_\theta(\bx, q)=\mathcal{PI}_\mu(\bx, q)=\mathcal{PI}_\nu(\bx, q)$ when $q$ is not low since $\mathcal{PI}_\theta(\bx, q)$ with a moderate or high coverage probability is very likely to include the estimated mean and the estimated median. Certainly, these prediction intervals are also expected to be more similar when the conditional distribution for the response is less skewed. 

\begin{figure} 
\centering
\setlength{\linewidth}{0.45\textwidth}
\subfigure[]{ \includegraphics[width=\linewidth]{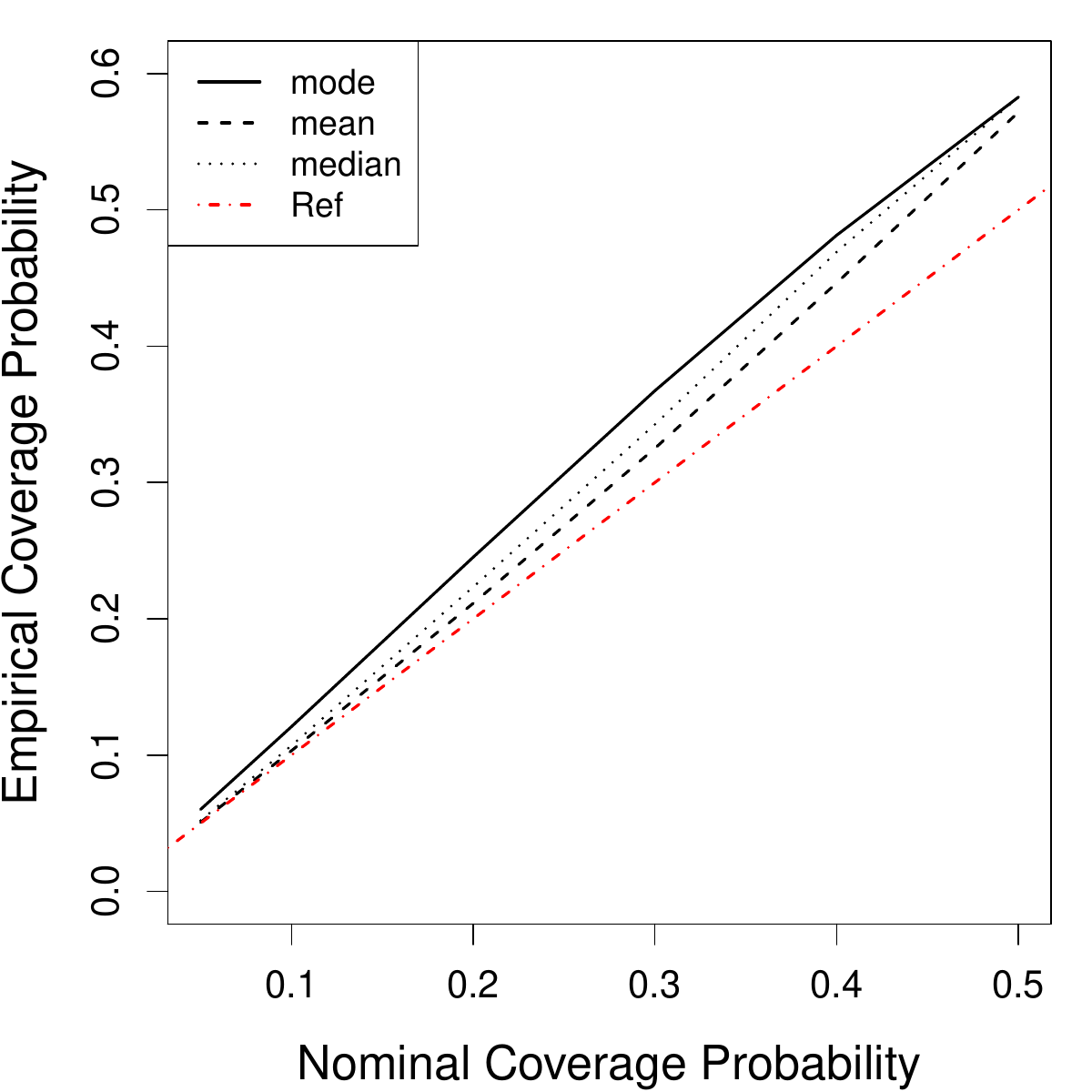} }
\subfigure[]{ \includegraphics[width=\linewidth]{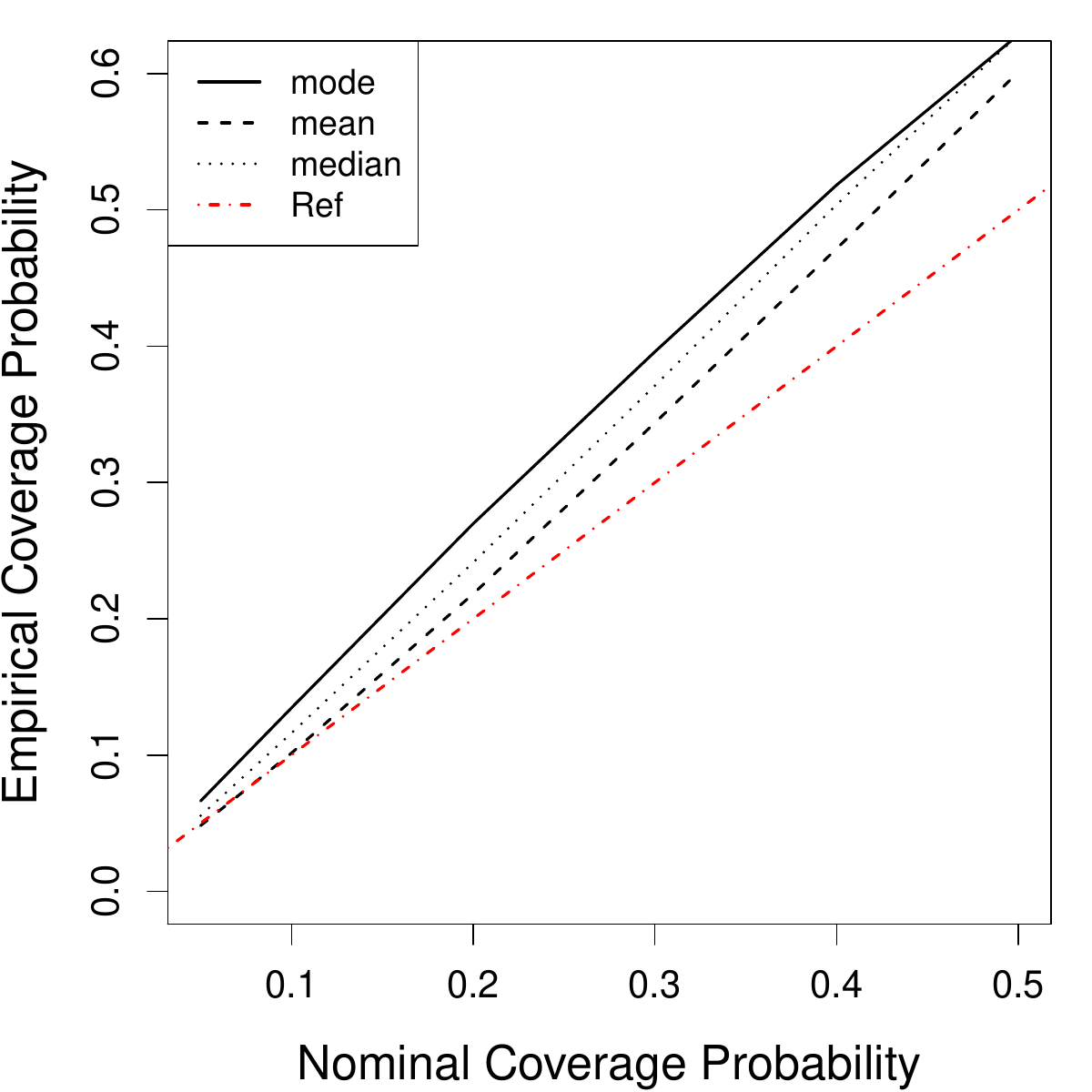} }\\
\subfigure[]{ \includegraphics[width=\linewidth]{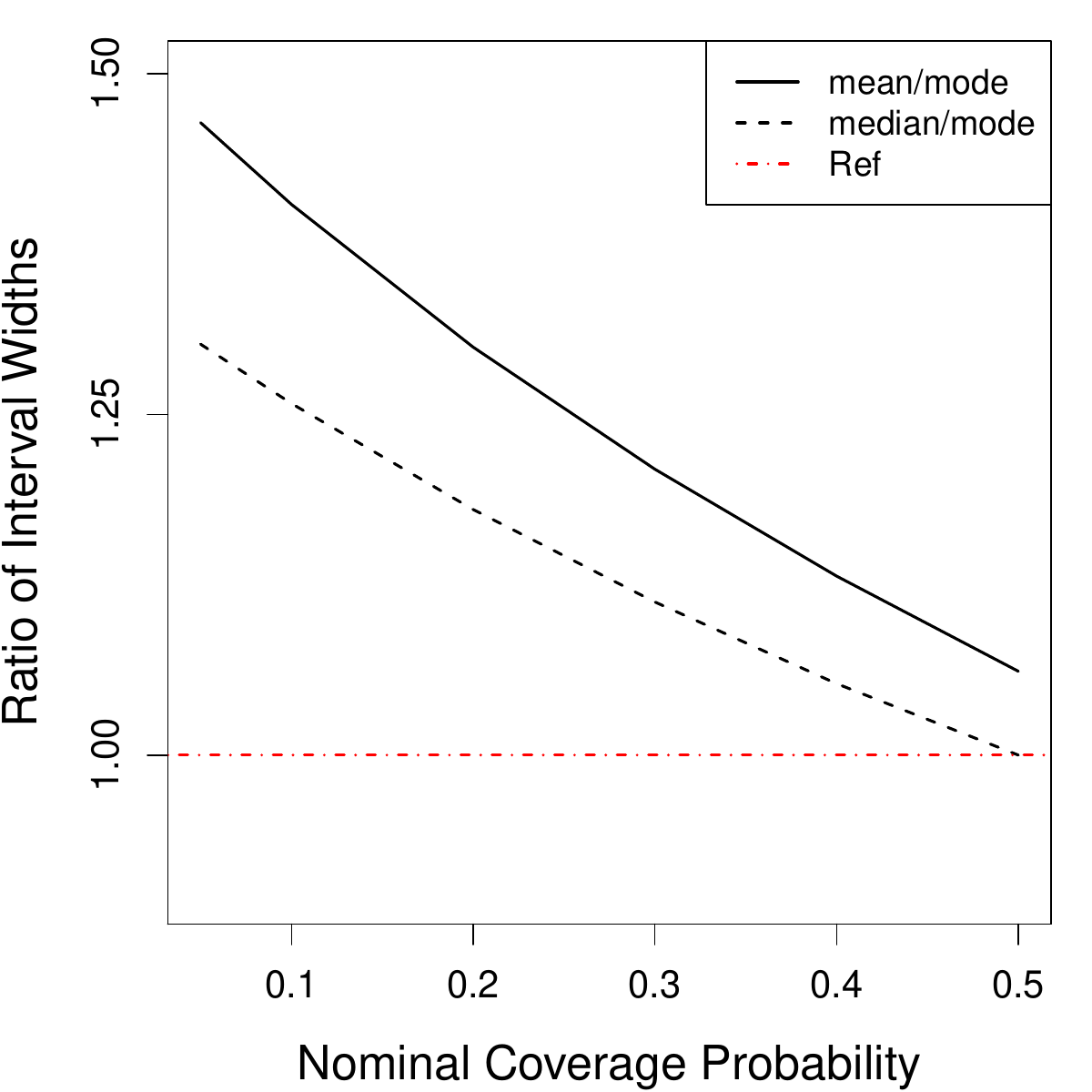} }
\subfigure[]{ \includegraphics[width=\linewidth]{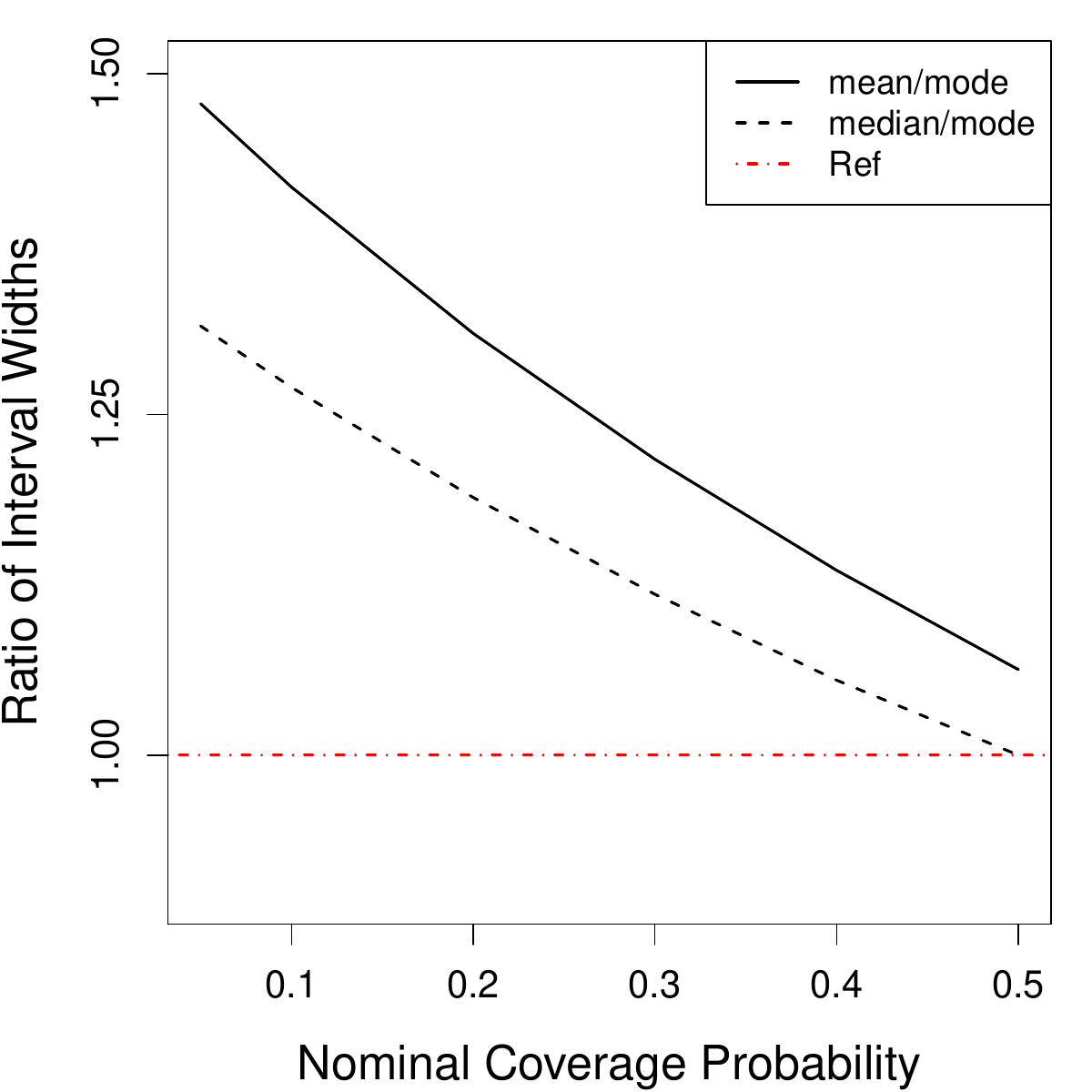} }
\caption{\label{f:ecpwidthbetaoutlier}{\small Prediction intervals based on beta mode regression using data from the beta mode model in (B1), with outliers replacing 5\% of the original data. Top panels (a) and (b) depict average empirical coverage probabilities (across 300 Monte Carlo replicates) of mode-based prediction intervals (solid lines), those of mean-based prediction intervals (dashed lines), and those of median-based prediction intervals (dotted lines) versus nominal coverage probabilities. Red dash-dotted lines are $45^\circ$ reference lines. Lower panels (c) and (d) depict ratios of the average width of mean-based prediction intervals over that of mode-based prediction intervals (solid lines) and ratios of the average width of median-based prediction intervals over that of mode-based prediction intervals (dashed lines) versus nominal coverage probabilities. Red dash-dotted horizontal lines are reference lines at value one. Panels (a) and (c) are for $n=50$. Panels (b) and (d) are for $n=100$.}}
\end{figure}

To demonstrate the impact of outliers on the aforementioned prediction intervals, we contaminate each of 300 Monte Carlo replicate data sets generated from the beta mode model by replacing 5\% of randomly chosen responses with random numbers simulated from uniform$(0, t)$, where $t$ is the 0.001-th quantile of the true conditional distribution of the response. This contamination produces data with a heavier (left) tail than the distribution specified in (B1). Despite the model misspecification, we fit the resultant data assuming a beta mode regression model and construct prediction intervals based on the three central tendency measures. Figure~\ref{f:ecpwidthbetaoutlier} shows the comparison between different types of prediction intervals in regard to empirical coverage probability and width. From there, one can see that a direct consequence of fitting (and making predictions based on) a beta mode model to data from an underlying distribution with a heavier (than assumed) tail is inflated coverage probabilities, despite the choice of central tendency measure for prediction. Interestingly, even though the empirical coverage probability of the mode-based prediction interval is higher than those of the other two types of prediction intervals, the mode-based prediction interval remains the narrowest among the three. In conclusion, even in the presence of severe outliers, the conditional mode still yields more reliable and precise predictions than the conditional mean or median does. Parallel pictures when data are generated from (G1), with or without outliers contamination, and one assumes a GBP mode model are provided in supplementary Figures~\mytextcolor{S4 and S5}. 

\section{Application to  ADNI data}
\label{s:realdata}
There has been a consensus among medical researchers that regional brain atrophy in the medial temporal lobe structures, such as the entorhinal cortex (ERC) and hippocampus (HPC), are correlated with clinical alterations in the pre-dementia phase of mild cognitive impairment (MCI) and various dementia phases of AD \citep{devanand2007hippocampal, jauhiainen2009discriminating}. While early detection and intervention in MCI subjects has been actively pursued by many researchers, there are mixed opinions among them regarding the roles volumetric measures of ERC and HPC play in predicting an MCI subject's risk of developing AD \citep{jack1999prediction, killiany2002mri, detoledo2004mri, hamalainen2007voxel, whitwell2008mri}. 

Using one's ADAS-11 score as a surrogate for one's severity of cognitive impairment, we apply the proposed mode regression models to data from the ADNI database (\url{http://adni.loni.usc.edu}) to study the association between one's ADAS-11 score at month 12 and the volumetric changes in ERC and HPC at month 6 compared to their baseline measures. In particular, the dataset we consider consists of a cohort of $245$ subjects who were diagnosed with LMCI when they entered the ADNI Phase 1 study and were followed up at least at both months 6 and 12. The original response variable is a subject's ADAS-11 score at month 12, which has a bounded support on $[0, 70]$. We rescale the support to the unit interval by dividing ADAS-11 scores by 70. 

Besides carrying out mode regression analysis, we also adopt the beta mean regression model for a rate or proportion response proposed by \citet{ferrari2004beta} to study the association. In their proposed regression model, the authors reparameterized the beta distribution by setting $\alpha_1=\mu\phi$ and $\alpha_2=(1-\mu)\phi$, where $\mu\in [0, 1]$ is the mean parameter and $\phi$ is a positive shape parameter, with a larger $\phi$ resulting in a smaller variance, and they incorporated the linear predictor $\eta(\bX)$ by letting $\mu=g\{\eta(\bX)\}$. 
In a preliminary analysis, we fit the beta mean model, beta mode model, and GBP mode model to the data using various link functions $g(t)$. Based on values of log-likelihood, we choose the log-log link in the regression models for further analysis.

\begin{figure} 
	\centering
	\setlength{\linewidth}{0.47\textwidth}
	\subfigure[]{ \includegraphics[width=\linewidth]{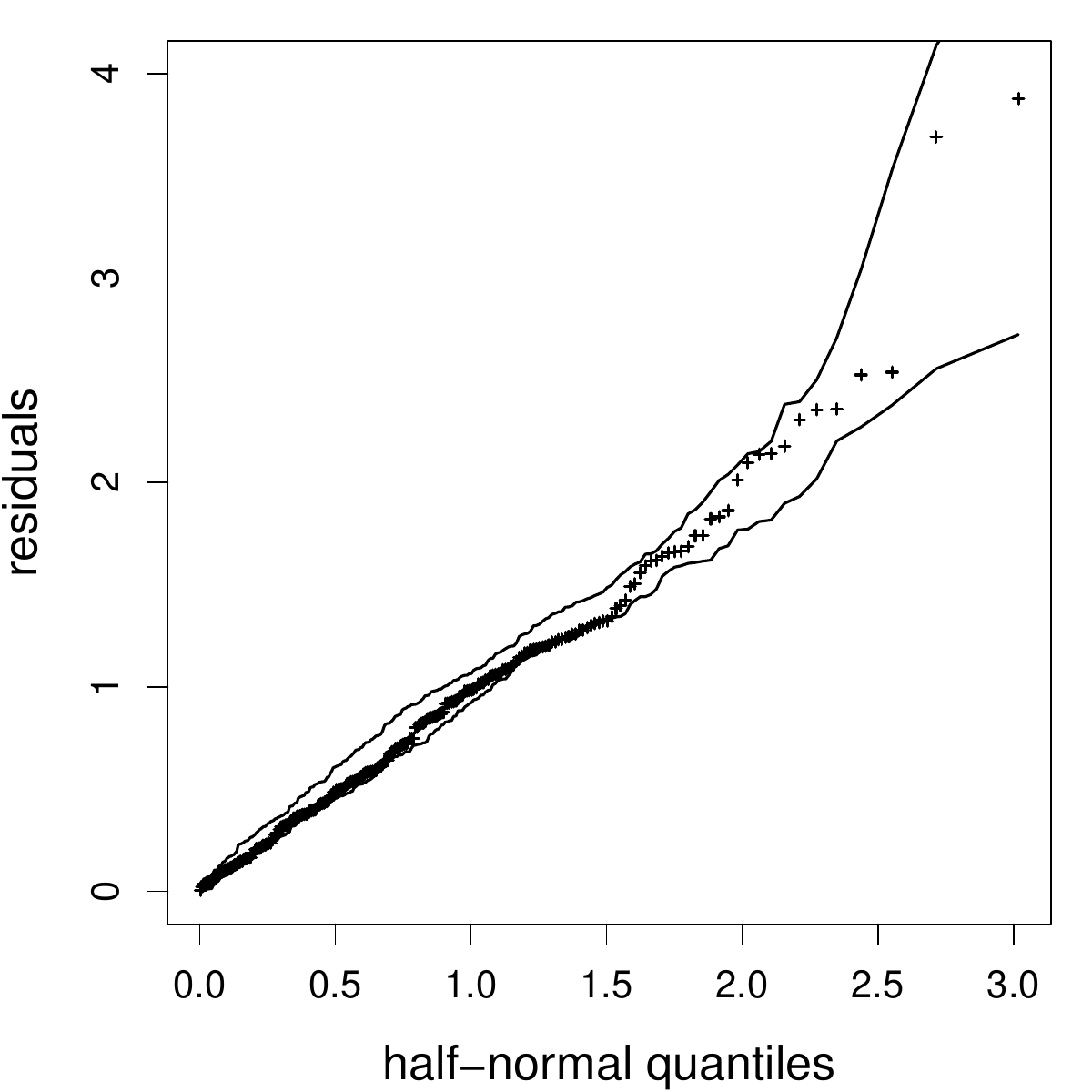} }
	\subfigure[]{ \includegraphics[width=\linewidth]{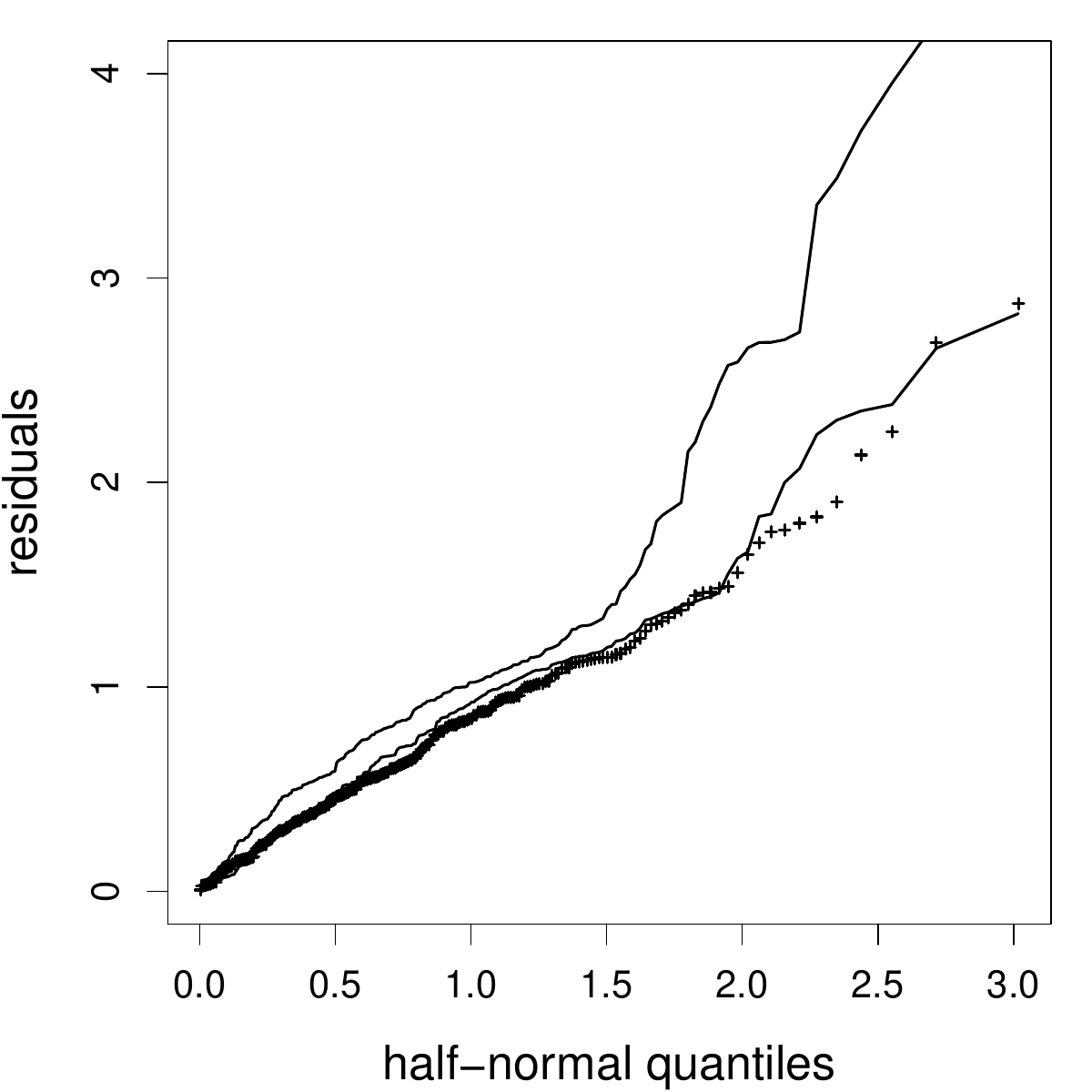} }
		\caption{Half-normal residual plots with simulated envelope for the ADNI data associated with the beta mode model (in (a)) and the GBP mode model (in (b)).}
	\label{f:realdata:envelope}
\end{figure}

Panels (a) and (b) in Figure~\ref{f:realdata:envelope} provide the half-normal residual plots associated with the beta mode model and GBP mode model based on this data set. Having majority of the residuals from GBP mode regression falling outside of the envelope suggests a poor fit of the GBP model for the data, and the beta mode model is more adequate for the current data. The score test when the null hypothesis states a beta mode model yields an estimated $p$-value of 0.45, while the score test when one assumes a GBP mode model gives an estimated $p$-value of 0. Gathering these graphical and numerical diagnosis, we conclude that the beta mode model potentially captures the underlying conditional distribution better than the GBP mode model does. 

Table~\ref{Qm:MLE} provides MLEs of unknown parameters in each of the three considered regression models. According to Table~\ref{Qm:MLE}, inference for the covariate effects from all three regression models suggest that the volumetric change in ERC is a statistically significant predictor for one's cognitive impairment. However, results from the GBP mode model does not indicate that the volumetric change in HPC is significantly associated with the response (with a $p$-value of 0.303), although inference from both beta mean and beta mode model imply a significant effect of the change in hippocampal volume on the ADAS-11 score (with $p$-values 0.042 and 0.043, respectively). 

\begin{table} 
	\caption{Maximum likelihood estimates corresponding to each regression model for the ADNI data. Numbers in parentheses are estimated standard errors associated with the MLEs.} 
	\label{Qm:MLE}
	\centering
	{
		\begin{tabular}{lccc}
			\hline\noalign{\smallskip}
			Parameter &  \mytextcolor{beta} mean model & \mytextcolor{beta} mode model & GBP mode model \\
			\noalign{\smallskip}\hline\noalign{\smallskip}
			$\beta_0$ (Intercept) & $-0.555$ (0.020) & $-0.697$ (0.026)  & $-0.971$ (0.017) \\
			$\beta_1$ (ERC.change)  & $-0.102$ (0.043) & $-0.125$ (0.052)  & $-0.117$ (0.028)\\
			$\beta_2$ (HPC.change) & $-0.170$ (0.084) & $-0.216$ (0.107)  & $-0.112$ (0.109) \\
			$\phi$ or $\log m$  & 17.990 (1.609) & 2.772 (0.099)  & 1.826 (0.068) \\
			\noalign{\smallskip}\hline
		\end{tabular}
	}
\end{table}

\begin{figure} 
	\centering
	\setlength{\linewidth}{0.47\textwidth}
	\subfigure[]{ \includegraphics[width=\linewidth]{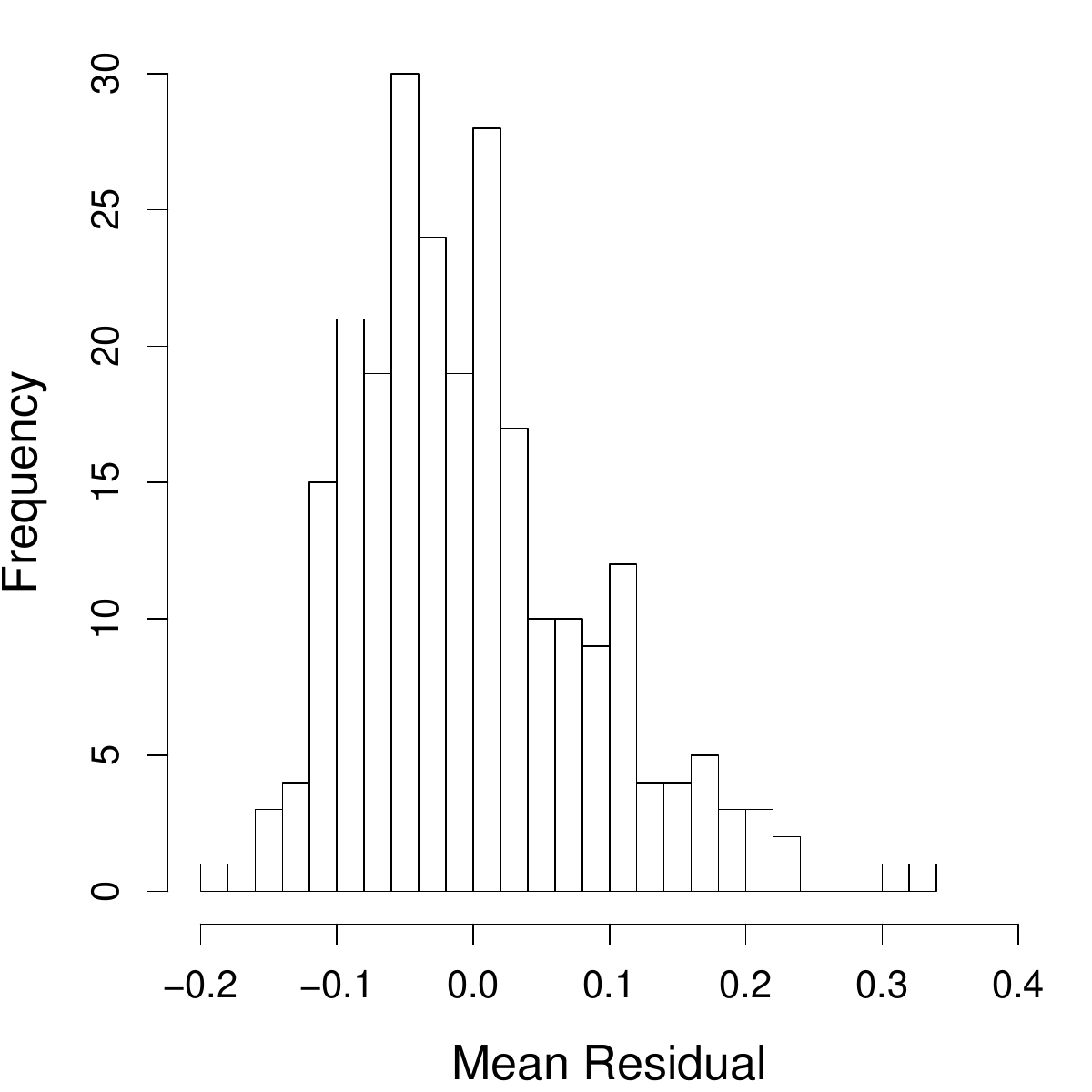} }
	\subfigure[]{ \includegraphics[width=\linewidth]{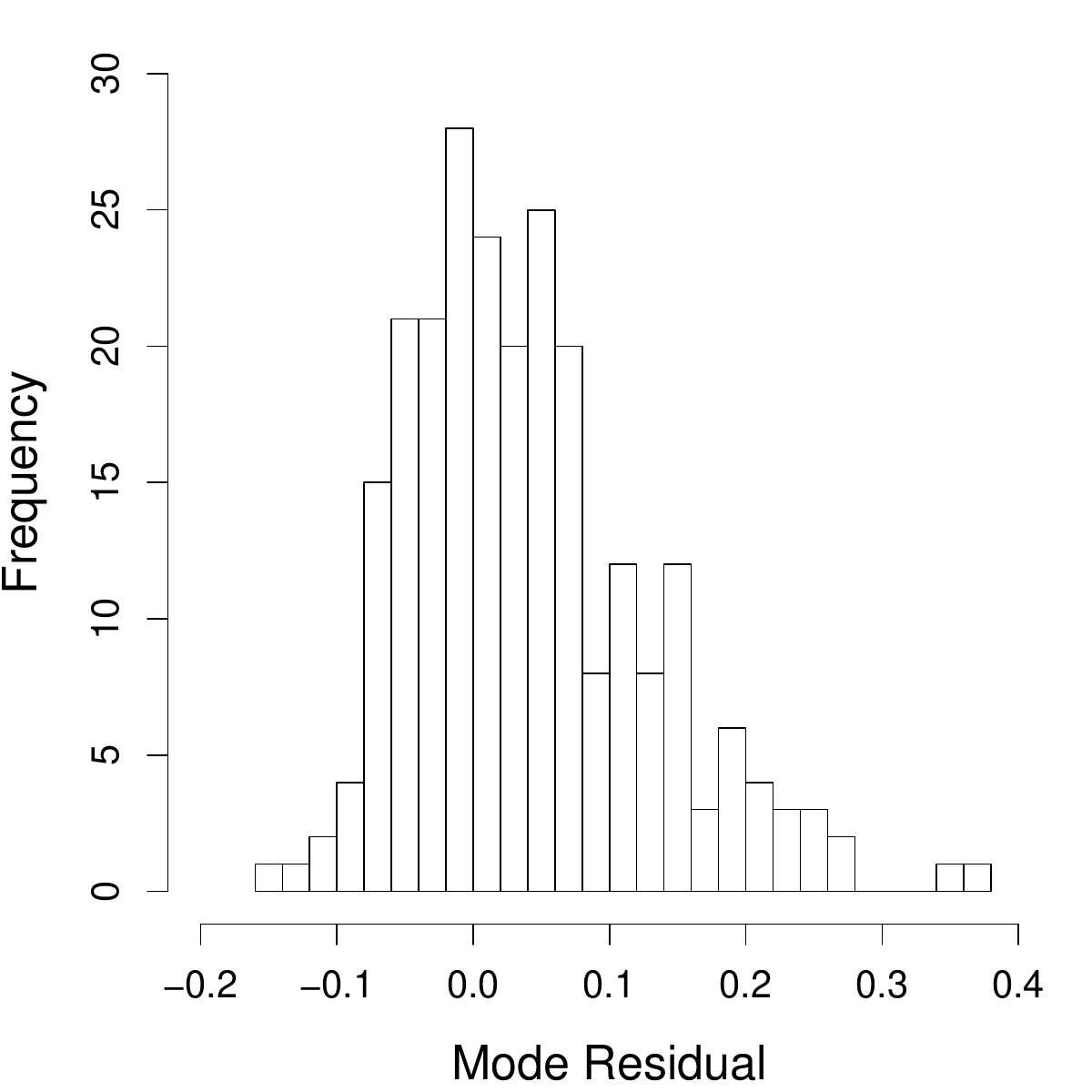} }
	\caption{Histograms of residuals when fitting a beta mean model (in (a)) and when fitting a beta mode model (in (b)) to the ADNI data.}
	\label{f:adniresid}
\end{figure}

Figure~\ref{f:adniresid} presents the histogram of mean residuals and the histogram of mode residuals resulting from beta mean regression and beta mode regression, respectively. Both histograms suggest a right-skewed distribution of the ADAS-11 score conditional on the two volumetric measures, and the two residual distributions are overall similar. Despite such similarity, it is worth stressing that, in general, interpretations of a covariate effect inferred by the two models are different even though we use the same notations in Table~\ref{Qm:MLE} for regression coefficients under different regression models. To avoid such abuse of notation, let us write the mean function under the beta mean regression model as $\mu(\bX)=g(b_0+\bb_1^\T \bX)$, in contrast to the mode function under the beta mode regression model, $\theta(\bX)=g(\beta_0+\bbeta_1^\T \bX)$. Under the parameterization leading to the beta mean model, the mode function is $\theta(\bX)=\{\phi\mu(\bX)-1\}/(\phi-2)$. This indicates that, when $\phi$ is large, in particular, much larger than one, one has $\theta(\bX)\approx \mu(\bX)$.  Similarly, under the parameterization for the beta mode model, the mean function is $\mu(\bX)=\{m\theta(\bX)+1\}/(m+2)$, suggesting that, when $m$ is much larger than one, $\mu(\bX)\approx \theta(\bX)$. According to Table~\ref{Qm:MLE}, $\hat \phi\approx 17.99$ and $\hat m \approx 15.99$, both fairly large compared to one, which can be the reason for the similarity between the MLEs of the two sets of covariate effects under the two models, $\hat \bb_1$ and $\hat\bbeta_1$, in terms of magnitude and statistical significance. This serves as an example where beta mean regression and beta model regression perform similarly in terms of identifying influential predictors. With the rich data information for a large collection of biomarkers available in ADNI, one may consider carrying out variable selection based on beta mode regression, which is beyond the scope of the current study. In the follow-up project, we will give variable selection based on mode regression a more careful treatment, where we formulate flexible regression models built upon the currently proposed beta or GBP mode model that allow heavier tails (to capture severe outliers) and multi-modality. By then we will \mytextcolor{have a fairer comparison} with the variable selection procedure proposed by \citet{NIPS2017_6743} applying to ADNI data that is based on nonparametric mode regression.

We next construct prediction intervals based on the estimated densities from beta mode regression and GBP mode regression following the method described in Section~\ref{s:pred}. For each regression model, we compare the mode-based prediction interval, the mean-based prediction interval, and the median-based prediction interval, that is, $\mathcal{PI}_\theta(\cdot,q)$, $\mathcal{PI}_\mu(\cdot,q)$, and $\mathcal{PI}_\nu(\cdot,q)$, for a given $q$ in regard to their empirical coverage probabilities and widths. A five-fold cross validation procedure is used to obtain the empirical coverage probability of a considered type of prediction interval. Among the three types of prediction intervals based on different central tendency measures, the narrower interval that also has an empirical coverage probability close to $q$ is more preferable. Table~\ref{adni:prediction} reports summary statistics of these prediction intervals. Comparing the three types of prediction intervals with a fixed $q$ under each mode regression model, one can see that  $\mathcal{PI}_\theta(\cdot,q)$ tends to be the narrowest among the three when $q$ is small, while all possessing empirical coverage probabilities close to the nominal level. Hence, in this application of assessing an LMCI subject's extent of cognitive impairment in the near future, using the mode tends to provide more accurate prediction than when using the mean or median. 

\begin{table}[h] 
	\caption{Five-fold cross-validated coverage probabilities of mode-based prediction intervals, $\mathcal{PI}_\theta(\cdot, q)$, mean-based prediction intervals, $\mathcal{PI}_\mu(\cdot, q)$, and median-based prediction intervals, $\mathcal{PI}_\nu(\cdot, q)$, under each regression model for the ADNI data. Numbers in parentheses are the average widths of the prediction intervals.} 
	\label{adni:prediction}
	\centering
	{
		\begin{tabular}{ccccccc}
			\hline\noalign{\smallskip}
			$q$ & \multicolumn{3}{c}{\mytextcolor{beta} mode model} & \multicolumn{3}{c}{GBP mode model}  \\
			\noalign{\smallskip}\hline\noalign{\smallskip}
			& $\mathcal{PI}_\theta$  & $\mathcal{PI}_\mu$ & $\mathcal{PI}_\nu$ & $\mathcal{PI}_\theta$ & $\mathcal{PI}_\mu$ & $\mathcal{PI}_\nu$\\ 
			0.1 & 0.127 (0.021) & 0.086 (0.022) & 0.094 (0.022) & 0.078 (0.020) & 0.098 (0.028) & 0.151 (0.025) \\
			0.2 & 0.233 (0.043) & 0.216 (0.044) & 0.224 (0.043) & 0.159 (0.041) & 0.253 (0.053) & 0.269 (0.048)\\
			0.5 & 0.531 (0.114) & 0.531 (0.114) & 0.531 (0.114) & 0.522 (0.114) & 0.527 (0.117) & 0.522 (0.114)\\
			\noalign{\smallskip}\hline
		\end{tabular}
	}
\end{table}

\section{Discussion}
\label{s:discussion}
We propose two classes of regression models for studying the association between a bounded response and covariates via inferring the conditional mode of the response. Among all existing regression methodology, only a small subset of them are designed for mode regression, and an even smaller collection of them are in the parametric paradigm. The two mode regression models proposed in our study contribute new regression platforms for association studies when a bounded response is of interest. Under each proposed mode regression model, we have developed model diagnostic tools to detect various forms of inadequate parametric assumptions. 

Besides allowing the mode to depend on covariates, one may consider covariate-dependent shape parameter $m(\bX)$ to expand the class of mode regression models. A more flexible family of mode regression models can be formulated as mixtures of beta or GBP distributions, or mixing beta or GBP with a uniform distribution by mimicking the construction of beta rectangular distributions \citep{hahn2008mixture}. \mytextcolor{These} mixture distributions will allow inclusion of multimodal distributions and distributions with heavier tails than those of beta or GBP distributions. 

The family of GBP distributions is a rare distribution family that directly includes the mode in the parameterization, which makes it especially suitable for mode regression. If, unlike responses in our current study, the support of the response is unknown, then we have additional parameter(s) in the GBP density relating to the support, resulting in a non-regular model. In this case, maximum likelihood estimation can break down, or leads to estimators that do not possess properties one usually sees in an MLE under a regular model \citep{cheng1983estimating}. Parameter estimations and properties of MLEs for parameters in a non-regular GBP regression model demand systematic investigations.

\begin{acknowledgement}
Data collection and sharing for this project was funded by the Alzheimer's Disease
Neuroimaging Initiative (ADNI) (National Institutes of Health Grant U01 AG024904) and
DOD ADNI (Department of Defense award number W81XWH-12-2-0012). ADNI is funded
by the National Institute on Aging, the National Institute of Biomedical Imaging and
Bioengineering, and through generous contributions from the following: AbbVie, Alzheimer’s Association; Alzheimer’s Drug Discovery Foundation; Araclon Biotech; BioClinica, Inc.; Biogen; Bristol-Myers Squibb Company; CereSpir, Inc.; Cogstate; Eisai Inc.; Elan Pharmaceuticals, Inc.; Eli Lilly and Company; EuroImmun; F. Hoffmann-La Roche Ltd and its affiliated company Genentech, Inc.; Fujirebio; GE Healthcare; IXICO Ltd.; Janssen Alzheimer Immunotherapy Research \& Development, LLC.; Johnson \& Johnson Pharmaceutical Research \& Development LLC.; Lumosity; Lundbeck; Merck \& Co., Inc.; Meso Scale Diagnostics, LLC.; NeuroRx Research; Neurotrack Technologies; Novartis Pharmaceuticals Corporation; Pfizer Inc.; Piramal Imaging; Servier; Takeda Pharmaceutical Company; and Transition Therapeutics. The Canadian Institutes of Health Research is providing funds to support ADNI clinical sites in Canada. Private sector contributions are facilitated by the Foundation for the National Institutes of Health (www.fnih.org). The grantee organization is the Northern California Institute for Research and Education, and the study is coordinated by the Alzheimer’s Therapeutic Research Institute at the University of Southern
California. ADNI data are disseminated by the Laboratory for Neuro Imaging at the
University of Southern California. 
\end{acknowledgement}
\vspace*{1pc}

\noindent {\bf{Conflict of Interest}}

\noindent {\it{The authors have declared no conflict of interest. }}

\bibliographystyle{apalike}
\bibliography{GBPref}


\end{document}



 \centerline{\large\bf Parametric mode regression for bounded responses}

\vspace{.25cm}
\centerline{Haiming Zhou and Xianzheng Huang}
\vspace{.4cm}
 \centerline{\it Northern Illinois University and University of South Carolina}
\vspace{.55cm}
 \centerline{\bf Supplementary Material}
\vspace{.55cm}
\fontsize{9}{11.5pt plus.8pt minus .6pt}\selectfont
\noindent
\par

\setcounter{section}{0}
\setcounter{equation}{0}
\def\theequation{S\arabic{section}.\arabic{equation}}
\def\thesection{S\arabic{section}}
\def\thefigure{S\arabic{figure}}
\def\thetable{S\arabic{table}}

\fontsize{12}{14pt plus.8pt minus .6pt}\selectfont

\section{Algorithm for estimating $p$-value for the GBP score test}\label{Append:test}
\setcounter{equation}{0}
This section provides additional information for Section 3.2 in the main paper.
\begin{itemize}
	\item[Step 1] Fit the assumed mode regression model to the observed data, $\calD=\{(Y_i, \bX_i), \, i=1, \ldots, n\}$. Denote by $\hat\bOmega$ the resultant MLE for $\bOmega$. 
	\item[Step 2] Compute $Q(\hat\bOmega; \calD)$ defined in Equation (3.3) in the main paper. 
	\item[Step 3] For $b=1, \ldots, B$, generate $Y_{i,b}$ from $\textrm{GBP}(\hat \theta(\bX_i), \, \hat m)$, where $\hat \theta(\bX_i)$ and $\hat m$ are MLEs for $\theta(\bX_i)$ and $m$, respectively. This produces $B$ sets of bootstrap data, $\calD_b=\{(Y_{i,b}, \bX_i), \, i=1, \ldots, n\}$, for $b=1, \ldots, B$.
	\item[Step 4] Fit the assumed mode regression model to each bootstrap data set, $\calD_b$, and obtain the MLE for $\bOmega$, denoted by $\hat \bOmega_b$, for $b=1, \ldots, B$. 
	\item[Step 5] Compute the test statistic, $Q(\hat \bOmega_b; \calD_b)$, according to Equation (3.3) in the main paper, for $b=1, \ldots, B$. 
	\item[Step 6] Compute the estimated $p$-value defined by $B^{-1}\sum_{b=1}^B I\{ Q(\hat \bOmega_b; \calD_b)>Q(\hat\bOmega; \calD)\}$.
\end{itemize}

\section{Additional results for simulation studies}
\subsection{\mytextcolor{Additional results for Section 4.3}}
\begin{figure} 
	\centering
	\setlength{\linewidth}{0.47\textwidth}
	\subfigure[]{ \includegraphics[width=\linewidth]{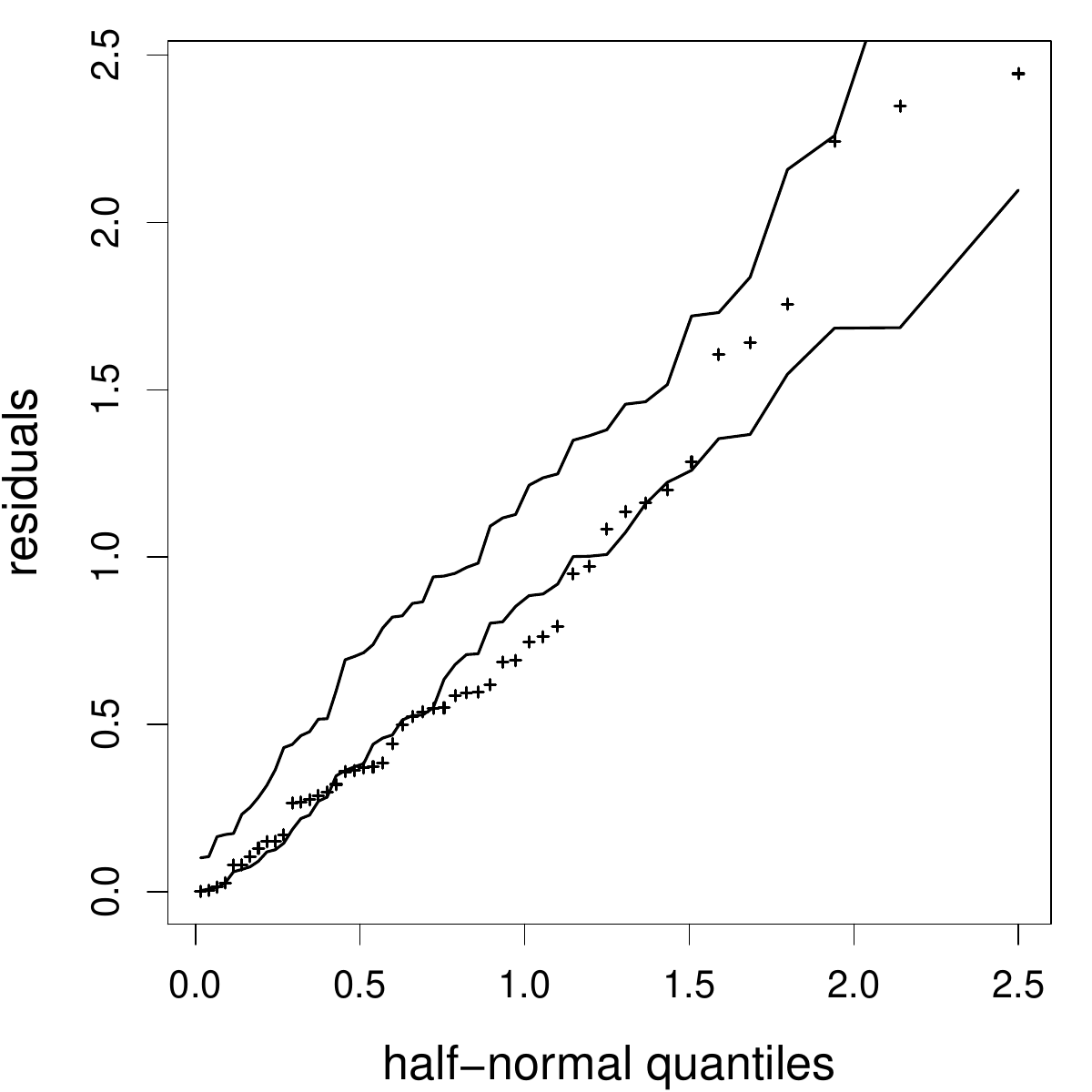} }
	\subfigure[]{ \includegraphics[width=\linewidth]{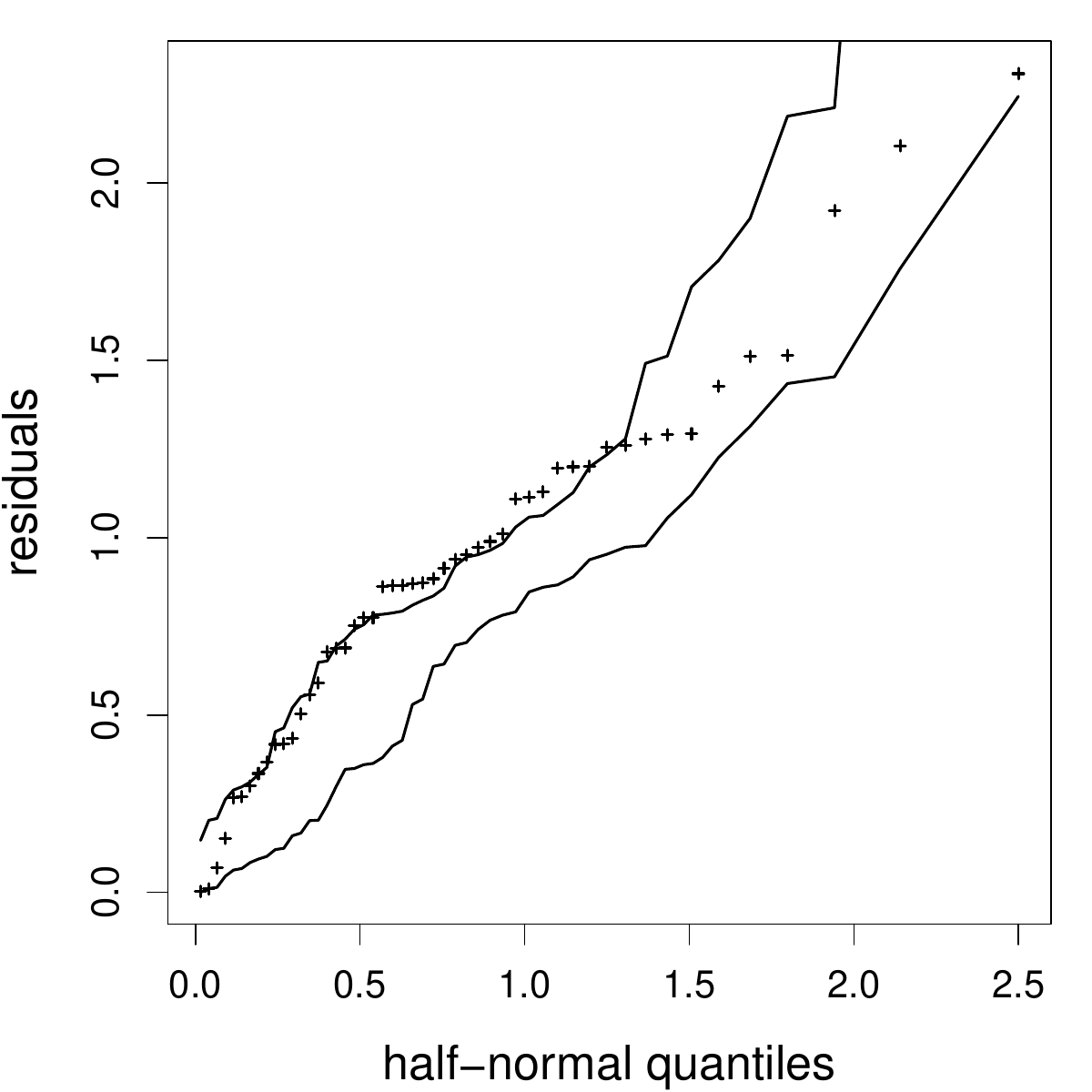} }\\
	\subfigure[]{ \includegraphics[width=\linewidth]{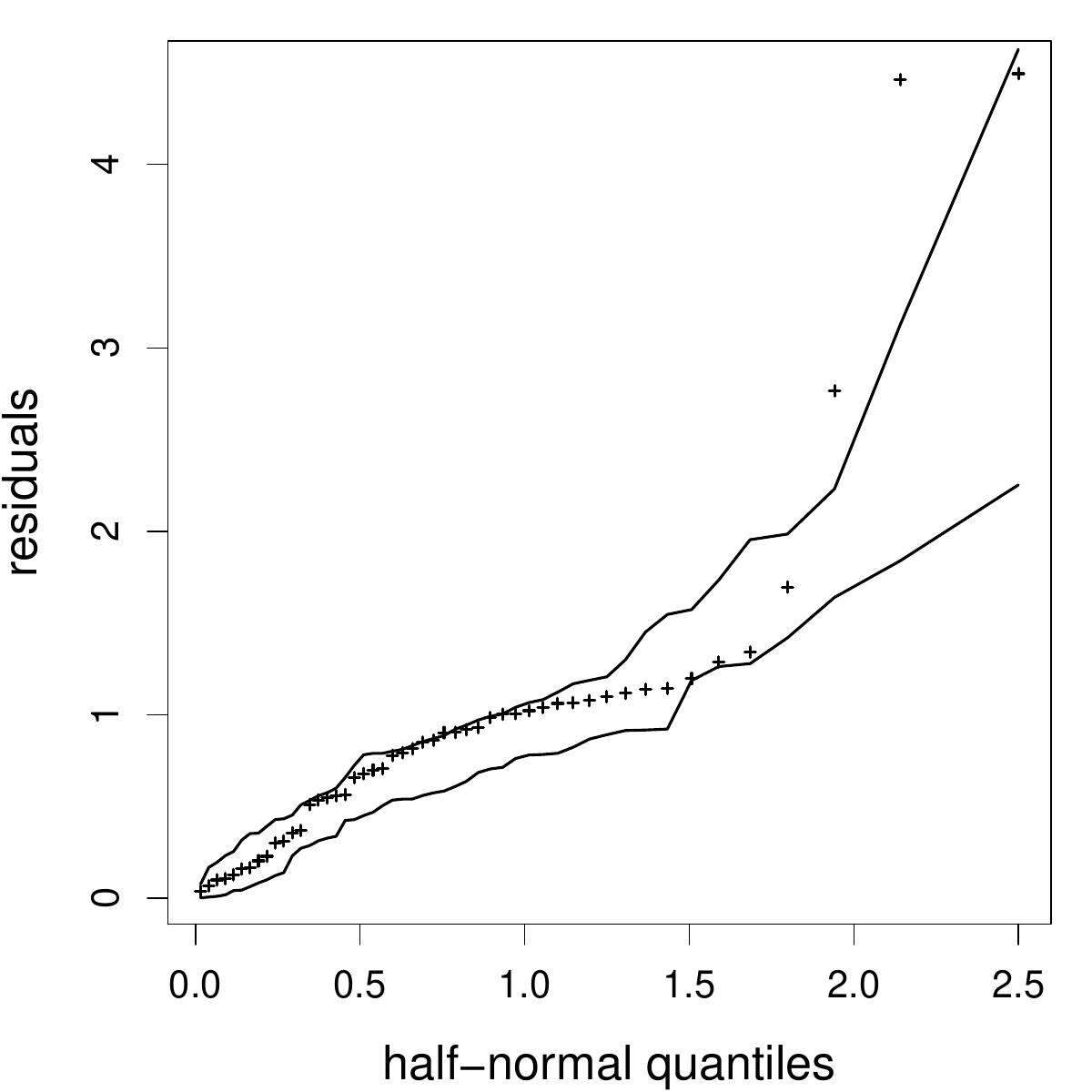} }
	\subfigure[]{ \includegraphics[width=\linewidth]{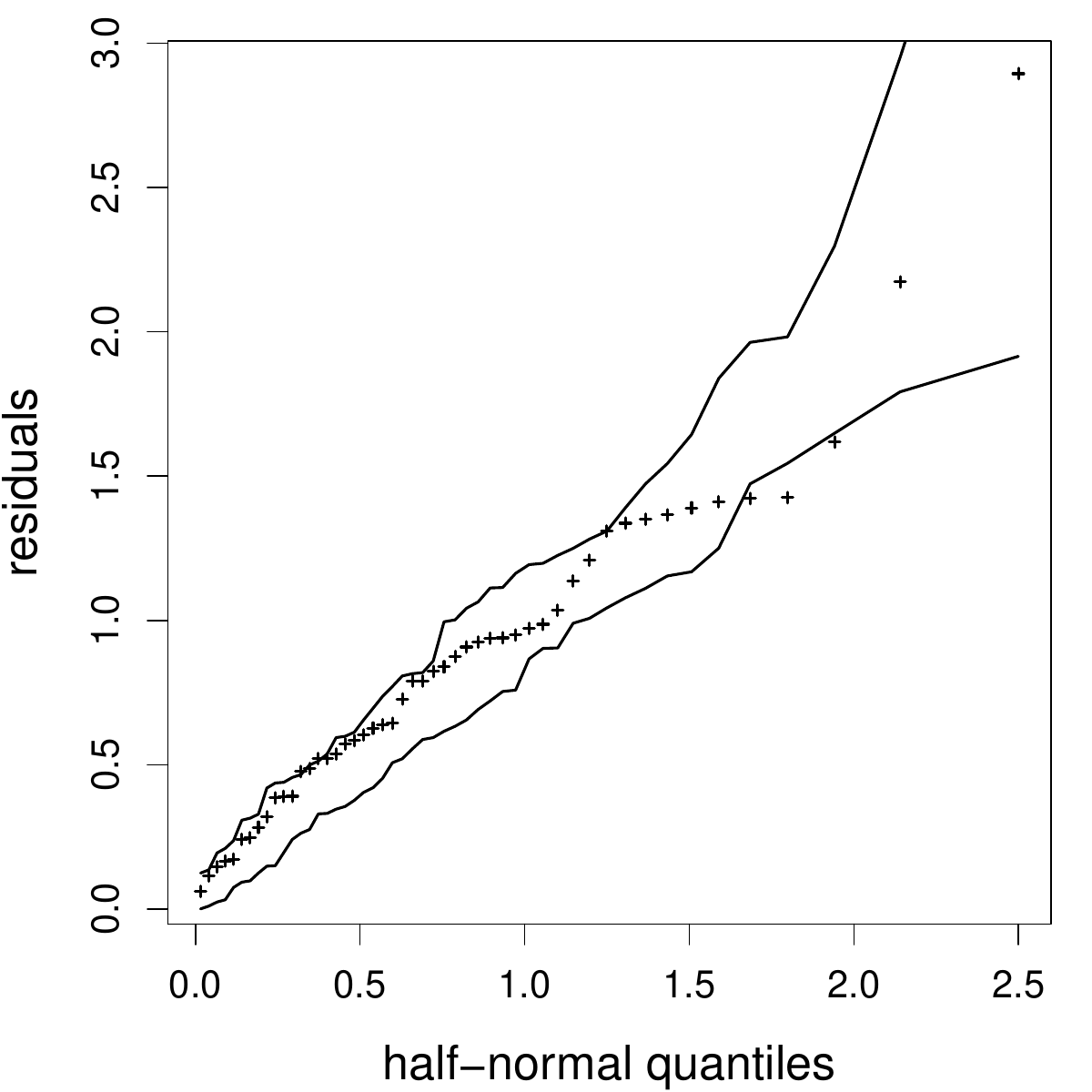} }
	\caption{Half-normal residual plots with simulated envelopes based on one random sample of size $n=50$ when one assumes a beta mode model for data generated from (B1)--(B4), shown in (a)--(d), respectively.}
	\label{Sim:envelope-beta-n50}
\end{figure}

\begin{figure} 
	\centering
	\setlength{\linewidth}{0.47\textwidth}
	\subfigure[]{ \includegraphics[width=\linewidth]{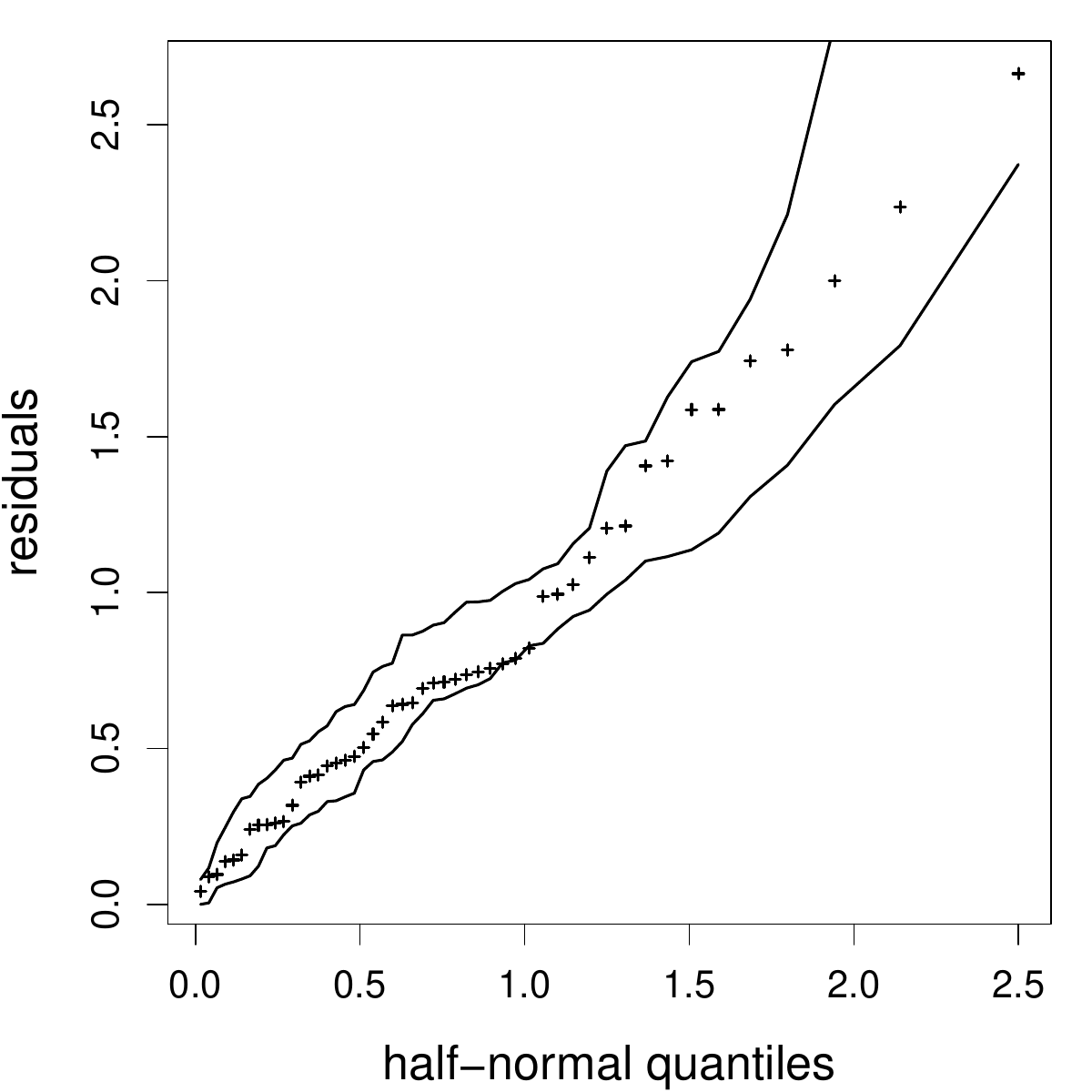} }
	\subfigure[]{ \includegraphics[width=\linewidth]{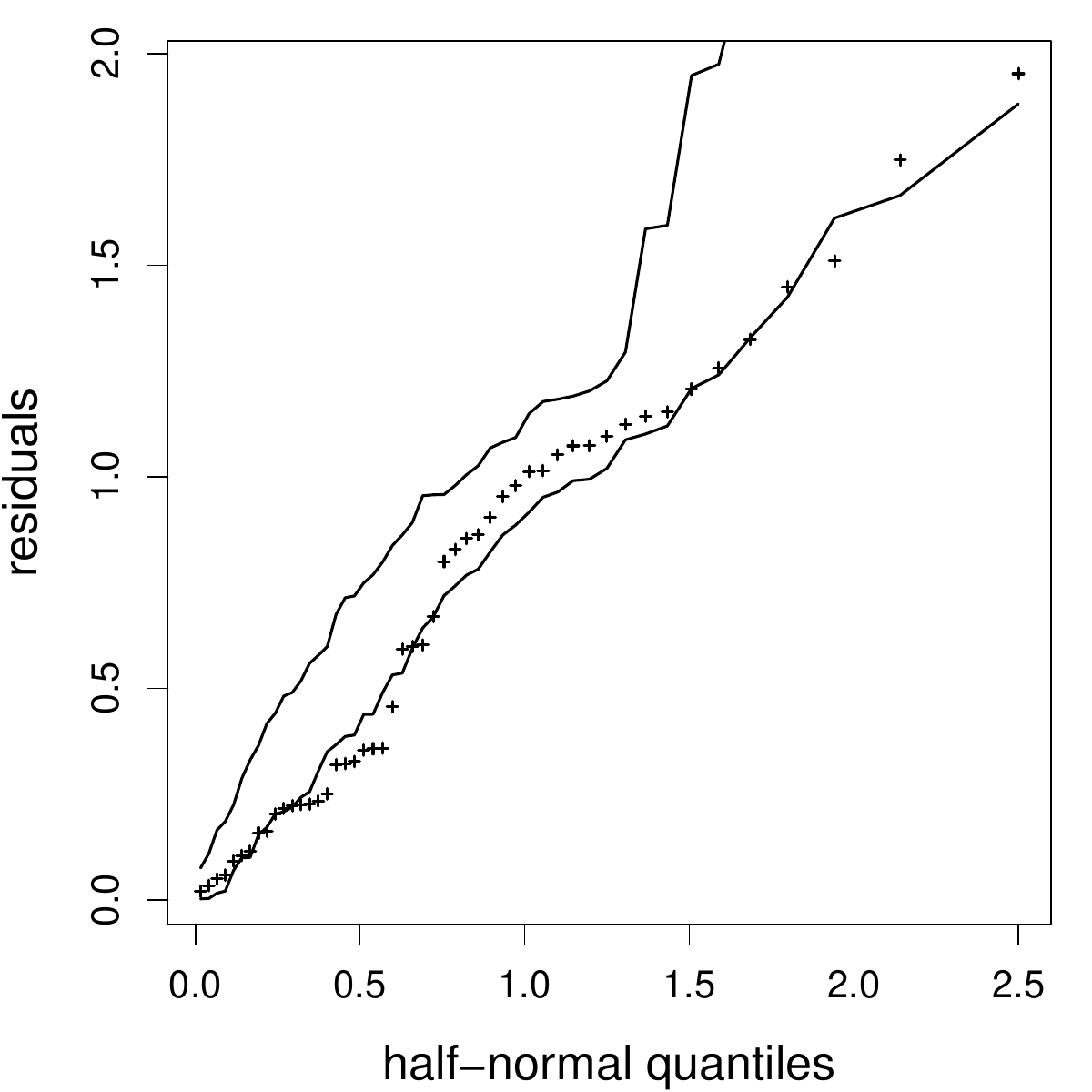} }\\
	\subfigure[]{ \includegraphics[width=\linewidth]{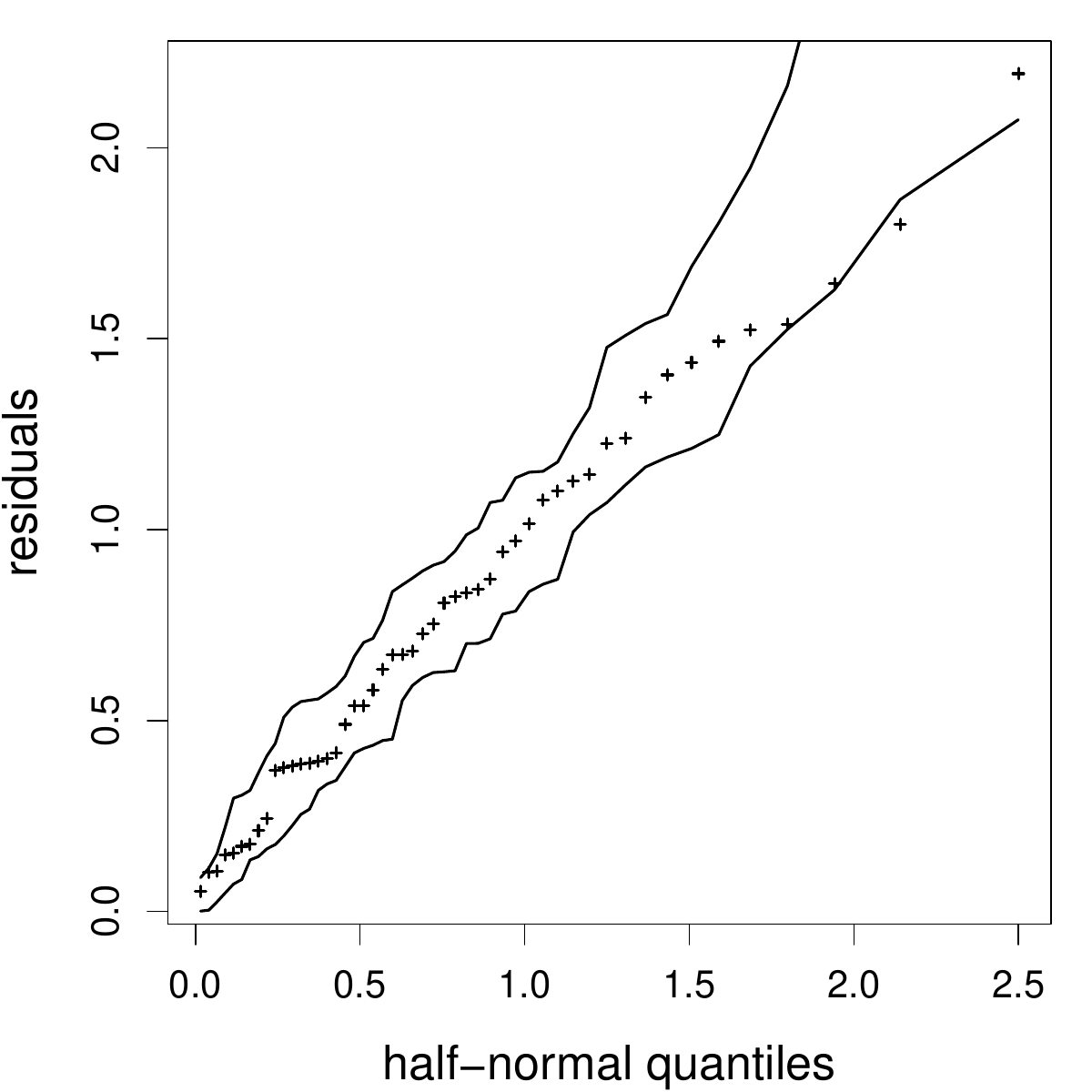} }
	\subfigure[]{ \includegraphics[width=\linewidth]{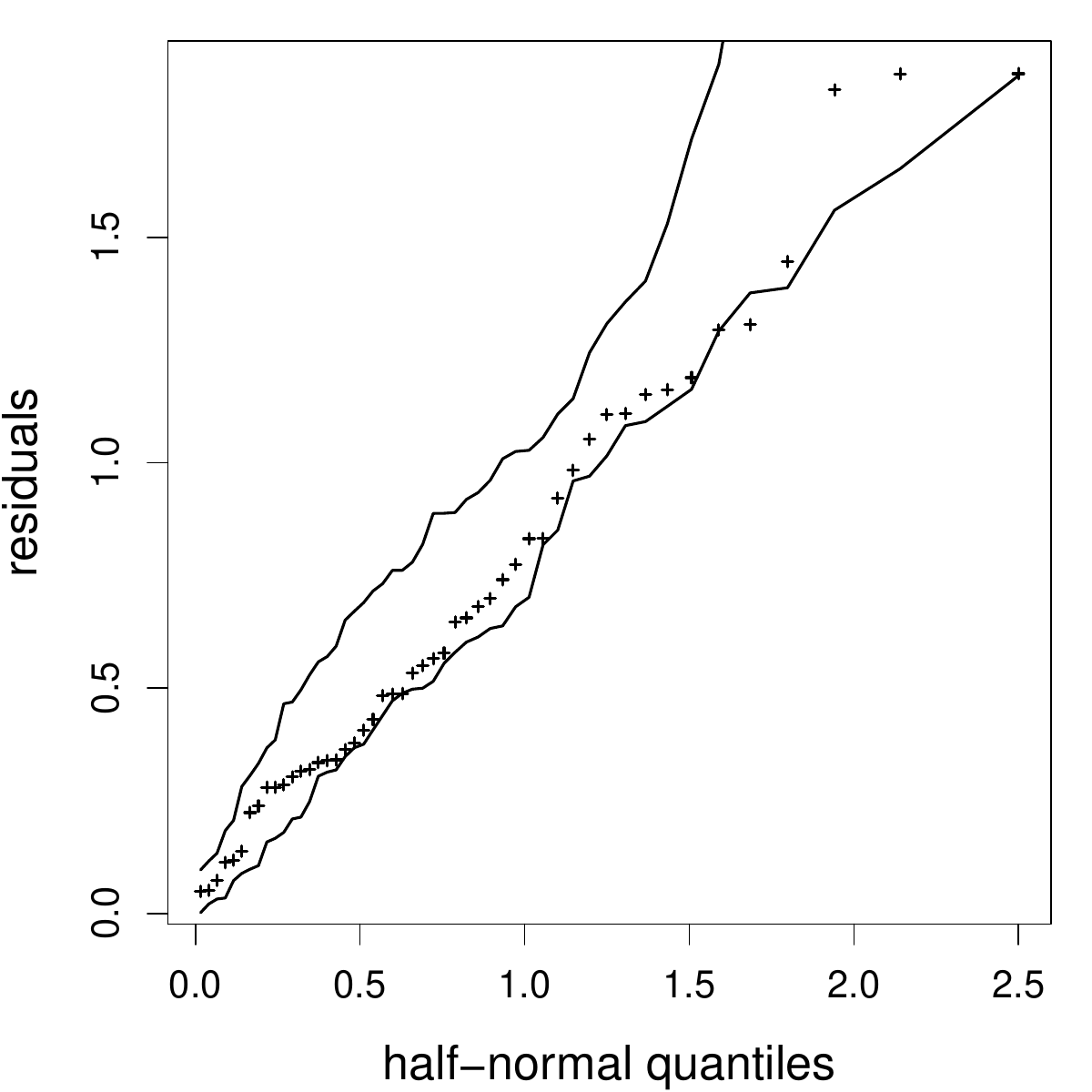} }
	\caption{Half-normal residual plots with simulated envelopes based on one random sample of size $n=50$ when one assumes a GBP mode model for data generated from (G1)--(G4), shown in (a)--(d), respectively.}
	\label{Sim:envelope-gbp-n50}
\end{figure}

\mytextcolor{This section provides additional information for Section 4.3 in the main paper. Figure~\ref{Sim:envelope-beta-n50} demonstrates the half-normal residual plots obtained based on one data set of size $n=50$ generated from each of (B1)--(B4), where $m=80$ in (B1)--(B3), and $m=10$ in (B4).  Figure~\ref{Sim:envelope-gbp-n50} demonstrates the half-normal residual plots using one data set of size $n=50$ generated from each of (G1)--(G4), where $m=10$ in (G1)--(G3), and $m=80$ in (G4). Comparing the results with those obtained in Section 4.3 of the main paper, expectedly, we see that the envelops provide relatively weaker indication of lack-of-fit in the presence of misspecification.}

\begin{figure} 
	\centering
	\setlength{\linewidth}{0.47\textwidth}
	\subfigure[]{ \includegraphics[width=\linewidth]{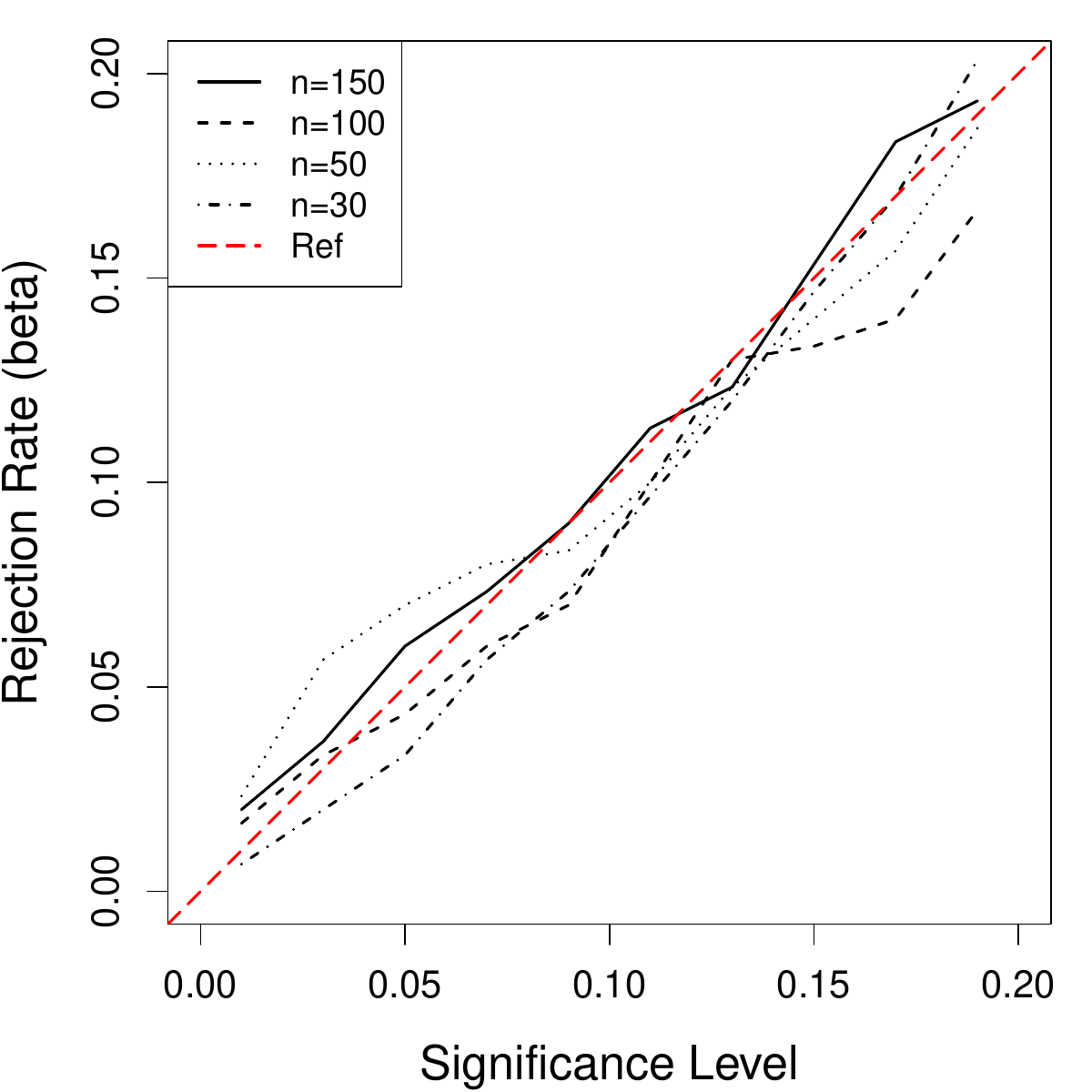} }
	\subfigure[]{ \includegraphics[width=\linewidth]{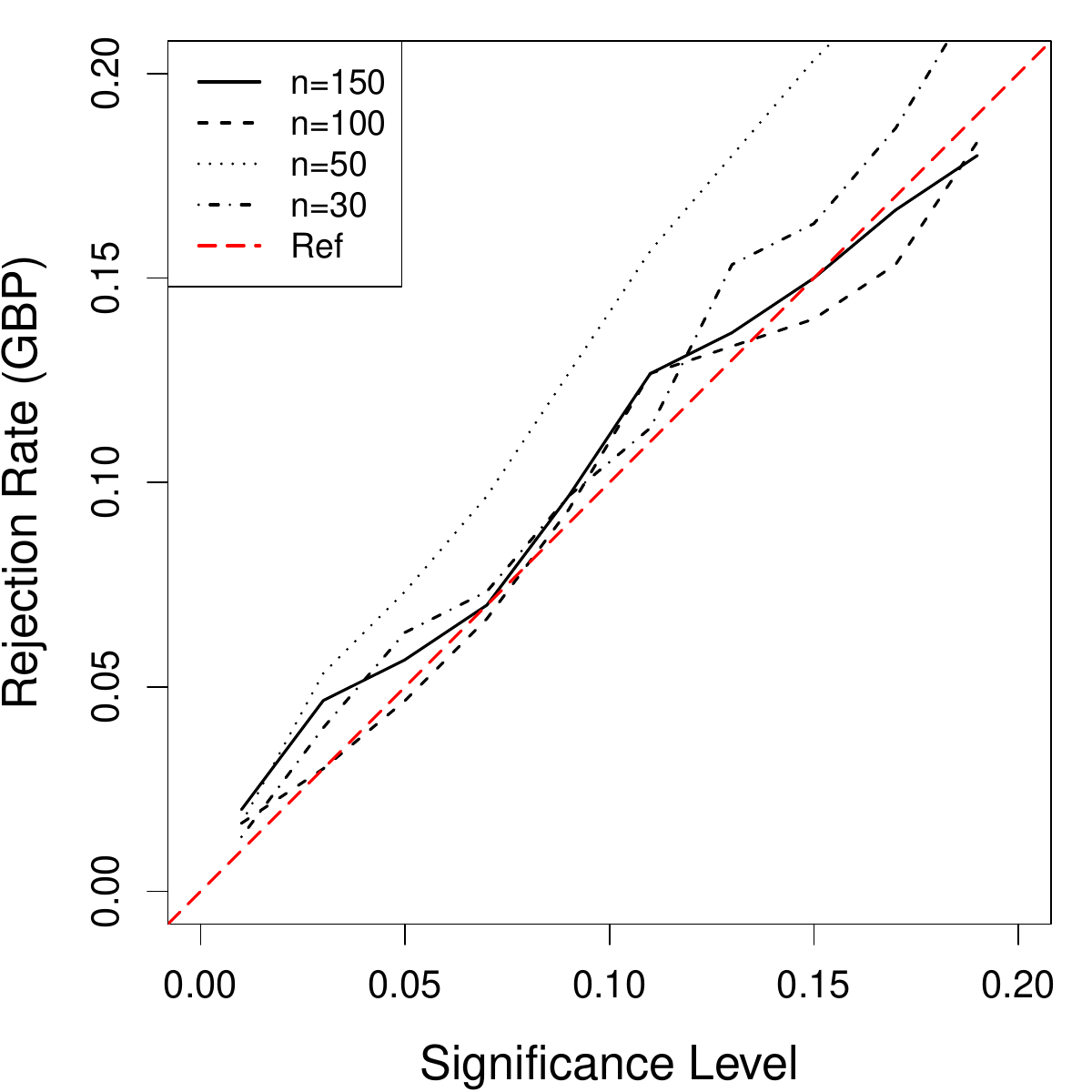} }
	\caption{\label{f:score:tails}Rejection rates across 300 Monte Carlo replicates associated with the score tests under (B1) when a beta mode model is assumed (panel a), and under (G1) when a GBP mode model is assumed (panel b).  Red long-dashed lines are $45^\circ$ reference lines.}
\end{figure}

\mytextcolor{Figure~\ref{f:score:tails} presents the empirical type I error rate of the score tests when a beta mode model or a GBP mode model is assumed, where the empirical type I error rate of a test is defined as the rejection rate of the test for a given significance level $\alpha$ across 300 Monte Carlo replicates under each true model specification. We can see that as sample size increases, the empirical type I rate is approaching to the significance level, indicating that the bootstrap procedure can estimate the tail of the null distribution of test statistic well enough to preserve the right size of the proposed score tests. }

\subsection{\mytextcolor{Additional results for Section 4.4}}
\begin{figure} 
	\centering
	\setlength{\linewidth}{0.45\textwidth}
	\subfigure[]{ \includegraphics[width=\linewidth]{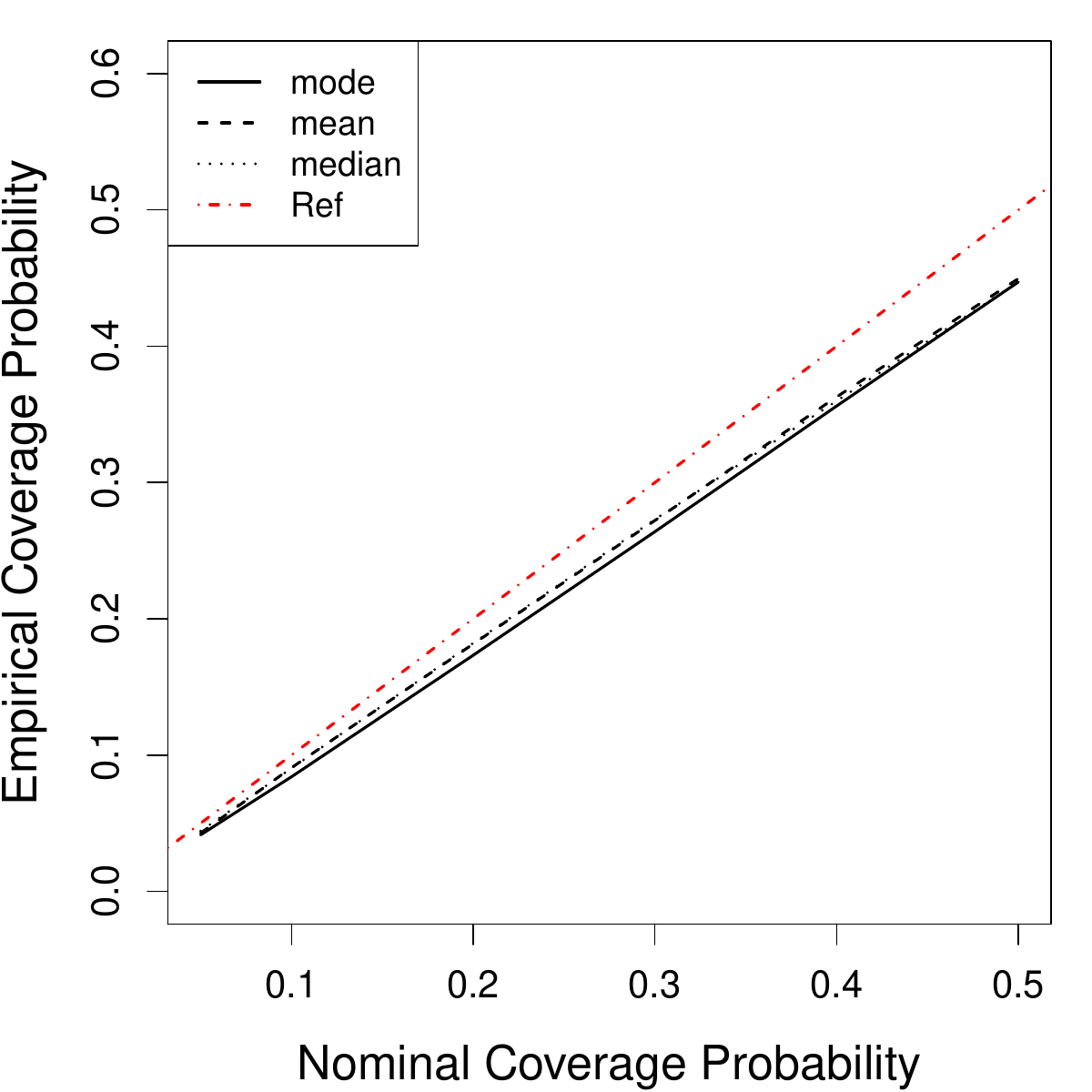} }
	\subfigure[]{ \includegraphics[width=\linewidth]{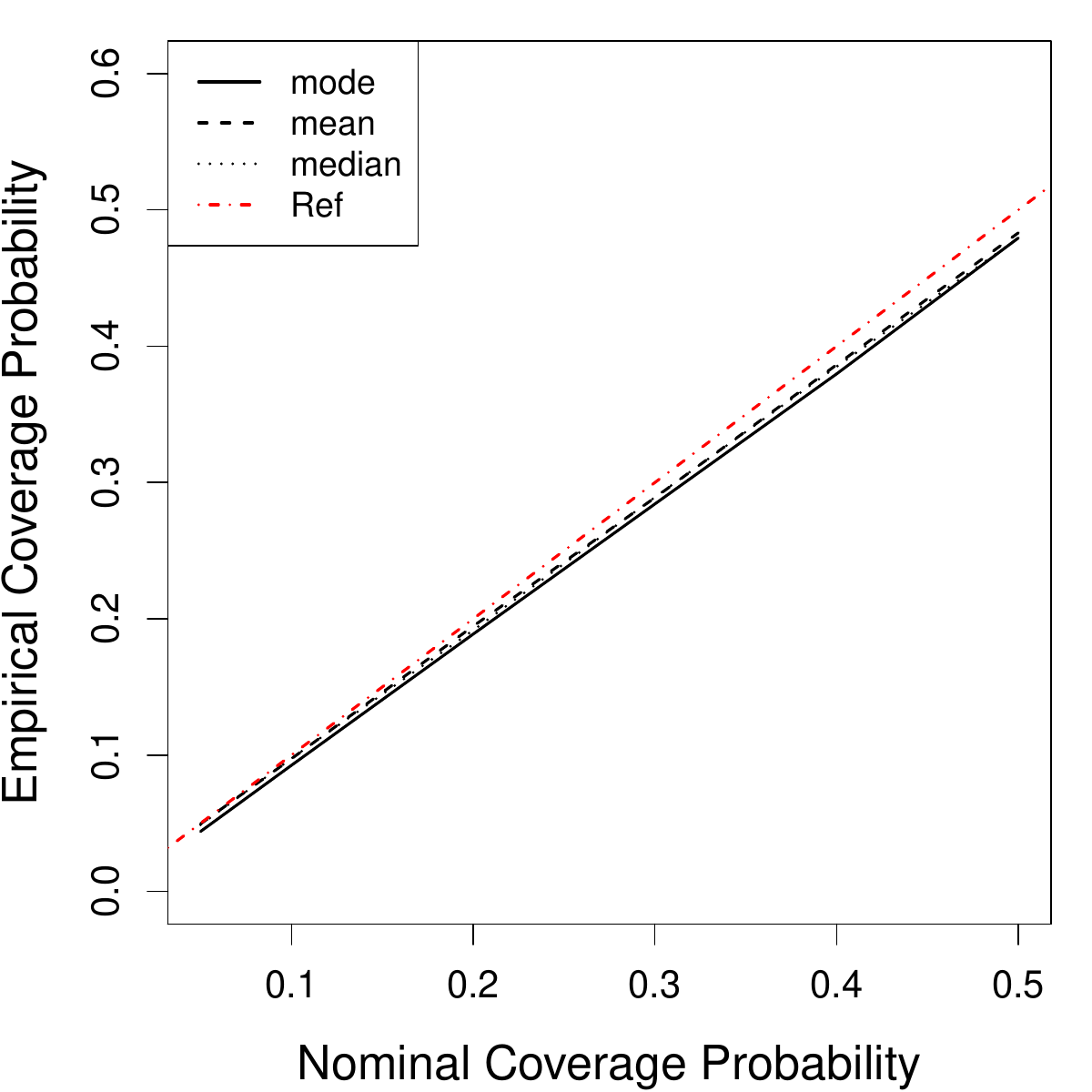} }\\
	\subfigure[]{ \includegraphics[width=\linewidth]{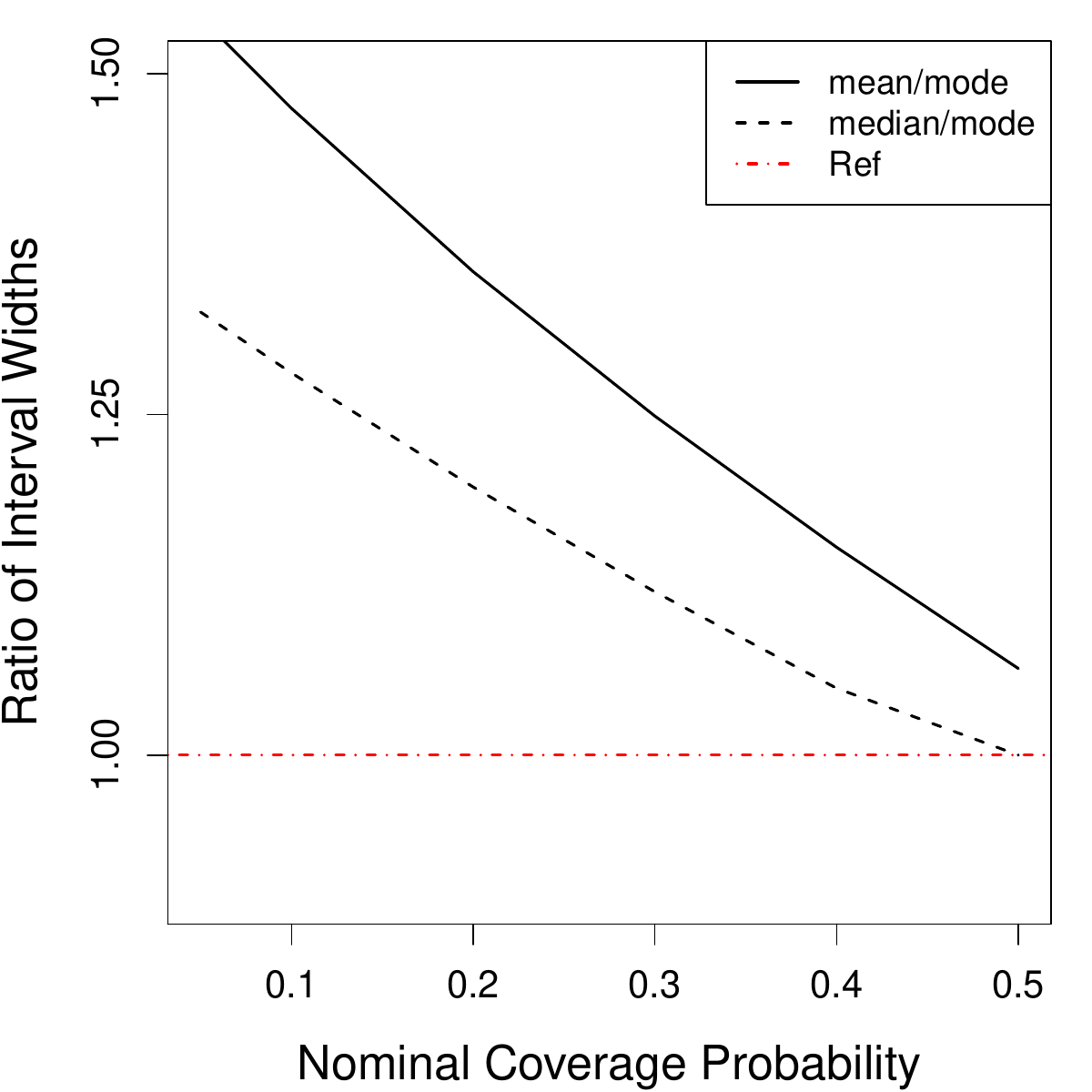} }
	\subfigure[]{ \includegraphics[width=\linewidth]{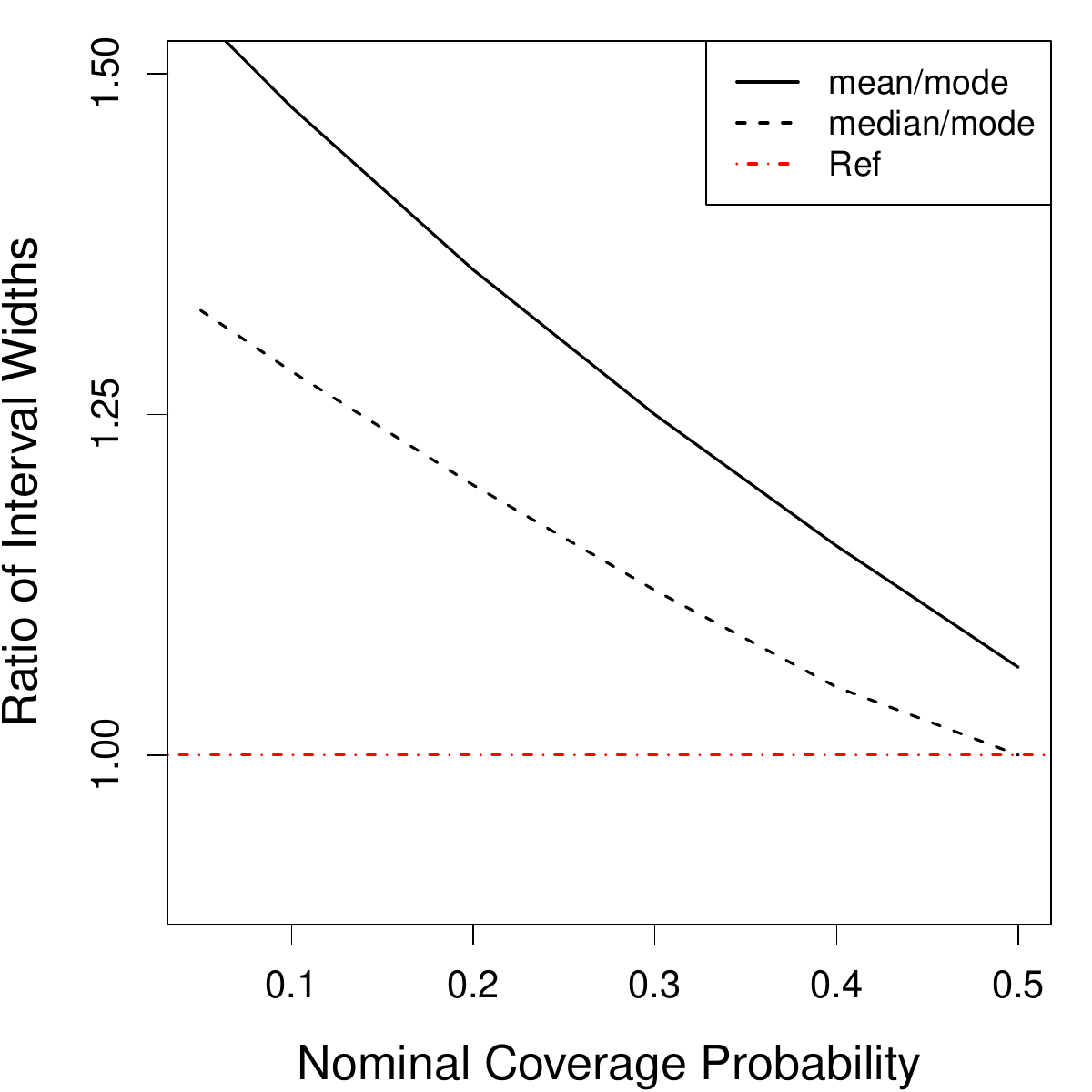} }
	\caption{\label{f:ecpwidthGBP}{\small Prediction intervals based on data from the beta mode model in (G1). Top panels (a) and (b) depict average empirical coverage probabilities (across 300 Monte Carlo replicates) of mode-based prediction intervals (solid lines), those of mean-based prediction intervals (dashed lines), and those of median-based prediction intervals (dotted lines) versus nominal coverage probabilities. Red dash-dotted lines are $45^\circ$ reference lines. Lower panels (c) and (d) depict ratios of the average width of mean-based prediction intervals over that of mode-based prediction intervals (solid lines) and ratios of the average width of median-based prediction intervals over that of mode-based prediction intervals (dashed lines) versus nominal coverage probabilities. Red dash-dotted horizontal lines are reference lines at value one. Panels (a) and (c) are for $n=50$. Panels (b) and (d) are for $n=100$.}}
\end{figure}

This section provides additional information for Section 4.4 in the main paper. Figure~\ref{f:ecpwidthGBP} provides prediction intervals when one assumes a GBP mode model and data are generated from (G1) with $\bbeta=(3, 1, 1)$, $m=10$ and $n=50, \mytextcolor{100}$, where $q$ ranges from 0.05 to 0.5. According to these figures, mode-based prediction intervals,  mean-based prediction intervals, and media-based prediction intervals achieve similar empirical coverage probabilities that become closer to the nominal coverage probability as $n$ increases. More importantly, the mode-based prediction interval tends to be narrowest among the three, and the mean-based prediction interval is the widest.

\begin{figure} 
	\centering
	\setlength{\linewidth}{0.45\textwidth}
	\subfigure[]{ \includegraphics[width=\linewidth]{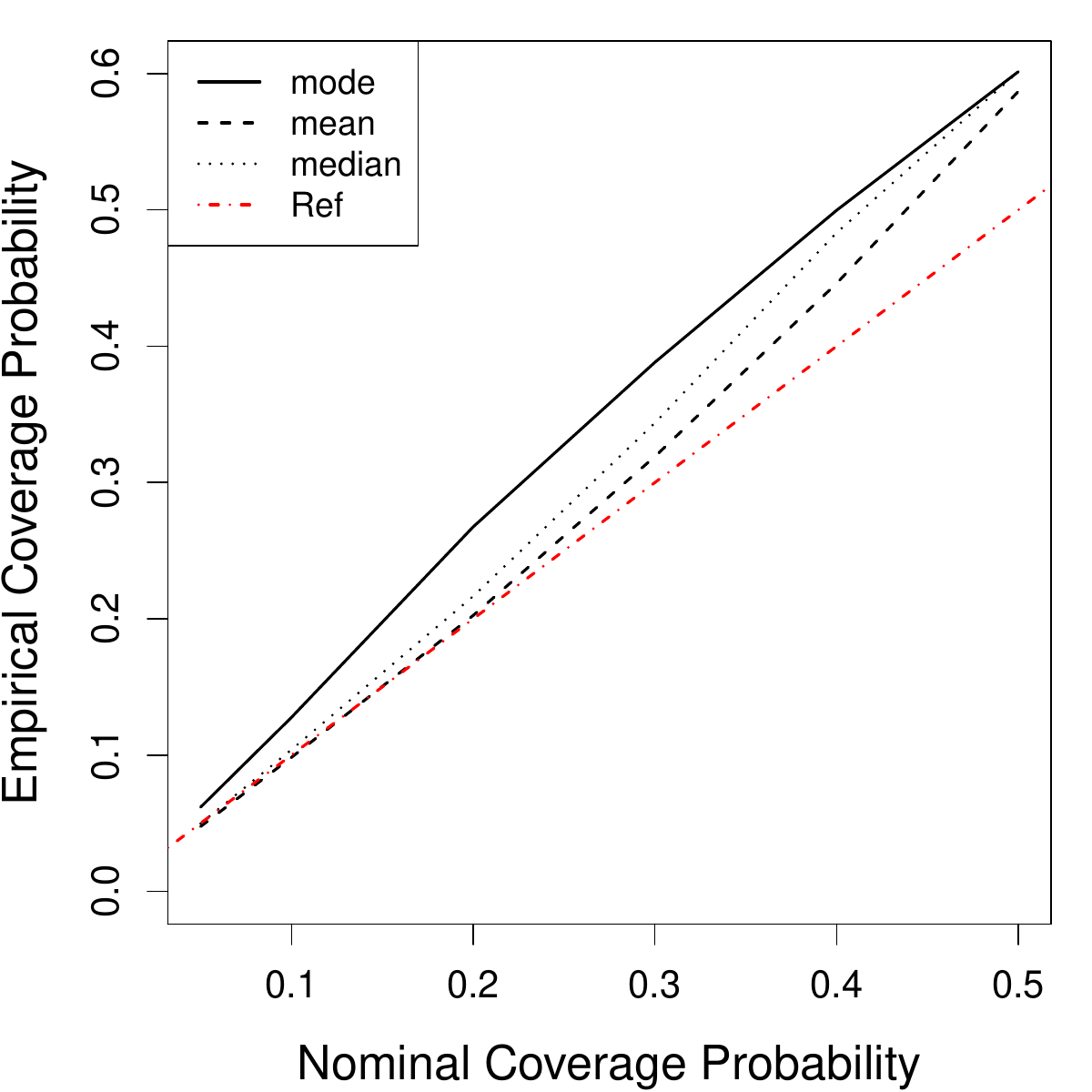} }
	\subfigure[]{ \includegraphics[width=\linewidth]{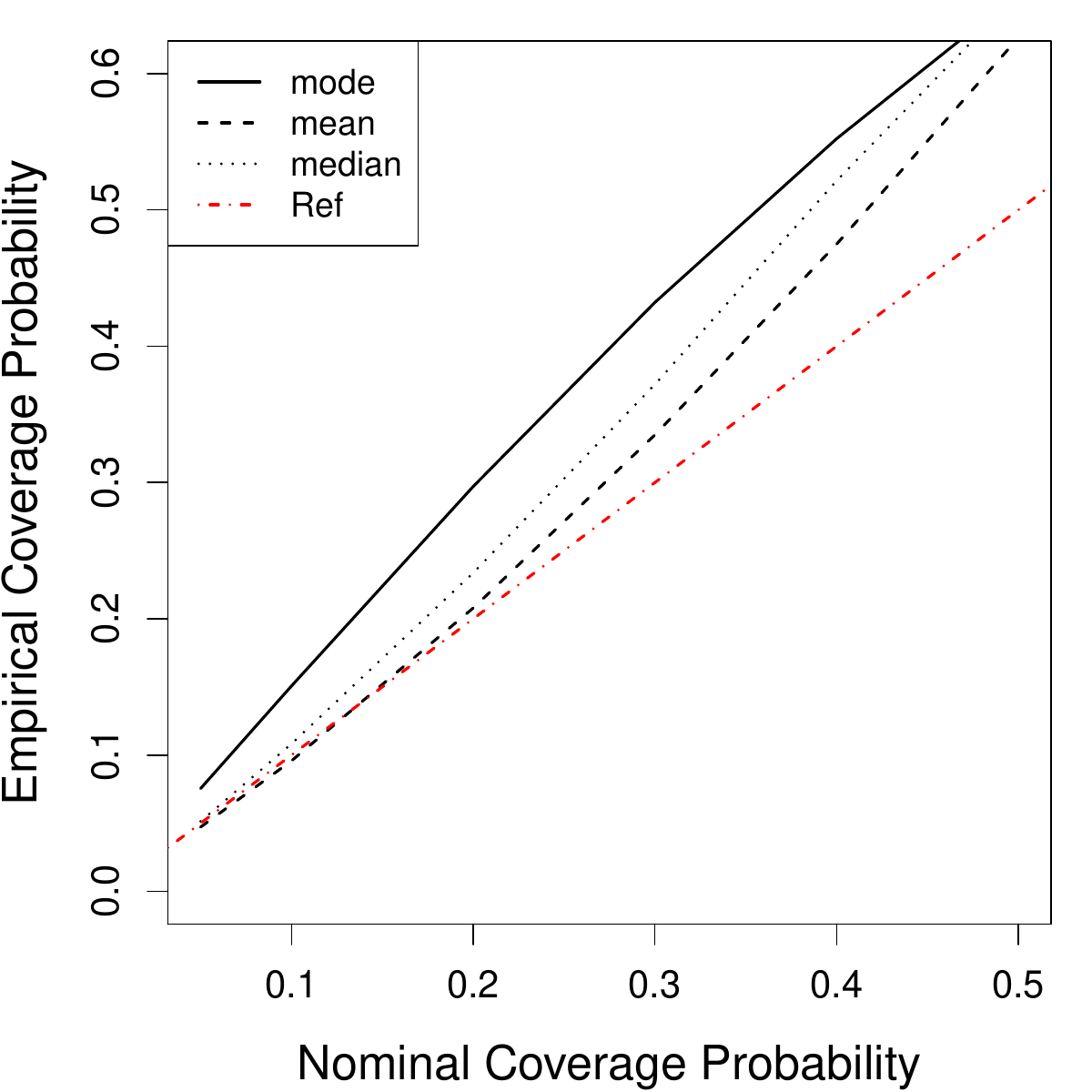} }\\
	\subfigure[]{ \includegraphics[width=\linewidth]{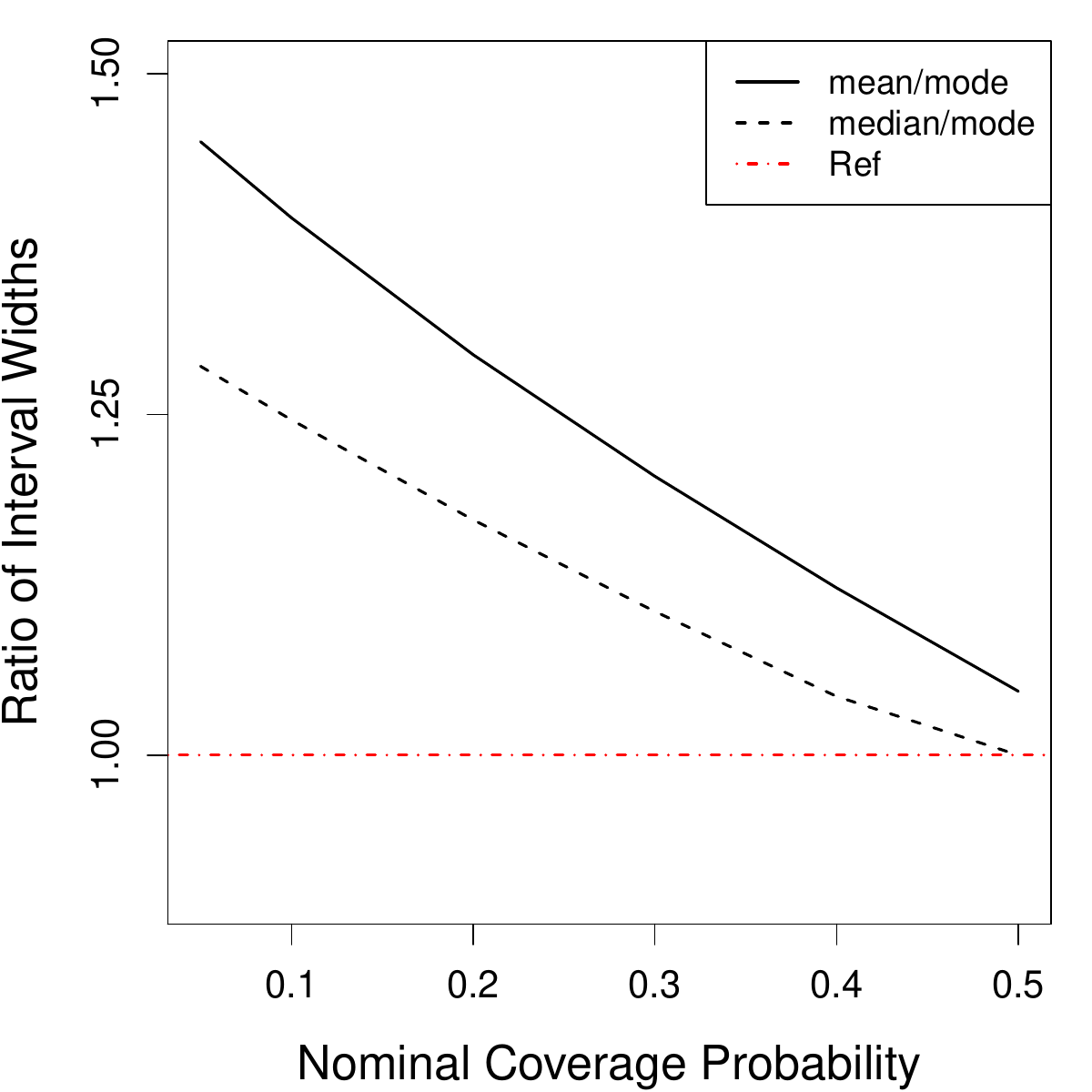} }
	\subfigure[]{ \includegraphics[width=\linewidth]{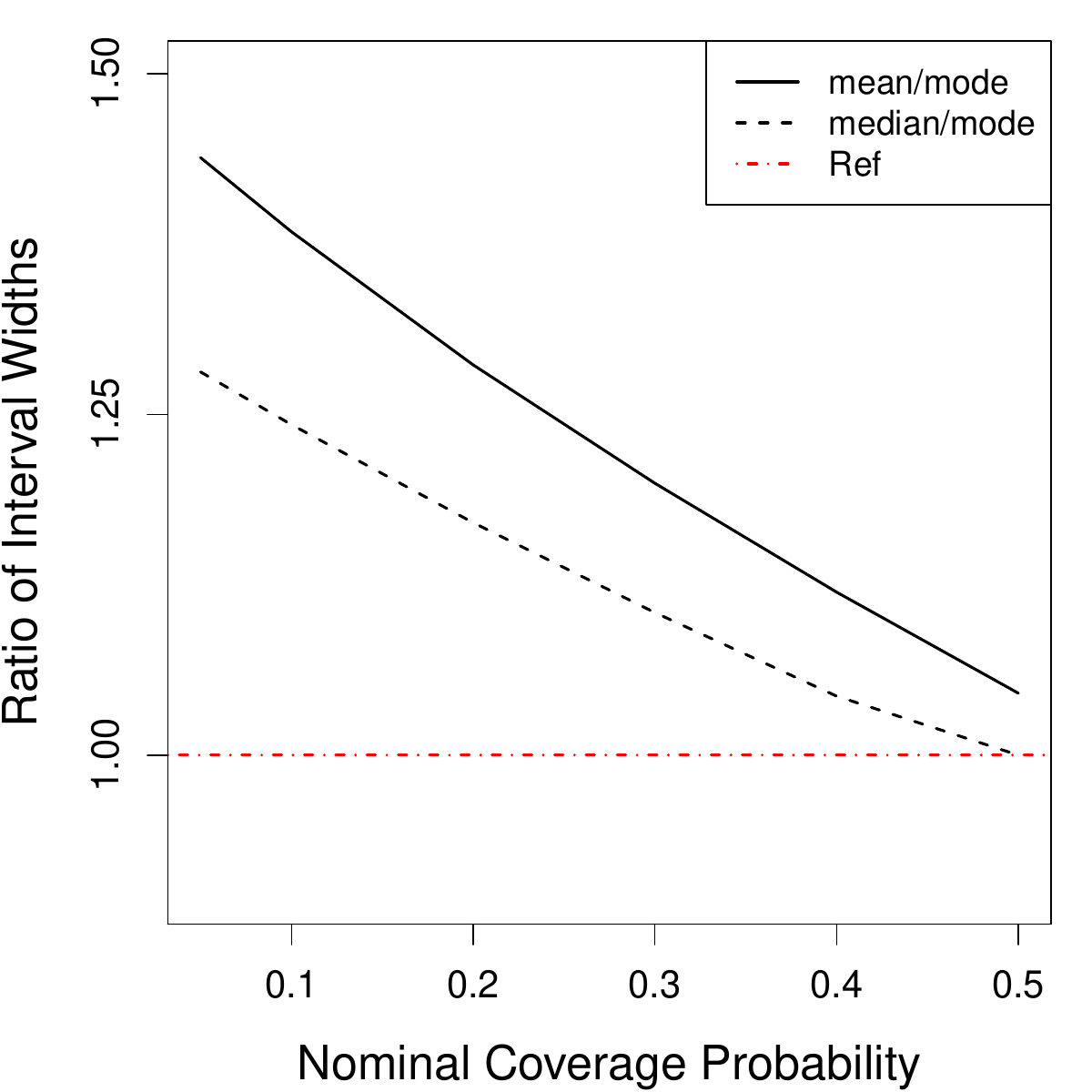} }
	\caption{\label{f:ecpwidthGBPoutlier}{\small Prediction intervals based on beta mode regression using data from the beta mode model in (G1), with outliers replacing 5\% of the original data. Top panels (a) and (b) depict average empirical coverage probabilities (across 300 Monte Carlo replicates) of mode-based prediction intervals (solid lines), those of mean-based prediction intervals (dashed lines), and those of median-based prediction intervals (dotted lines) versus nominal coverage probabilities. Red dash-dotted lines are $45^\circ$ reference lines. Lower panels (c) and (d) depict ratios of the average width of mean-based prediction intervals over that of mode-based prediction intervals (solid lines) and ratios of the average width of median-based prediction intervals over that of mode-based prediction intervals (dashed lines) versus nominal coverage probabilities. Red dash-dotted horizontal lines are reference lines at value one. Panels (a) and (c) are for $n=50$. Panels (b) and (d) are for $n=100$.}}
\end{figure}

To demonstrate the impact of outliers on the aforementioned prediction intervals, we contaminate each of 300 Monte Carlo replicate data sets generated from the GBP mode model by replacing 5\% of randomly chosen responses with random numbers simulated from uniform$(0, t)$, where $t$ is the 0.001-th quantile of the true conditional distribution of the response. This contamination produces data with a heavier (left) tail than the distribution specified in (G1). Despite the model misspecification, we fit the resultant data assuming a GBP model regression model and construct prediction intervals based on the three central tendency measures. Figure~\ref{f:ecpwidthGBPoutlier} shows the comparison between different types of prediction intervals in regard to empirical coverage probability and width. From there, one can see that a direct consequence of fitting (and making predictions based on) a GBP mode model to data from an underlying distribution with a heavier (than assumed) tail is inflated coverage probabilities, despite the choice of central tendency measure for prediction. Interestingly, even though the empirical coverage probability of the mode-based prediction interval is higher than those of the other two types of prediction intervals, the mode-based prediction interval remains the narrowest among the three. In conclusion, even in the presence of severe outliers, the conditional mode still yields more reliable and precise predictions than the conditional mean or median does.

